\documentclass[11pt,a4paper]{article}
\pdfoutput=1

\usepackage[british]{babel}

\usepackage{amsmath}
\usepackage{amssymb}
\usepackage{amsmath,amsthm,epsfig,euscript,array,cancel}
\usepackage[nosort]{cite}
\usepackage{mathtools}
\usepackage{slashed}
\usepackage{bbm}
\usepackage[textwidth = 430 pt, textheight = 630 pt]{geometry}

\usepackage{color}
\definecolor{MyDarkBlue}{rgb}{0.15,0.25,0.45}

\usepackage{mathrsfs}
\usepackage{mathabx}

\usepackage{bm}
\usepackage{braket}
\usepackage{dsfont}

\if{}
\fi

\usepackage[linktocpage=true,hypertexnames=false]{hyperref} %
\hypersetup{
colorlinks=true,
citecolor=MyDarkBlue,
linkcolor=MyDarkBlue,
urlcolor=MyDarkBlue,
pdfauthor={},
pdftitle={},
pdfsubject={}
breaklinks=true
}

\let\fn\footnote
\renewcommand{\footnote}[1]{\linespread{1.1}\fn{#1}\linespread{1.29}}

\makeatletter\renewcommand{\section}{\@startsection
{section}{1}{\z@}{-3.5ex plus -1ex minus
    -.2ex}{2.3ex plus .2ex}{\bf\mathversion{bold} }}
\makeatletter\renewcommand{\subsection}{\@startsection{subsection}{2}{\z@}{-3.25ex
plus -1ex minus
   -.2ex}{1.5ex plus .2ex}{\bf\mathversion{bold} }}
\makeatletter\renewcommand{\subsubsection}{\@startsection{subsubsection}{3}{-2.45ex}{-3.25ex
plus -1ex minus -.2ex}{1.5ex plus .2ex}{\it }}
\renewcommand{\thesection}{\arabic{section}}
\renewcommand{\thesubsection}{\arabic{section}.\arabic{subsection}}
\renewcommand{\@seccntformat}[1]{\@nameuse{the#1}.~~}

\renewcommand{\theequation}{\thesection.\arabic{equation}}
\makeatletter \@addtoreset{equation}{section}

\renewcommand*\l@section{\@dottedtocline{1}{0em}{2em}}
\renewcommand*\l@subsection{\@dottedtocline{2}{2em}{2.4em}}
\renewcommand*\l@subsubsection{\@dottedtocline{4}{3.8em}{3.7em}}

\renewcommand\tableofcontents{%
    \section*{\large\contentsname
        \@mkboth{%
          \MakeUppercase\contentsname}{\MakeUppercase\contentsname}}%
       {\baselineskip=15pt plus 2pt minus 1pt
    \@starttoc{toc}}%
}

\renewenvironment{thebibliography}[1]
     {\baselineskip=16pt plus 2pt minus 1pt
      \section*{\large\refname
        \@mkboth{\MakeUppercase\refname}{\MakeUppercase\refname}}%
     \list{\@biblabel{\@arabic\c@enumiv}}%
           {\settowidth\labelwidth{\@biblabel{#1}}%
            \leftmargin\labelwidth
            \advance\leftmargin\labelsep
            \@openbib@code
            \usecounter{enumiv}%
            \let\p@enumiv\@empty
            \renewcommand\theenumiv{\@arabic\c@enumiv}}%
      \sloppy
      \clubpenalty4000
      \@clubpenalty \clubpenalty
      \widowpenalty4000%
      \sfcode`\.\@m
 \catcode`\^^M=10%
}

\newcommand{\acknowledgements}{\section*{Acknowledgements}
\addcontentsline{toc}{section}{Acknowledgements}}

\newcommand{\datamanagement}{\section*{Data Management}
\addcontentsline{toc}{section}{Data Management}}

\newcommand{\appendices}{
\section*{Appendices}\label{appendices}\setcounter{subsection}{0}
\addcontentsline{toc}{section}{Appendices}
\setcounter{equation}{0}
\makeatletter
\renewcommand{\theequation}{\Alph{subsection}.\arabic{equation}}
\renewcommand{\thesubsection}{\Alph{subsection}}
\@addtoreset{equation}{subsection}
\makeatother
}

\newcommand{\be}{\begin{equation}}
\newcommand{\ee}{\end{equation}}

\def\bea{\begin{eqnarray}}
\def\eea{\end{eqnarray}}

\def\ha{\hat a}
\def\hb{\hat b}
\def\hc{\hat c}
\def\hd{\hat d}

\RequirePackage{tikz} 
\definecolor{darkred}{rgb}{0.5, 0, 0}
\definecolor{darkblue}{rgb}{0, 0, 0.5}

\begin{document}

\begin{titlepage}

\setcounter{page}{0}
\renewcommand{\thefootnote}{\fnsymbol{footnote}}

\vspace*{-2cm}
\begin{flushright}
QGASLAB--15--05\\[-5pt]
DMUS--MP--15/11\\[-5pt]
Imperial--TP--LW--2015--03
\end{flushright}

\vspace{-10pt}

\begin{center}

\textbf{\LARGE\mathversion{bold} T-Duality of Green--Schwarz Superstrings on \\[1mm] AdS$_d \times S^d \times M^{10-2d}$}

\vspace{5mm}

{\large\hspace{-1em} Michael C. Abbott$^a$, Jeff Murugan$^a$, Silvia Penati$^b$, Antonio Pittelli$^c$, Dmitri Sorokin$^d$,\hspace{-1em}\mbox{}\linebreak Per Sundin$^b$, Justine Tarrant$^a$, Martin Wolf$^c$, and Linus Wulff$^e$ %
\footnote{{\it E-mail addresses:\/}
\href{mailto:michael.abbott@uct.ac.za}{\ttfamily michael.abbott@uct.ac.za},
\href{mailto:jeff@nassp.uct.ac.za}{\ttfamily jeff@nassp.uct.ac.za},
\href{mailto:silvia.penati@mib.infn.it}{\ttfamily silvia.penati@mib.infn.it},
\href{mailto:a.pittelli@surrey.ac.uk}{\ttfamily a.pittelli@surrey.ac.uk},
\href{mailto:dmitri.sorokin@pd.infn.it}{\ttfamily dmitri.sorokin@pd.infn.it},
\href{mailto:nidnus.rep@gmail.com}{\ttfamily nidnus.rep@gmail.com},
\href{mailto:tarrant.justine@gmail.com}{\ttfamily tarrant.justine@ gmail.com},
\href{mailto:m.wolf@surrey.ac.uk}{\ttfamily m.wolf@surrey.ac.uk},
\href{mailto:l.wulff@imperial.ac.uk}{\ttfamily l.wulff@imperial.ac.uk}
}}
\vspace{5mm}

{\it
$^a$
Laboratory for Quantum Gravity \& Strings,\\[-1mm]
Department of Mathematics, University of Cape Town\\[-1mm] Rondebosch 7701, Cape Town, South Africa\\[.1cm]

$^b$
Dipartimento di Fisica, Universit\`a degli studi di Milano--Bicocca and INFN, \\[-1mm] Sezione di Milano--Bicocca, Piazza della Scienza 3, 20126 Milano, Italy\\[.1cm]

$^c$
Department of Mathematics, University of Surrey, Guildford GU2 7XH, U.K.\\[.1cm]

$^d$
INFN, Sezione di Padova, via F. Marzolo 8, 35131 Padova, Italy\\[.1cm]

$^e$
The Blackett Laboratory, Imperial College, London SW7 2AZ, U.K.\\[.5cm]
}

\vspace{1mm}

{\bf Abstract}
\end{center}
\vspace{-.3cm}
\begin{quote}
We verify the self-duality of Green--Schwarz supercoset sigma models on AdS$_d \times S^d $ backgrounds ($d=2,3,5$) under combined bosonic and fermionic T-dualities without gauge fixing kappa symmetry. We also prove this property for superstrings on AdS$_d \times S^d \times S^d$ $(d=2,3)$ described by supercoset sigma models with the isometries governed by the exceptional Lie supergroups $D(2,1;\alpha)$ ($d=2$) and $D(2,1;\alpha)\times D(2,1;\alpha)$ ($d=3$), which requires an additional T-dualisation along one of the spheres. Then, by taking into account the contribution of non-supercoset fermionic modes (up to the second order), we provide evidence for the T-self-duality
of the complete type IIA and IIB Green--Schwarz superstring theory on AdS$_d\times S^d \times T^{10-2d}$ ($d=2,3$) backgrounds with Ramond--Ramond fluxes. Finally, applying the Buscher-like rules to T-dualising supergravity fields, we prove the T-self-duality of the whole class of the AdS$_d\times S^d \times M^{10-2d}$ superbackgrounds with Ramond--Ramond fluxes in the context of supergravity.
\vfill
\vspace{3mm}
15th December 2015
\null
\end{quote}

\setcounter{footnote}{0}\renewcommand{\thefootnote}{\arabic{thefootnote}}

\end{titlepage}

\tableofcontents

\bigskip
\bigskip
\hrule
\bigskip
\bigskip


\section{Introduction}

The AdS/CFT correspondence (see {\it e.g.}~\cite{Aharony:1999ti,Maldacena:2003nj,Nastase:2007kj,Beisert:2010jr} for reviews) relates a string theory on a $d$-dimensional anti-de Sitter space (AdS$_d$) to a conformal field theory (CFT) on its $(d-1)$-dimensional conformal boundary, $\partial$AdS$_d$. This correspondence is a particular realization of the more general concept known as holographic duality. In general, the latter relates gravity theories with gauge field theories in a large-$N$ limit \cite{'tHooft:1973jz}, where $N$ is associated with the rank of a gauge group. Powerful tools introduced by holographic duality have found interesting applications, not only in string theory but also in as diverse fields as nuclear and condensed matter physics.

Essentially, the power of holographic duality lies in the fact that it often relates a perturbative regime of a theory on one side of the correspondence to a strongly coupled regime of a theory on the other side and vice versa. This, in turn, allows one to extract information about the behavior of the theories at strong coupling, to which the perturbative methods do not normally apply.

Instances of holographic dualities include the AdS$_d$/CFT$_{d-1}$ correspondences \cite{Maldacena:1997re} for $d=2,3,4,5$. The most developed and the best understood example is the AdS$_5$/CFT$_4$ correspondence between type IIB superstring theory in an AdS$_5\times S^5$ background and the $SU(N)$,  $\mathcal N=4$  supersymmetric Yang--Mills (SYM) theory on the  four-dimensional conformal boundary of AdS$_5$. A striking feature of this correspondence,  based on the 4-dimensional superconformal group $PSU(2,2|4)$, is the integrability of both the AdS$_5\times S^5$ superstring theory and the planar limit ($N\rightarrow \infty$)  of the SYM theory \cite{Minahan:2002ve,Bena:2003wd}. It connects the regime of perturbative gauge theory with the regime of perturbative string theory. Integrability has also been observed in other instances of the AdS/CFT correspondence.

Integrability manifests itself in various features of the theory. For instance, it is believed to be at the core of the relation between planar scattering amplitudes and Wilson loops at strong and weak gauge coupling in the SYM theory, and is related to the existence of a hidden dual superconformal symmetry of gauge theory scattering amplitudes which acts on the momenta as ordinary conformal symmetry acts on coordinates and associates each amplitude to a string worldsheet in a dual AdS space (see \cite{Beisert:2010jr} for a review and references).

On the string theory side, the existence of the dual superconformal symmetry is attributed to the self-duality of the superstring sigma model under (Buscher-like) T-duality transformations of fermionic and bosonic string modes on the worldsheet associated with certain (anti-)commuting isometries of the AdS$_5\times S^5$ background \cite{Berkovits:2008ic,Beisert:2008iq} (see also \cite{Ricci:2007eq}). In turn, this self-duality is an immediate consequence of the important property that the combined bosonic and fermionic T-dualities do not change the values of the AdS$_5\times S^5$ background fields, in particular the Ramond--Ramond flux and the dilaton (see \cite{OColgain:2012si} for review and references).

Fermionic T-duality and its relation to the dual superconformal symmetry are pretty well understood and studied in detail in the case of the AdS$_5\times S^5$ superstring and corresponding dual $\mathcal N=4$ SYM theory~\cite{Berkovits:2008ic,Beisert:2008iq,OColgain:2012si}. However, the manifestation and role in the AdS/CFT correspondence of the fermionic T-duality of the sigma models describing superstrings in less supersymmetric integrable\footnote{The classical integrability of the (full) superstring in these backgrounds has been analyzed in \cite{Babichenko:2009dk,Sorokin:2010wn,Sorokin:2011rr,Cagnazzo:2011at,Sundin:2012gc,Wulff:2014kja} and recently a general construction for all symmetric space Ramond--Ramond backgrounds preserving some amount of supersymmetry was given in \cite{Wulff:2015mwa}.} AdS backgrounds which give rise to other examples of AdS$_d$/CFT$_{d-1}$ correspondence, such as AdS$_2\times S^2\times M^6$, AdS$_3\times S^3\times M^4$ (where $M^{10-2d}$ is a compact manifold, \emph{e.g.} $T^{10-2d}$ or $S^d\times T^{10-3d}$) and, especially, in AdS$_4\times \mathbbm{C}P^3$ are much less understood.

In particular, the AdS$_4\times \mathbbm{C}P^3$ background, which preserves 24 out of 32 supersymmetries of the type IIA superstring theory remains the most challenging case, since it seems to face obstructions in performing the fermionic T-duality of the corresponding superstring sigma model \cite{Adam:2009kt,Grassi:2009yj} and the supergravity background itself \cite{Grassi:2009yj,Bakhmatov:2009be,Godazgar:2010ph,Bakhmatov:2010fp}. On the other hand, results in the dual field theory indicate that the AdS$_4\times \mathbbm{C}P^3$ string model should be self-dual under bosonic and fermionic T-duality transformations. In fact, dual superconformal symmetry appears in the planar amplitude sector of the ABJM model both at the tree level \cite{Huang:2010qy, Gang:2010gy} and at the loop level \cite{Chen:2011vv, Bianchi:2011fc}, Yangian invariance has been observed at the tree level \cite{Bargheer:2010hn}, and the amplitudes/Wilson loop duality has been found up to two loops \cite{Bianchi:2011rn, Bianchi:2011dg}. In \cite{Grassi:2009yj} it was assumed that an obstruction  in performing the fermionic T-duality may be caused by the presence of worldsheet fermionic fields associated with 8 broken supersymmetries \cite{Gomis:2008jt} in the complete superstring Lagrangian. Indeed, the role of the `broken supersymmetry' fermions still needs to be better understood and reconciled with other issues caused by a singularity of the bosonic and fermionic T-duality transformations along $\mathbbm{C}P^3$ isometries, as observed for example in \cite{Adam:2009kt,Bakhmatov:2010fp,Adam:2010hh}.

We will leave aside the AdS$_4\times \mathbbm{C}P^3$ case in this paper, concentrating rather on the study of remaining issues of the T-duality of superstrings on AdS$_d \times S^d \times M^{10-2d}$ backgrounds, with the hope that the better understanding of the latter may also provide new insights into the issues of AdS$_4\times \mathbbm{C}P^3$.

So far T-(self-)duality has been demonstrated for supercoset sigma models associated with strings propagating in AdS$_d \times S^d$ ($d=2,3,5$) upon imposing a partial gauge fixing of the kappa symmetry of the sigma model actions by putting to zero a quarter (complex spinors) of the supercoset fermionic modes \cite{Berkovits:2008ic,Beisert:2008iq,Adam:2009kt}. In \cite{Berkovits:2008ic}, the T-self-duality of the AdS$_5\times S^5$ superstring was demonstrated in the pure spinor formulation, which does not possess  kappa symmetry but is instead BRST invariant. The proof used BRST cohomology arguments to extend the kappa-gauge fixed result of the Green--Schwarz formulation to the whole set of the fermionic modes of the pure spinor string. As was mentioned in  \cite{Berkovits:2008ic}, if the T-dualised Green--Schwarz action could be written in a kappa-invariant form, in order to directly prove the T-self-duality of the pure spinor action, one could use the prescription of \cite{Oda:2001zm} which relates the Green--Schwarz kappa symmetry transformations with the pure spinor BRST transformations.

In the cases of the AdS$_d \times S^d$ supercoset models (with $d=2,3$) an additional issue arises. It is related to the fact that the supercoset models describe only  particular sectors of the complete superstring theories on the AdS$_d \times S^d\times M^{10-2d}$ backgrounds. In the $d=3$ case, these backgrounds preserve 16 of the 32  supersymmetries in ten dimensions, while in the $d=2$ case the number of preserved supersymmetries reduces to 8. Therefore, respectively, only 16 and 8 fermionic modes on the string worldsheet can be associated with the fermionic directions of the corresponding coset superspace, while the remaining 16 and 24 fermionic modes correspond to broken supersymmetries. The supercoset sectors of the theory are non-trivially coupled to the non-supercoset directions $M^{10-2d}$ via these fermionic modes.

In the $d=3$ case, one can use kappa symmetry to put all the 16 non-supercoset (`non-supersymmetric') fermionic modes to zero, but this gauge fixing is not admissible for a wide class of classical string configurations (including those when the string moves only in AdS$_3\times S^3$, \cite{Rughoonauth:2012qd}). Moreover, though the AdS$_3\times S^3$ supercoset sigma model with 16 fermions possesses kappa symmetry with 8 independent parameters (see {\it e.g.}~\cite{Adam:2009kt}), this kappa symmetry is broken when the supercoset model is coupled via the Virasoro constraints to the $T^4$ sector of the complete superstring action in AdS$_3\times S^3 \times T^4$ \cite{Babichenko:2009dk} in which the 16 non-supercoset fermions have already been kappa gauge fixed to zero. In other words, the kappa symmetry of the AdS$_3\times S^3$ supercoset subsector is part of the kappa symmetry of the complete 10-dimensional superstring and is lost when the latter is completely gauge fixed.

 In the $d=2$ case kappa symmetry allows one to remove (for certain classical string solutions) only 16 of the 24 non-supersymmetric fermions, so at least 8 non-supercoset fermionic modes are always present in the AdS$_2\times S^2 \times M^6$ string spectrum (see {\it e.g.}~\cite{Babichenko:2009dk,Sorokin:2011rr} and references therein for the discussion of these issues). In these cases the self T-duality of the corresponding supercoset models has been proved in a (partially) fixed kappa symmetry gauge where some of the fermionic coset coordinates are set to zero \cite{Adam:2009kt}. However, when the supercoset models are used to describe a gauge-fixed sector of the superstring sigma model where kappa symmetry has already been used to remove (part of) the non-supersymmetric fermions, one cannot use kappa symmetry anymore for proving the  self T-duality of the corresponding supercoset sectors of the AdS$_d \times S^d\times M^{10-2d}$ superstrings.

In view of the above mentioned issues, it is important to demonstrate explicitly  the T-self-duality of superstring theory on the AdS$_d \times S^d\times M^{10-2d}$ backgrounds without fixing kappa symmetry and taking into account the non-supercoset fermionic modes. This is the main goal of this paper. Specifically, we verify the combined bosonic and fermionic T-self-duality of Green--Schwarz supercoset sigma models on AdS$_d \times S^d $ backgrounds ($d=2,3,5$)  without fixing a kappa symmetry gauge. Furthermore, we prove the same for AdS$_d \times S^d \times S^d$ backgrounds ($d=2,3$) described by supercoset sigma models with the isometries governed by the exceptional supergroups $D(2,1;\alpha)$ (for $d=2$) and $D(2,1;\alpha)\times D(2,1;\alpha)$ (for $d=3$). In these supercoset models (which, by the way, do not possess kappa symmetry), in order to map the dualised actions to the original ones, the T-dualisation of $d-1$ directions in AdS$_d$ and of $2(d-1)$ fermionic directions should be accompanied by T-dualisation of (complexified) $d-1$ directions of one sphere.

We also prove the T-(self-)duality of complete type IIA and IIB Green--Schwarz superstring actions on AdS$_3\times S^3 \times T^4$ and AdS$_2\times S^2 \times T^6$ backgrounds with different Ramond--Ramond fluxes, by taking into account (up to the second order) the contribution of their non-supercoset fermionic modes. An important consequence of the presence of these non-supersymmetric fermions is that for the actions to be invariant under the combined fermionic and bosonic T-duality transformations, the latter should involve the dualisation of half of the torus directions. This is in accordance with results of \cite{OColgain:2012ca} in which the combined bosonic-fermionic T-duality of some of AdS$_d\times S^d \times M^{10-2d}$ superbackgrounds was studied from the supergravity perspective.

\vspace{1cm}
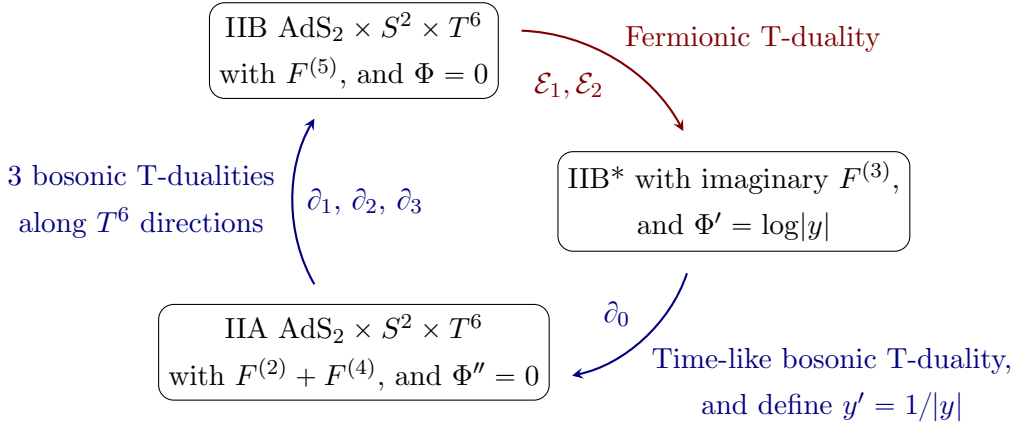
\begin{figure}[h]
\centering

\newcommand{\cE}{\mathcal{E}}
\begin{tikzpicture}[scale=1, rounded corners=2mm, auto, bend left, shorten >= 3mm, shorten <= 3mm, >=stealth]

\node (top) [rectangle,draw,align=center] at (0,4) {IIB AdS$_2\times S^2\times T^6$ \\ with $F^{(5)}$, and $\Phi=0$};

\node (right) [rectangle,draw,align=center] at (5,2) {IIB* with imaginary $F^{(3)}$, \\ and $\Phi'=\log \lvert y\rvert$};

\node (bot) [rectangle,draw,align=center] at (0,0) {IIA AdS$_2\times S^2\times T^6$ \\ with $F^{(2)} + F^{(4)}$, and $\Phi''=0$};

\draw [->,thick,darkred] (top) to node {Fermionic T-duality} node [swap] {$\cE_1,\cE_2$} (right);

\draw [->,thick,darkblue] (right) to node [align=center] {Time-like bosonic T-duality, \\ and define $y'=1/\lvert y\rvert$} node [swap] {$\partial_0$} (bot);

\draw [->,thick,darkblue] (bot) to node [align=center] {3 bosonic T-dualities \ \\ along $T^6$ directions \ } node [swap] {$\partial_1$, $\partial_2$, $\partial_3$} (top);

\end{tikzpicture}

\begin{minipage}{14cm}
\caption{\small The idea of self-duality we study is that a sequence of fermionic
and bosonic T-dualities returns us to the same background. This is
depicted here for the case in which we start with type IIB AdS$_{2}\times S^{2}\times T^{6}$
supported by $F^{(5)}$ Ramond--Ramond flux, in which the bosonic duality is along
the only boundary direction of AdS and along three torus directions.
The other cases are similar, although the number of Killing spinors
varies, and for AdS$_{d}\times S^{d}\times S^{d}\times T^{10-3d}$
cases some of the four bosonic dualities are along complexified Killing
vectors of one sphere.\label{fig:IIB-diagram}}
\end{minipage}
\vspace{1cm}
\end{figure}

In this respect, for completeness, in Section \ref{sugra} we extend the results of \cite{Berkovits:2008ic,OColgain:2012ca} (see also \cite{Bakhmatov:2009be,Godazgar:2010ph,Bakhmatov:2010fp} for the AdS$_4\times\mathbbm{C}P^3$ case) and prove (using the T-duality rules \cite{Buscher:1987sk,Buscher:1987qj,Simon:1998az}) the invariance under the combined T-duality of the  whole class of the AdS$_d\times S^d \times T^{10-2d}$ superbackgrounds with Ramond--Ramond fluxes (see Figure \ref{fig:IIB-diagram}). The combined T-duality involves the directions along the $(d-1)$-dimensional Minkowski boundary, half of the $T^d$ torus directions and $2(d-1)$ complex fermionic T-dualities. For the AdS$_d\times S^d \times S^d \times T^{10-3d}$ cases we also find it necessary to perform bosonic T-duality along complexified directions of one sphere.

\section{General setup}\label{generic}

In this section, we recall some basic facts about  superstring sigma models and their T-dualisation.

{The conventional form of the Green--Schwarz action describing the propagation of a superstring in a generic 10-dimensional type II background is \cite{Grisaru:1985fv}
\be\label{GSaction}
S\ =\ -\tfrac{T}{2}\int_\Sigma (*\mathcal E^A\wedge\mathcal E^B\eta_{AB}+2\kappa B_2)~.
\ee
Here, $T$ denotes the string tension and $\Sigma$ is a 2-dimensional worldsheet with a curved metric $h_{pq}(\tau, \sigma)$ of Lorentz signature so that the corresponding worldsheet Hodge duality operation $*$ squares to one ($*^2=1$) when acting on one--forms.\footnote{Explicitly, in local coordinates $(\tau,\sigma)$ on $\Sigma$, $ *\mathcal E^A\wedge\mathcal E^B=\sqrt{-\det(h_{rs})}\, h^{pq} \mathcal E^A_p \mathcal E^B_q$.} The $\mathcal E^A=\mathcal E^A(X,\Theta)$ with $A,B,\ldots=0,\ldots,9$ are vector supervielbeins where $(X,\Theta)$ are target space coordinates (10 Gra{\ss}mann-even (bosonic) coordinates $X$ and 32 Gra{\ss}mann-odd (fermionic) coordinates $\Theta$) and  $(\eta_{AB})={\rm diag}(-1,1,\ldots,1)$ is the 10-dimensional target tangent space Minkowski metric. In addition to $\mathcal E^A=\mathcal E^A(X,\Theta)$, the description of the geometry also involves spinor supervielbeins  $\mathcal E^{\hat\alpha}=\mathcal E^{\hat\alpha}(X,\Theta)$ with $\hat\alpha,\hat \beta,\ldots=1,\ldots,32$. Furthermore, $B_2(X,\Theta)$ is the worldsheet pullback of the Neveu--Schwarz--Neveu--Schwarz 2-form gauge superfield. In the models in which we are interested, it has vanishing field strength at $\Theta=0$, that is, ${\rm d}B_2|_{\Theta=0} =0$. Kappa symmetry invariance requires the coupling constant $\kappa$ to be $\pm1$. In what follows, we shall choose $\kappa=1$. Note that for generic supergravity backgrounds, the action \eqref{GSaction} is known explicitly up to fourth order in $\Theta$ \cite{Wulff:2013kga}.}

We will be interested in (bosonic) symmetric space backgrounds of the type AdS$_d \times S^d \times T^{10-2d}$, $d=2,3,5$ and AdS$_d \times S^d \times S^d\times T^{10-3d}$, $d=2,3$.
As shown in \cite{Sorokin:2011rr} for $d=2$ and in \cite{Wulff:2015mwa} in general, the full type II superspace corresponding to these backgrounds contains a sub-superspace which is a supercoset space $G/H=\{gH\,|\,g\in G\}$, $G$ being the superisometry group and $H$ the isotropy subgroup of the background in question. For a background with no Neveu--Schwarz--Neveu--Schwarz flux, $G/H$ is, in fact, a so-called semi-symmetric superspace meaning that the Lie algebra of $G$ admits a $\mathbbm Z_4$-automorphism $\Omega:G\to G$ whose fixed point set is $H$, that is, $\Omega^4=1$ and $\Omega(H)=H$. Correspondingly, there exists a truncation of the Green--Schwarz string action to a supercoset sigma model. If the background admits at least 16 supersymmetries, this sigma model can be viewed as a kappa symmetry gauge fixing of the full superstring (for configurations where this gauge fixing is consistent). Below we give the coset superspaces relevant for our discussion.

\paragraph{\mathversion{bold}$\mathbbm{Z}_4$-graded coset superspaces.}
For the AdS$_d \times S^d\times T^{10-2d}$ backgrounds, we have the following supercosets\footnote{Note that since all these coset superspaces are smooth supermanifolds, they naturally fiber over their bosonic part, that is, they are smooth vector bundles with bosonic base and fermionic fibers.}
\begin{subequations}\label{eq:Cosets}
\be\label{SC}
\begin{gathered}
d=5: \quad \frac{PSU(2,2|4)}{SO(1,4)\times SO(5)}\ \hat =\ {\rm AdS}_5 \times S^5\quad + \quad 32 ~{\rm fermionic~directions}~,\\[5pt]
d=3: \quad \frac{PSU(1,1|2)\times PSU(1,1|2)}{SU(1,1)\times SU(2)}\ \hat =\  {\rm AdS}_3 \times S^3\quad + \quad 16 ~{\rm fermionic ~directions}~,\\[5pt]
d=2: \quad \frac{PSU(1,1|2)}{SO(1,1)\times U(1)}\ \hat =\ {\rm AdS}_2 \times S^2\quad + \quad 8 ~{\rm fermionic ~directions}~.
\end{gathered}
\ee
while for  AdS$_d \times S^d\times S^d\times T^{10-3d}$, we deal with
\be\label{eq:supercosetsD21}
\begin{gathered}
d=3: \quad \frac{D(2,1;\alpha)\times D(2,1;\alpha)}{SO(1,2)\times SO(3)\times SO(3)}\ \hat =\  {\rm AdS}_3 \times S^3\times S^3\quad + \quad 16 ~{\rm fermionic ~directions}~,\\[5pt]
d=2: \quad \frac{D(2,1;\alpha)}{SO(1,1)\times SO(2)\times SO(2)}\ \hat =\ {\rm AdS}_2 \times S^2\times S^2\quad + \quad 8 ~{\rm fermionic ~directions}~,
\end{gathered}
\ee
\end{subequations}
where $0\leq\alpha\leq 1$. Note that while for $d=5$ the coset superspace describes the full superstring theory, for $d=2,3$, the listed coset superspaces describe only those subsectors of the full superstring theories in which the non-supersymmetric fermions have been removed by truncation/gauge-fixing and the string does not fluctuate along the torus directions.

\paragraph{Maurer--Cartan form.}
The $\mathbbm{Z}_4$-automorphism $\Omega:G\to G$ induces a corresponding automorphism on the Lie superalgebra $\mathfrak{g}$ of $G$, which we shall again denote by $\Omega:\mathfrak{g}\to\mathfrak{g}$ (see {\it e.g.}~\cite{Serganova:1983vp} for a classification). We therefore have a decomposition $\mathfrak{g}\otimes\mathbbm{C}\cong\bigoplus_{m=0}^3\mathfrak{g}_{(m)}$ into the eigenspaces of $\Omega$, that is, $\Omega(V_{(m)})={\rm i}^m V_{(m)}$ for $V_{(m)}\in \mathfrak{g}_{(m)}$. In addition, $[\mathfrak{g}_{(m)},\mathfrak{g}_{(n)}]\subseteq \mathfrak{g}_{(m+n \mbox{ \footnotesize mod } 4)}$ and $\mathfrak{g}_{(0)}$ is the Lie algebra of $H$. Furthermore, $\mathfrak{g}$ comes with a $\mathbbm{Z}_2$-grading, and the generators of $\mathfrak{g}_{(0)}$ and $\mathfrak{g}_{(2)}$ are bosonic while the generators of  $\mathfrak{g}_{(1)}$ and $\mathfrak{g}_{(3)}$ are fermionic. For various general properties of the Lie superalgebras associated with the Lie supergroups appearing in \eqref{eq:Cosets}, we refer the reader to {\it e.g.}~\cite{Frappat:1996pb}.

Next, we consider maps $g:\Sigma\to G$ from a 2-dimensional worldsheet Riemann surface $\Sigma$ with an (arbitrarily chosen) Lorentzian metric  into $G$ and  introduce the (pull-back to $\Sigma$ via $g$ of the) Maurer--Cartan form
\be\label{eq:J}
J\ :=\ g^{-1}{\rm d}g~.
\ee
Here, ${\rm d}$ denotes the exterior derivative on $\Sigma$.\footnote{In our conventions, d acts from the right.} By construction, the $\mathfrak{g}$-valued differential 1-form $J$ is invariant under global left $G$-transformations $g\mapsto g_0 g$ for $g_0\in G$ and satisfies the Maurer--Cartan equation, ${\rm d}J - J \wedge J =0$. Using the $\mathbbm{Z}_4$-automorphism $\Omega:\mathfrak{g}\to\mathfrak{g}$, we may decompose $J$ into the eigenspaces of $\Omega$ according to
\be\label{eq:GenCosCur}
J\ =\ J_{(0)}+J_{(1)}+J_{(2)}+J_{(3)}\quad\mbox{with}\quad \Omega\,(J_{(m)})\ =\ {\rm i}^m J_{(m)}.
\ee
It is then straightforward to check that under local right $H$-transformations $g\mapsto gh$ for $h\in H$, the part $J_{(0)}$ behaves as a $\mathfrak{g}_{(0)}$-valued connection 1-form, $J_{(0)}\mapsto h^{-1}J_{(0)} h+h^{-1}{\rm d}h$, while the $J_{(m)}$s for $m=1,2,3$ transform adjointly, $J_{(m)}\mapsto h^{-1}J_{(m)} h$. Since the physical fields will take values in the coset superspace $G/H=\{gH\,|\,g\in G\}$ for \eqref{eq:Cosets}, the corresponding action must be invariant under such  local right $H$-transformations. This, in turn, implies that the action will involve only the $J_{(m)}$ for $m=1,2,3$. Correspondingly, $G/H$ is parametrised by $d_{\rm b}$ bosonic local coordinates $\mathbb X$  and $d_{\rm f}$ fermionic local coordinates $\vartheta$,  where $d_{\rm b}+d_{\rm f}:=\dim(G/H)=\dim(G)-\dim(H)$, so that we will be dealing with maps $({\mathbb X},\vartheta):\Sigma\to G/H$. Furthermore, $J_{(2)}$ play the role of bosonic supervielbeins while $J_{(1)}$ and $J_{(3)}$ play the role of fermionic supervielbeins.

\paragraph{Supercoset action.}
The supercoset string action for a $\mathbbm{Z}_4$-graded $G/H$ coset superspace is constructed from the 1-forms $J_{(m)}$  for $m=1,2,3$, and it has the following form (see \cite{Metsaev:1998it,Rahmfeld:1998zn,Zhou:1999sm,Berkovits:1999zq,Arutyunov:2008if,Stefanski:2008ik,Babichenko:2009dk} and references therein)
\begin{equation}\label{GHaction0}
\begin{aligned}
S\ =\ -T\int_\Sigma\mathcal L_{G/H} \ &=\ -\tfrac{T}{2}\int_\Sigma {\rm Str}({*J_{(2)}}\wedge {J_{(2)}}+ J_{(1)}\wedge  J_{(3)})~.
\end{aligned}
\end{equation}
where Str denotes the supertrace compatible with the $\mathbbm{Z}_4$-grading,
\be
{\rm Str}(V_{(m)}V_{(n)})\ =\ 0\quad{\rm for}\quad V_{(m)}\ \in\ \mathfrak{g}_{(m)}\quad\mbox{and}\quad m+n\ \neq\ 0\mbox{ mod } 4~,
\ee
As in \eqref{GSaction}, in the non-exceptional cases the relative coefficient of the two terms in \eqref{GHaction0} is fixed by kappa symmetry, while in the exceptional cases the action \eqref{GHaction0} is not kappa symmetry invariant \cite{Zarembo:2010sg,Dekel:2011qw}. In the latter cases, the relative coefficient gets fixed by their relation to the original Green--Schwarz action and/or by integrability of the sigma-models.  Clearly, the action \eqref{GHaction0} is invariant under rigid left $G$-transformations and local right $H$-gauge transformations. The Wess--Zumino term (the second term in this action) was first given in the above form in \cite{Berkovits:1999zq}. Comparison with the Green--Schwarz action \eqref{GSaction} tells us that  $B_2 = \frac12 {\rm Str}(J_{(1)}\wedge J_{(3)})$.

\paragraph{Schematic form of the superconformal algebra.}
The T-dualisation of the action \eqref{GHaction0} is performed along certain bosonic and fermionic directions of the $G/H$ supercoset which correspond to an (anti-)commuting (that is, Abelian) subgroup of the isometries of the underlying coset superspace. To identify these isometries, one chooses a basis of the Lie superalgebra $\mathfrak{g}$ of $G$ which is associated with the superconformal group on the Minkowski (conformal) boundary $\mathbbm{R}^{1,d-2}$ of the AdS$_d$ space. In this basis, $\mathfrak{g}$ is described schematically as follows. The bosonic conformal algebra and the $R$-symmetry on $\mathbbm{R}^{1,d-2}$ are given by (we only display non-vanishing commutators)
\bea\label{conf}
\begin{gathered}
 [P,K]\ \sim\  D+M~,\\
  [D,P]\ \sim\ P~,\quad [D,K]\ \sim\ K~,\quad
[M,P]\ \sim\ P~,\quad[M,K]\ \sim \ K~,\\
[M,M]\ \sim\ M~,\quad
[R,R]\ \sim\ R~,
\end{gathered}
\eea
where $P$ are the $(d-1)$ translation generators, $M$ are the $\frac12(d-1)(d-2)$ Lorentz generators, $K$ are the $(d-1)$ conformal boost generators, and $D$ is the dilatation generator. In the AdS$_d\times S^d$ case the $R$-symmetry generators $R$ are associated with the $SO(d+1)$ isometries of $S^d$, while in the case of AdS$_d\times S^d\times S^d$ they correspond to the $SO(d+1)\times SO(d+1)$ isometries of $S^d\times S^d$.

The superconformal extension of the algebra \eqref{conf} contains the fermionic generators $Q$, $\hat Q$, $S$, and $\hat S$ which are  the complex supersymmetry and superconformal generators  related by Hermitian conjugation (the specific form of the conjugation rules depends on the chosen superalgebra), each being $2(d-1)$-dimensional. The additional non-vanishing (anti-)commutation relations have the following schematic form
\begin{subequations}\label{eq:SCext}
\be\label{qs}
\begin{gathered}
[D,Q]\ \sim \  Q~,\quad [M,Q]\ \sim\ Q~,\quad  [K,Q]\ \sim\  \hat S~, \quad [R,Q]\ \sim\ Q+\alpha\hat Q~,\\
[D,S]\ \sim \ S~,\quad [M,S]\ \sim\ S~,\quad  [P,S]\ \sim\ \hat Q~, \quad [R,S]\ \sim\ S+\alpha\hat S~,
\end{gathered}
\ee
and similarly for $\hat Q$ and $\hat S$, plus
\be\label{qhats}
\begin{gathered}
\{Q,\hat Q\}\ \sim\ P~, \quad \{S,\hat S\}\ \sim\ K~,\quad\{Q,\hat S\}\ \sim\ \alpha R~,\quad \{\hat Q,S\}\ \sim\ \alpha R\\
\{Q,S\}\ \sim\ D+M+R~,\quad  \{\hat Q,\hat S\}\ \sim\ D+M+R~.
\end{gathered}
\ee
\end{subequations}
In these relations, $\alpha$ is the parameter appearing in the coset superspaces \eqref{eq:supercosetsD21}. Note that the $d=2,3$ coset superspaces in \eqref{SC} are obtained from those in \eqref{eq:supercosetsD21} by taking the limit $\alpha\to 0$. Hence, in the case of the AdS$_d\times S^d\times T^{10-2d}$ backgrounds  we simply set $\alpha=0$.

In summary, the Lie superalgebra $\mathfrak{g}$ is generated by $\mathfrak{g}=\langle P,K,D,M,R,Q,\hat Q,S,\hat S\rangle$ and described by the (anti-)commutation relations \eqref{conf} and \eqref{eq:SCext}.

\paragraph{Choice of $\mathbbm{Z}_4$-grading.}
As we shall see below, the specific choice of a $\mathbbm{Z}_4$-grading and its superconformal splitting onto Abelian sub-isometries are crucial when performing the T-duality transformations --- an inappropriate choice would make the proof of the self-duality of the complete superstring actions much more complicated if at all possible. Decomposing the $R$-symmetry generators $R$ as $R=(R_{(0)},R_{(2)})$ with $R_{(0)}\in\mathfrak{g}_{(0)}$ and $R_{(2)}\in\mathfrak{g}_{(2)}$, the $\mathbbm{Z}_4$-grading we shall be using is formally of the form
\be \label{eq:GenZ4}
\begin{gathered}
 \mathfrak{g}_{(0)}\ :=\ \langle P+K,M,R_{(0)}\rangle~,\quad
 \mathfrak{g}_{(2)}\ :=\ \langle P- K,D,R_{(2)}\rangle~,\\
 \mathfrak{g}_{(1)}\ :=\ \langle Q-S,\hat Q - \hat S\rangle~,\quad
 \mathfrak{g}_{(3)}\ :=\ \langle Q+S,\hat Q + \hat S\rangle~.
 \end{gathered}
\ee
We emphasize that the specific form of the decomposition $R=(R_{(0)},R_{(2)})$ will depend on the particular form of the superconformal algebra, and we shall say a few things about this in the next paragraph.

\paragraph{Coset representative and associated current.}
In the AdS$_d\times S^d\times T^{10-2d}$ case, the form of the superalgebra \eqref{conf} and \eqref{eq:SCext} (with $\alpha=0$) implies that the $(d-1)$ generators $P$ and  $2(d-1)$ complex supercharges $Q$ are in involution, and, hence, a maximal Abelian subalgebra of $\mathfrak{g}$ is simply $\langle P,Q\rangle$. Thus, the  (anti-)commuting isometries of the $G/H$ Green--Schwarz sigma model can be associated with $\langle P,Q\rangle$.

In the AdS$_d\times S^d\times S^d$ case, the situation is somewhat more complicated, and as we shall see in Section \ref{sec:d21}, a maximal Abelian subalgebra of $\mathfrak{g}$ is again generated by $P$ and $Q$ but also by some of the $R$-symmetry generators which we denote formally by $L_+$. To jump ahead of our story a bit, we will have one complex generator $L_+\equiv L_+^1$ for $d=2$ and two complex generators $L_+\equiv L_+^{1,2}$ for $d=3$. Hence, the (anti-)commuting isometries are associated with $\langle P,Q,L_+\rangle$ in this case. In the following, we shall denote the Hermitian conjugate of $L_+$ by $L_-$ and we have $[L_+^1,L_-^1]\sim  L_3 \sim [L_+^2,L_-^2]$. In view of the $\mathbbm{Z}_4$-grading \eqref{eq:GenZ4}, it turns out that $L_+^1+L_-^1\in\mathfrak{g}_{(0)}$, $L_+^1-L_-^1\in\mathfrak{g}_{(2)}$, $L_+^2-L_-^2\in\mathfrak{g}_{(0)}$, $L_+^2+L_-^2\in\mathfrak{g}_{(2)}$, and $L_3\in\mathfrak{g}_{(2)}$ will be the appropriate choice. See Section \ref{sec:d21} for details.

Motivated by this discussion, to perform the T-dualisation of the action \eqref{GHaction0} along these isometries, it is convenient to take the supercoset representative $g$ in a form similar to that of \cite{Berkovits:2008ic,Beisert:2008iq,Dekel:2011qw}
\begin{equation}\label{g}
g\ :=\ {\rm e}^{xP+\theta Q+{\sqrt \alpha}\, \lambda_+ L_+}{\rm e}^B{\rm e}^{\xi S}~, \quad
{\rm e}^B\ :=\ {\rm e}^{\hat\theta\hat Q+\hat\xi\hat S}|y|^D{\rm e}^{-{\sqrt \alpha}\,\lambda_3 L_3} \Lambda_{\alpha }(y)~,
\end{equation}
where $x$ are the coordinates of the Minkowski boundary and $|y|$ is associated with the radial (bulk) direction in AdS$_d$. In the AdS$_d \times S^d\times T^{10-2d}$ case ($\alpha =0$) the coordinates $y$ parametrize $S^d$, whereas in the AdS$_d \times S^d \times S^d$ background ($\alpha  \neq 0$) one $S^d$ is parametrized by $y$ and the second one is described by $\lambda_+$ and $\lambda_3$. The latter coordinates are assumed to be complex (we will explain this in more detail in Section \ref{sec:d21}). Moreover, the specific form of $\Lambda_{\alpha }=\Lambda_{\alpha }(y)$ will depend on the chosen background. The set of $2(d-1)$ (complex conjugate) fermionic coordinates $(\theta,\hat\theta,\xi,\hat\xi)$ parametrize the Gra{\ss}mann-odd directions of the coset superspace. In order to achieve the form \eqref{g} of the representative, we have employed local right $H$-transformations, and since $P$, $Q$, and $L_+$ are in involution, this choice of the representative ensures that the action \eqref{GHaction0} will depend on $x$, $\theta$, and $\lambda_+$ only through their derivatives ${\rm d}x$, ${\rm d}\theta$, and ${\rm d}\lambda_+$.

So far, the proof of self-duality of supercoset sigma models \eqref{GHaction0} under bosonic and fermionic T-duality has been performed in a fixed kappa symmetry gauge, the most convenient choice being $\xi=0$ \cite{Berkovits:2008ic,Beisert:2008iq,Dekel:2011qw}. However, as already explained, if the supercoset model describes the gauge-fixed version of the corresponding superstring action, the kappa symmetry has been already used to (partially) gauge away the non-supersymmetric fermions, and cannot be used once again in the T-dualisation procedure. Moreover, for sigma models based on the exceptional Lie supergroups \eqref{eq:supercosetsD21}, the rank of the kappa symmetry is zero \cite{Babichenko:2009dk,Zarembo:2010sg,Dekel:2011qw} and one cannot put any of the fermionic coordinates to zero. Therefore, in what follows we are not going to (partially) fix kappa symmetry to get rid of some of the fermionic coordinates. All the fermionic coordinates in \eqref{g} will be taken into account.

\vskip 10pt
In the realization \eqref{g} of the coset element, the current \eqref{eq:J} has the following form
\be\label{Jg}
J\ =\ g^{-1}{\rm d}g\ =\ {\rm e}^{-\xi S}J^{(0)}{\rm e}^{\xi S}+{\rm d}\xi S~,
\ee
where $J^{(0)}$ is the current at $\xi=0$.  Writing the currents $J$ and $J^{(0)}$ as
\be \label{currentcomponents}
J\ =\ J_PP+J_KK+\cdots \qquad {\rm and}  \qquad  J^{(0)}\ =\ J_P^{(0)}P+J_K^{(0)}K+\cdots
\ee
the components of $J^{(0)}$ are given by
\begin{subequations}\label{eq:FormalCurrents}
\be
\begin{gathered}
J_{P}^{(0)}\ =\ \big[{\rm e}^{-B}({\rm d}x P+{\rm d}\theta Q+{\sqrt \alpha}\,{\rm d}\lambda_+ L_+){\rm e}^{B}\big]_{P}~,\quad
J_{K}^{(0)}\ =\ 0~,\\
J_{D}^{(0)}\ =\ \big[{\rm e}^{-B}{\rm de}^B\big]_{D}~,\quad
J_{M}^{(0)}\ =\ \big[{\rm e}^{-B}{\rm de}^B\big]_{M}~,\\
J_{R}^{(0)}\ =\ \big[{\rm e}^{-B}{\rm de}^B\big]_{R}~,\\
J_{L_+}^{(0)}\ =\ \big[{\rm e}^{-B}({\rm d}x P+{\rm d}\theta Q+{\sqrt \alpha}\,{\rm d}\lambda_+ L_+){\rm e}^{B}\big]_{L_+}~,\quad
J_{L_-}^{(0)}\ =\ 0~,\quad J_{L_3}^{(0)}\ =\ \big[{\rm e}^{-B}{\rm de}^B\big]_{L_3}~,\\
J_{Q}^{(0)}\ =\ \big[{\rm e}^{-B}({\rm d}x P+{\rm d}\theta Q+{\sqrt \alpha}\, {\rm d}\lambda_+L_+){\rm e}^{B}\big]_{Q}~,\quad
J_{\hat Q}^{(0)}\ =\ \big[{\rm e}^{-B}{\rm de}^B\big]_{\hat Q}~,\\
J_{S}^{(0)}\ =\ 0~,\quad
J_{\hat S}^{(0)}\ =\ \big[{\rm e}^{-B}{\rm de}^B\big]_{\hat S}~,
\end{gathered}
\ee
where $[\cdots]_{P}$ etc.~indicates the projection onto the generators $P$ etc., while the components of $J$ read schematically as
\be
\begin{gathered}
 J_P\ =\ J_{P}^{(0)}~,\quad J_Q\ =\ J_Q^{(0)}~,\quad J_{L_+}\ =\ J_{L_+}^{(0)}~,\\
 J_D\ =\ J_D^{(0)}+J_Q^{(0)}\xi~,\quad J_M\ =\ J_M^{(0)}+J_Q^{(0)}\xi~,\\
  J_R\ =\ J_R^{(0)}+J_Q^{(0)}\xi~,\quad J_{L_3}\ =\ J_{L_3}^{(0)}+\alpha  J_Q^{(0)}\xi~,\\
 J_{\hat Q}\ =\ J_{\hat Q}^{(0)}+J_P^{(0)}\xi~,\quad J_{\hat S}\ =\ J_{\hat S}^{(0)}+\alpha  J_{L_+}^{(0)}\xi~,\\
 J_K\ =\ J_{\hat S}^{(0)}\xi+\alpha  J_{L_+}^{(0)}\xi^2~,\quad J_{L_-}\ =\ \alpha  J_{\hat Q}^{(0)}\xi+\alpha  J_P^{(0)}\xi^2~,\\
 J_S\ =\ {\rm d}\xi+(J_D^{(0)}+J_M^{(0)}+J_R^{(0)}+\alpha  J_{L_3}^{(0)})\xi+J_Q^{(0)}\xi^2~.
\end{gathered}
\ee
\end{subequations}
We note that thanks to the appropriate choice of the coset representative \eqref{g}, the current $J$ depends on $\xi$ at most quadratically. This will drastically simplify the T-dualisation procedure.

Finally, decomposing the current $J$ according to the $\mathbbm{Z}_4$-grading \eqref{eq:GenZ4}, the supercoset sigma model action \eqref{GHaction0} takes the following schematic form
\begin{equation}\label{GHaction1}
\begin{aligned}
S\ &=\ -\tfrac T2 \int_\Sigma \Big[*(J_{P}- J_{K})\wedge(J_{P}- J_{K})+*J_D\wedge J_D+*J_{R_{(2)}}\wedge J_{R_{(2)}}\,+\\
&\kern1.5cm+\sum_{k=1}^{d-1}{*(J_{L_+^k}+(-)^k J_{L_-^k})}\wedge(J_{L_+^k}+(-)^k J_{L_-^k})+ *J_{L_3}\wedge J_{L_3}\,+\\[5pt]
&\kern2.5cm+\big(J_Q\wedge \gamma J_Q-J_{\hat S}\wedge  \gamma J_{\hat S}-J_{S}\wedge \gamma  J_{S}+J_{\hat Q}\wedge \gamma  J_{\hat Q}\big)\big]~.
\end{aligned}
\end{equation}
Here,  $\gamma$ in the last four terms stands for a constant symmetric matrix being part of the $G$-invariant bilinear form contracting the spinor indices of the fermionic currents. The form of this matrix is related to the value of Ramond--Ramond fluxes supporting the corresponding AdS$_d\times S^d\times M^{10-2d}$ background.

\paragraph{T-duality procedure.}
In the $\alpha =0$ case, we T-dualise the action \eqref{GHaction1} along $x$ and $\theta$, following the discussion of \cite{Berkovits:2008ic,Beisert:2008iq}. For the $\alpha \neq 0$ case, we dualise also along $\lambda_+$, following ideas of  \cite{Dekel:2011qw}.

According to the standard  procedure \cite{Buscher:1987sk,Buscher:1987qj,Simon:1998az}, starting from the action \eqref{GHaction1} we first make the substitution $({\rm d}x,{\rm d}\theta,{\rm d}\lambda_+)\mapsto(A_{\rm b},A_{\rm f},A_+)$  and modify it according to
\begin{equation}\label{eq:GeneralFOA}
\begin{aligned}
  S\ \mapsto\ S_{\rm f.o.}\ & =\ S[({\rm d}x,{\rm d}\theta,{\rm d}\lambda_+)\ \mapsto\ (A_{\rm b},A_{\rm f},A_+)]+\int_\Sigma \big(\tilde x {\rm d}A_{\rm b}+\tilde \theta {\rm d}A_{\rm f}+{\sqrt \alpha}\,  \tilde \lambda_+ {\rm d}A_+\big)
  \end{aligned}
\end{equation}
Here, $\{A_{\rm b},A_{\rm f},A_+\}$ are auxiliary differential 1-forms and $\{\tilde x,\tilde\theta,\tilde\lambda_+\}$ are Lagrange multipliers. The latter enforce  the constraints ${\rm d}A_{\rm b}=0$, ${\rm d}A_{\rm f}=0$, and ${\rm d}A_+=0$ or, equivalently,  $A_{\rm b}={\rm d}x$, $A_{\rm f}={\rm d}\theta$, and $A_+={\rm d}\lambda_+$. Consequently, upon integrating $\{\tilde x,\tilde\theta,\tilde\lambda_+\}$ out, we recover the original action \eqref{GHaction1}.

To derive the dualised action $\tilde S$, we instead need to integrate out the differential 1-forms $\{A_{\rm b},A_{\rm f},A_+\}$. Once done, the Lagrange multipliers $\{\tilde x,\tilde\theta,\tilde\lambda_+\}$ shall be interpreted as T-dual coordinates. In order to perform this operation, we make a simplification by noticing that
\begin{equation}\label{A'}
  {\rm e}^{-B} \big(A_{\rm b} P +A_{\rm f} Q+{\sqrt \alpha}\,  A_+L_+\big) {\rm e}^{B}\ =\ A'_{\rm b} P +A'_{\rm f} Q+{\sqrt \alpha}\,  A'_+L_+~
\end{equation}
since the Abelian algebra $\langle P,Q,L_+\rangle$ is invariant under conjugation by the group element ${\rm e}^{B}$. Equivalently,
\begin{equation}
A_{\rm b} P +A_{\rm f} Q+{\sqrt \alpha}\,  A_+L_+\ =\   {\rm e}^{B} \big(A'_{\rm b} P +A'_{\rm f} Q+{\sqrt \alpha}\, A'_+ L_+\big) {\rm e}^{-B}~.
\end{equation}
Thus, we may consider the field re-definition $(A_{\rm b},A_{\rm f},A_+)\mapsto (A'_{\rm b},A'_{\rm f},A'_+)$. Note that the on-shell relations ${\rm d}x=A_{\rm b}$, ${\rm d}\theta=A_{\rm f}$, and ${\rm d}\lambda_+=A_+$ together with  \eqref{eq:FormalCurrents} imply the on-shell relations $A'_{\rm b}=J_P$, $A'_{\rm f}=J_{Q}$, and $A'_+=J_{L_+}$.

Upon substituting
\be
\begin{aligned}
 A_{\rm b}\ &=\ \big[{\rm e}^{B} \big(A'_{\rm b} P +A'_{\rm f} Q+{\sqrt \alpha}\, A'_+ L_+\big) {\rm e}^{-B}\big]_P~,\\
 A_{\rm f}\ &=\ \big[{\rm e}^{B} \big(A'_{\rm b} P +A'_{\rm f} Q+{\sqrt \alpha}\, A'_+ L_+\big) {\rm e}^{-B}\big]_Q~,\\
 A_+\ &=\ \big[{\rm e}^{B} \big(A'_{\rm b} P +A'_{\rm f} Q+{\sqrt \alpha}\, A'_+ L_+\big) {\rm e}^{-B}\big]_{L_+}\\
\end{aligned}
\ee
into the action \eqref{eq:GeneralFOA} and integrating out $\{A'_{\rm b},A'_{\rm f},A'_+\}$, one obtains the dualised action $\tilde S$. The main goal is to show that the action $\tilde S$ (upon certain field re-definitions) is again of the Green--Schwarz form \eqref{GHaction0}, however, in a coordinate system which is associated with a different choice of the coset representative
\be\label{tildeg1}
\tilde g\ :=\ {\rm e}^{\tilde xK+ \tilde\theta M^{-1} S+{\sqrt \alpha}\, \tilde\lambda_+L_-}{\rm e}^{B}{\rm e}^{F(\xi)}~, \quad
{\rm e}^{B}\ :=\ {\rm e}^{\hat\theta\hat Q+\hat\xi\hat S}|y|^D{\rm e}^{-{\sqrt \alpha}\,  \lambda_3 L_3} \Lambda_{\alpha }(y)~,
\ee
where $M:={\rm Str}(QS)$. Note that ${\rm e}^B$ in the representative \eqref{tildeg1} is the same as given in \eqref{g}. Furthermore, in the AdS$_5\times  S^5$ case, $F(\xi)$ is of the schematic form
\be\label{F}
F(\xi)\ \sim\ -\big[\xi+\xi^5\big]Q+ \big[\xi^3+\xi^7\big]S~,
\ee
while for AdS$_2\times S^2$ and AdS$_2\times S^2\times S^2$, $F(\xi)$ contains only the first linear term in $\xi$ and for AdS$_3\times S^3$ and  AdS$_3\times S^3\times S^3$ it consists of both linear and cubic terms. Because of the presence of $F(\xi)$, the current $\tilde J=\tilde g^{-1}{\rm d}\tilde g$ arising from the representative \eqref{tildeg1}, will, in general, not be quadratic in the fermionic coordinates $\xi$. However, as we will show, upon further complicated field re-definitions $(X,\Theta)\to (X',\Theta')$, the dual coset element \eqref{tildeg1} can be nevertheless brought to a form similar to that of \eqref{g}, {\it i.e.}
\be\label{tildeg111}
\tilde g\ =\ {\rm e}^{\tilde x'K+ \tilde\theta' M^{-1} S+{\sqrt \alpha}\, \tilde\lambda'_+L_-}{\rm e}^{B'}{\rm e}^{-\xi'Q}\,.
\ee

To see the explicit result of T-dualisation, in the subsequent sections we turn to the detailed consideration of the superstring in different backgrounds.  We begin by considering the AdS$_5 \times S^5$ background. This is the most involved example since, without partially gauge fixing kappa symmetry by putting $\xi=0$, we can have terms up to $\mathcal{O}(\xi^8)$.

The results of the T-dualisation of the less supersymmetric cases AdS$_3\times S^3$ and AdS$_2\times S^2$ can be then obtained upon an appropriate truncation of the AdS$_5\times S^5$ supercoset sigma model. The T-duality of superstrings in AdS$_d\times S^d\times T^{10-2d}$ in the presence of fluctuations along $T^{10-2d}$ and non-coset fermionic modes $\upsilon$ will also be considered.

Finally, we discuss the T-duality procedure for the superstring sigma models on AdS$_d\times S^d\times S^d$ for $d=2,3$, which turns out to be also technically quite involved.

\paragraph{Comment on the self-duality at the quantum level.}
Since the duality transformations can be performed via a Gau{\ss}ian path integral, they can be promoted to a duality of the quantum sigma model. A priori, the path integral measure could change upon integrating out the auxiliary fields $\{A'_{\rm b},A'_{\rm f},A'_+\}$. However, this is not the case, since the corresponding Berezinian is equal to one provided one also regularises the bosonic and fermionic determinants in the same way ({\it e.g.}~by using heat kernel methods as in \cite{Buscher:1987qj,Schwarz:1992te}). Therefore, there will be no shift in the dilaton (see also Section \ref{sugra}) and we may thus conclude that the self-duality of the Green--Schwarz sigma models under consideration also holds at the quantum level.

\section{Self-duality of AdS$_5\times S^5$ superstrings}\label{ADS5}

\subsection{Supercoset action on  AdS$_5\times S^5$}

We begin by focusing on the AdS$_5\times S^5$ superstring sigma model.
The coset superspace $\frac{PSU(2,2|4)}{SO(1,4)\times SO(5)}$
solves the type IIB 10-dimensional supergravity constraints and hence, it describes the full type IIB AdS$_5\times S^5$ background parametrized by ten bosonic coordinates $(X^M)=(x^m,|y|,y^{\ha})$ with $m,n,\ldots=0,\ldots,3$ and $\hat a,\hat b,\ldots=5,\ldots,9$ and two 16-component 10-dimensional Majorana--Weyl spinor coordinates $\Theta^{i}=\frac 12(1+\Gamma^{11})\Theta^i$ with $i,j,\ldots=1,2$ of the same chirality \footnote{See Appendix \ref{A} for conventions on $(16 \times 16)$ gamma matrices used in type IIB superspace.}.
In this parametrization, the AdS$_5\times S^5$ line element has the form \cite{Berkovits:2008ic}
\begin{equation}\label{AdsxS}
({\rm d}s)^2\ =\ \frac 1{|y|^2}\big({\rm d}x^m{\rm d}x^n\eta_{mn} +{\rm d}y^{\ha}{\rm d}y^{\ha}+{\rm d}\hat y {\rm d} \hat y\big)\,,
\end{equation}
where $(y^{\hat a}y^{\hat a}  + \hat y \hat y)=|y|^2$. For simplicity, we have set the radii of AdS$_5$ and $S^5$ to one.

The AdS$_5\times S^5$ background is supported by the self-dual 5-form flux $F_5$ with the non-zero components
\begin{subequations}\label{F55}
\be
F_{01234}\ =\ { -}F_{56789}\ =\ 4~,
\ee
or, equivalently,
\be \label{flux5}
 F_5\ =\ { 4}(1+*)\,{\rm Vol}_{{\rm AdS}_5}\ =\ { 4}\,(1+*)\,e^{0}\wedge\,\cdots\wedge e^4~.
\ee
\end{subequations}
and by the dilaton field that for simplicity we choose to be zero.

\paragraph{\mathversion{bold}Lie superalgebra $\mathfrak{psu}(2,2|4)$ and Cartan forms.}
The general form of the superisometry algebra for a symmetric space supergravity solution was determined in \cite{Wulff:2015mwa}. Inserting the form of the fluxes for the AdS$_5\times S^5$ background given above we obtain the following form of the $\mathfrak{psu}(2,2|4)$ Lie superalgebra
\be\label{psu22-0}
\begin{gathered}
[M_{AB},M_{CD}]\ =\ \eta_{AC}M_{BD}-\eta_{AD}M_{BC}-\eta_{BC}M_{AD}+\eta_{BD}M_{AC}~,\\
[P_A,P_B]\ =\ -\tfrac12R_{AB}{}^{CD}M_{CD}~,\\
[M_{AB},P_C]\ =\ \eta_{AC}P_B-\eta_{BC}P_A\,,\qquad [M_{AB}, \mathcal Q_{\alpha i}]\ =\ -\tfrac12(\mathcal Q\Gamma_{AB})_{\alpha i}~,\\
[P_A,\mathcal Q_{\alpha i}]\ =\ -\tfrac12(\mathcal Q\varepsilon\Gamma^{01234}\Gamma_A)_{\alpha i}~,\\
\{\mathcal Q_{\alpha i},\mathcal Q_{\beta j}\}\ =\ {\rm i}\delta_{ij}(\Gamma^A)_{\alpha\beta}\,P_A-\tfrac{\rm i}{2}\varepsilon_{ij}(\Gamma^A\Gamma^{01234}\Gamma^B)_{\alpha\beta}\,M_{AB}~,
\end{gathered}
\ee
where $\varepsilon^{ij}=-\varepsilon^{ji}$ ($\varepsilon^{12}=1$), $(M_{AB})=(M_{ab},M_{\ha\hb})$ with $a,b,\ldots=0,\ldots,4$ and $\ha,\hb,\ldots=5,\ldots,9$ generate the $SO(4,1)\times SO(5)$ rotations, $(P_A)=(\mathcal P_a,P_{\ha})$ generate AdS$_5\times S^5$ translations, $R_{ab}{}^{cd}=2\delta_{[a}^c\,\delta_{b]}^d$ and $R_{{\ha}{\hb}}{}^{{\hc}\hd}=-2\delta_{[{\ha}}^{{\hc}}\,\delta_{{\hb}]}^{\hd}$ are, respectively, AdS$_5$ and $S^5$ curvatures, and $\mathcal Q_{\alpha i}$ are the  supercharges. The corresponding Maurer--Cartan form
\be\label{CartanSG}
J(X,\Theta)\ =\ \tfrac 12 \Omega^{AB}M_{AB}+E^AP_A+E^{\alpha i} \mathcal Q_{\alpha i}
\ee
is made of the superconnection $\Omega^{AB}(X,\Theta)$ and the supervielbeins $E^A(X,\Theta),$ $E^{\alpha i}(X,\Theta)$ that satisfy the type IIB supergravity constraints of \cite{Wulff:2013kga}.

To obtain the  $\mathfrak{psu}(2,2|4)$ Lie superalgebra in the superconformal form \eqref{conf} and \eqref{eq:SCext}, we define new bosonic generators as follows ($a=m,4$ with $m=0,\ldots,3$):
\be\label{NE}
\begin{gathered}
D\ :=\ \mathcal P_4~,\quad P_m\ :=\ \mathcal P_m+M_{m4}~,\quad K_m\ :=\ -\mathcal P_m+M_{m4}~,\\ M_{mn}~, \quad R_{{\ha}}\ :=\ P_{{\ha}}~,\quad R_{{\ha}{\hb}}\ :=\ -M_{{\ha}{\hb}}~.
\end{gathered}
\ee
The non-vanishing commutators of these generators are
\be\label{eq:PSU22-12-Bos}
\begin{gathered}
[P_m,K_n]\ =\ -2\eta_{mn} D+{ 2}M_{mn}~, \quad [D,P_m]\ =\ P_m~,\quad [D,K_m]\ =\ -K_m~,\\
[P_m,M_{np}]\ =\ -\eta_{mn}P_p+\eta_{mp}P_n~,\quad[K_m,M_{np}]\ =\ -\eta_{mn}K_p+\eta_{mp}K_n~,\\
[M_{mn},M_{pq}]\ =\ \eta_{mp}M_{nq}-\eta_{mq}M_{np}-\eta_{np}M_{mq}+\eta_{nq}M_{mp}~,\\
[R_{\ha},R_{\hb}]\ =\ -R_{\ha\hb}~,\quad
[R_{\ha\hb},R_{\hc\hd}]\ =\ -\delta_{\ha\hc}R_{\hb\hd}+\delta_{\ha\hd}R_{\hb\hc}+\delta_{\hb\hc}R_{\ha\hd}-\delta_{\hb\hd}R_{\ha\hc}~,\\
[R_{\ha},R_{\hb\hc}]\ =\ \delta_{\ha\hb}R_{\hc}-\delta_{\ha\hc}R_{\hb}~.
\end{gathered}
\ee
The new independent fermionic generators are defined by setting
\be\label{PRO}
\begin{gathered}
Q\ :=\ -\tfrac{1}{\sqrt2}(\mathcal Q^1{ -}{\rm i}\mathcal Q^2)\mathbb P_+~,\quad \hat Q\ :=\ -\tfrac{1}{\sqrt2}(\mathcal Q^1{ +}{\rm i}\mathcal Q^2)\mathbb P_-~,\\
S\ :=\ \tfrac{1}{\sqrt2}(\mathcal Q^1{ +}{\rm i}\mathcal Q^2)\mathbb P_+~,\quad \hat S\ :=\ \tfrac{1}{\sqrt2}(\mathcal Q^1{ -}{\rm i}\mathcal Q^2)\mathbb P_-~,
\end{gathered}
\ee
where we have introduced the projectors acting on the indices $\alpha$  of the supercharge $\mathcal Q_{ \alpha i}$
\begin{equation}\label{eq:Projector}
\mathbb P_\pm\ :=\ \tfrac12(1\pm {\rm i}\Gamma^{0123})~.
\end{equation}
The commutators involving these supercharges are
\begin{subequations}\label{eq:PSU22-12}
\be\label{psu22-1}
\begin{gathered}
[D,Q_{\alpha }]\ =\ \tfrac12Q_{\alpha }~,\quad[D,S_{\alpha }]\ =\ -\tfrac12S_{\alpha }~,\\
[K_m,Q_{\alpha }]\ =\ (\hat S\Gamma_{m4})_{\alpha }~,\quad[P_m,S_{\alpha }]\ =\ (\hat Q\Gamma_{m4})_{\alpha }~,\\
[M_{mn},S_{\alpha }]\ =\ -\tfrac12(S\Gamma_{mn})_{\alpha }~,\quad[M_{mn},Q_{\alpha }]\ =\ -\tfrac12(Q\Gamma_{mn})_{\alpha }~,\quad[R_{\ha},S_{\alpha }]\ =\ \tfrac12(S\Gamma_{\ha 4})_{\alpha }~\\
[R_{\ha},Q_{\alpha }]\ =\ -\tfrac12(Q\Gamma_{\ha 4})_{\alpha }~,\quad[R_{\ha\hb},S_{\alpha }]\ =\ \tfrac12(S\Gamma_{\ha\hb})_{\alpha }~,\quad[R_{\ha\hb},Q_{\alpha }]\ =\ \tfrac12(Q\Gamma_{\ha\hb})_{\alpha }~,
\end{gathered}
\ee
and the same with $(Q,S)\leftrightarrow(\hat Q,\hat S)$. We also have
\be\label{psu22-2}
\begin{gathered}
\{\hat Q_{\alpha },Q_{\beta }\}\ =\ {\rm i}(\Gamma^m\mathbb P_+ )_{\alpha \beta }\,P_m~,\quad
\{\hat S_{\alpha },S_{\beta }\}\ =\ -{\rm i}(\Gamma^m\mathbb P_+ )_{\alpha \beta }\,K_m~,\\
\{S_{\alpha },Q_{\beta }\}\ =\ -{\rm i}(\Gamma^4\mathbb P_+ )_{\alpha \beta }\,D
-{\rm i}(\Gamma^{\ha}\mathbb P_+ )_{\alpha \beta }\,R_{\ha}
-\tfrac{\rm i}{2}(\Gamma^{mn}\Gamma_4\mathbb P_+ )_{\alpha \beta }\,M_{mn}
-\tfrac{\rm i}{2}(\Gamma^{\ha\hb}\Gamma^4\mathbb P_+ )_{\alpha \beta }\,R_{\ha\hb}~,\\
\{\hat S_{\alpha },\hat Q_{\beta }\}\ =\
-{\rm i}(\Gamma^4\mathbb P_- )_{\alpha \beta }\,D
-{\rm i}(\Gamma^{\ha}\mathbb P_- )_{\alpha \beta }\,R_{\ha}
-\tfrac{\rm i}{2}(\Gamma^{mn}\Gamma_4\mathbb P_- )_{\alpha \beta }\,M_{mn}
-\tfrac{\rm i}{2}(\Gamma^{\ha\hb}\Gamma^4\mathbb P_- )_{\alpha \beta }\,R_{\ha\hb}~.
\end{gathered}
\ee
\end{subequations}

Finally, the non-vanishing components of the invariant form on $\mathfrak{psu}(2,2|4)$ that is compatible with the above choice of the basis is
\begin{equation}\label{eq:psuIF}
{\rm Str}(K_nP_m)\ =\ -2\eta_{mn}~,\quad {\rm Str}(DD)\ =\ 1~,\quad {\rm Str}(S_{\alpha}Q_{\beta})\ =\ 2{\rm i}(\Gamma^4\mathbb P_+)_{\alpha\beta}~.
\end{equation}

\paragraph{Currents and supercoset action.}
In the parametrization we have chosen, the explicit form of the coset representative \eqref{g} is
\begin{equation}\label{g22}
g\ :=\ {\rm e}^{x^mP_m+\theta^{\alpha}Q_{\alpha}}{\rm e}^B{\rm e}^{\xi^{\alpha}S_{\alpha}}~,\quad
{\rm e}^B\ :=\ {\rm e}^{\hat\theta^{\alpha}\hat Q_{\alpha}+\hat\xi^{\alpha}\hat S_{\alpha}}|y|^D{\rm e}^{y^{\ha}R_{\ha}/|y|}~,
\end{equation}
where, as a consequence of the definitions \eqref{PRO}, the spinorial variables satisfy the projection relations $\mathbb P_+ \theta = \theta, \mathbb P_+ \xi = \xi, \mathbb P_- \hat{\theta} = \hat{\theta}$ and $\mathbb P_- \hat\xi = \hat\xi$.

Following the general procedure described in Section 2, we can derive the explicit form of the  $\mathfrak{psu}(2,2|4)$ currents \eqref{eq:FormalCurrents}. Concretely, using the (anti-)commutation relations \eqref{eq:PSU22-12-Bos} and \eqref{eq:PSU22-12}, we obtain the components of the current \eqref{Jg} which do not depend on $\xi^\alpha$,
\begin{subequations}\label{eq:cosetreppsu224}
\be\label{psu22c1}
\begin{gathered}
J_{P_m}\ =\ \big[{\rm e}^{-B}({\rm d}x^n P_n+{\rm d}\theta Q){\rm e}^{B}\big]_{P_m}~,\quad J_{Q_\alpha}\ =\ \big[{\rm e}^{-B}({\rm d}x^n P_n+{\rm d}\theta Q){\rm e}^{B}\big]_{Q_\alpha}~,\\
J_{\hat S_\beta}\ =\ \big[{\rm e}^{-B}{\rm de}^B\big]_{\hat S_\beta}~,
\end{gathered}
\ee
the currents which depend on $\xi^\alpha$ linearly (the label ${}^{(0)}$ indicates, as before, the current components which do not depend on $\xi$)
\be\label{psu22c2}
\begin{gathered}
J_{K_m}\ =\ -{\rm i}(\Gamma^m\xi)_{\alpha}\,J_{\hat S_{\alpha}}~,\quad
J_D\ =\ J^{(0)}_D-{\rm i}(\Gamma^4\xi)_{\alpha}J_{Q_{\alpha}}~,\quad
J_{R_{\ha}}\ =\ J_{R_{\ha}}^{(0)}-{i}(\Gamma^{\ha}\xi)_{\alpha}J_{Q_\alpha}~,\\
J_{\hat Q}\ =\ J_{\hat Q}^{(0)}+(\Gamma_{m4}\xi)^{\alpha}J_{P_m}~,\\
J_{M_{mn}}\ =\ J_{M_{mn}}^{(0)}-\tfrac{\rm i}{2}(\xi\Gamma^{mn}\Gamma_4)_{\alpha}\,J_{Q_{\alpha}}~,\quad
J_{R_{\ha\hb}}\ =\ J_{R_{\ha\hb}}^{(0)}-\tfrac{\rm i}{2}(\xi\Gamma^{\ha\hb}\Gamma^4)_{\alpha}\,J_{Q_{\alpha }}~,
\end{gathered}
\ee
and the current $J_S$ which has a quadratic dependence on $\xi^\alpha$
\be\label{psu22c3}
\begin{aligned}
J_{S_\alpha}&=\ {\rm d}\xi^\alpha-\tfrac12\xi^{\alpha}J_D^{(0)}
-\tfrac12(\Gamma_{mn}\xi)^{\alpha}J_{M_{mn}}^{(0)}
+\tfrac12(\Gamma_{\ha 4}\xi)^{\alpha}J_{R_{\ha}}^{(0)}
+\tfrac12(\Gamma_{\ha\hb}\xi)^{\alpha}J_{R_{\ha\hb}}^{(0)}+\mathcal S^{\alpha}{}_\beta J_{Q_\beta}\\
&=:\ J_{S_\alpha}^{(1)}+\mathcal S^{\alpha}{}_\beta J_{Q_\beta}~,
\end{aligned}
\ee
where we have defined
\be\label{MS}
\mathcal S^{\alpha}{}_\beta\ :=
\tfrac{\rm i}{4}\xi^{\alpha }(\xi\Gamma^4)_{\beta }+\tfrac{\rm i}{4}(\Gamma^4\Gamma_{\ha}\xi)^{\alpha }(\xi\Gamma^{\ha })_{\beta }
+\tfrac{\rm i}{8}(\Gamma_{mn}\xi)^{\alpha }(\xi\Gamma^{mn}\Gamma_4)_{\beta }
-\tfrac{\rm i}{8}(\Gamma_{\ha \hb}\xi)^{\alpha }(\xi\Gamma^{\ha \hb}\Gamma^4)_{\beta }~.
\ee
\end{subequations}
Note that $\mathcal S^T=-\Gamma^4\mathcal S\Gamma^4$.

Comparing the Maurer--Cartan current \eqref{CartanSG} with the corresponding coset expression \eqref{currentcomponents} and exploiting the definition of the superconformal generators \eqref{NE}, \eqref{PRO} in terms of the 10-dimensional ones, we can read off the relation between the 10-dimensional geometric objects and the components of the supercoset current $J$. Explicitly, we find
\begin{subequations}
\be\label{BE}
\begin{gathered}
E^m\ =\ J_{P_m} - J_{K_m}~, \quad E^4\ =\ J_D~ , \quad E^{\hat a}\ =\ J_{R_{\hat a}}~, \\
\Omega^{mn}\ =\ 2 J_{M_{mn}}~, \quad \Omega^{m4}\ =\ J_{P_m} + J_{K_m}~, \quad \Omega^{\hat a \hat b}\ =\ - 2 J_{R_{\hat a \hat b}}
\end{gathered}
\ee
and
\be\label{FE}
 E^1\ =\ \tfrac 1{\sqrt 2}(J_S+J_{\hat S}-J_{Q}-J_{\hat Q})~ , \quad  E^2\ =\ \tfrac {\rm i}{\sqrt 2}(J_S-J_{\hat S}+J_{Q}-J_{\hat Q})
\ee
\end{subequations}
Note that $E^1=J_{(1)}$ and $E^2=J_{(3)}$, {\it i.e.}~they have $\mathbbm Z_4$-grading 1 and 3, respectively.

\vskip 10pt
In terms of the currents \eqref{psu22c1}--\eqref{psu22c3}, the Lagrangians \eqref{GSaction} and \eqref{GHaction1} for the $PSU(2,2|4)$ supercoset sigma model take the following form
\be
\begin{aligned}
\mathcal L\ &=\  \tfrac 12 {* E^A}\wedge E^B\eta_{AB}-{\rm i}  E^1\wedge \Gamma^{01234} E^{2}\\
 &=\ \tfrac12{*(J_{P_m}-J_{K_m})}\wedge (J_{P_n}-J_{K_n})\eta_{mn}+\tfrac12{*J_D}\wedge J_D+\tfrac12{*J_{R_{\ha}}}\wedge J_{R_{\ha}}\, -\\
&\kern 1cm-\tfrac{\rm i}{2} J_S\wedge \Gamma^4J_S
-\tfrac{\rm i}{2} J_{\hat S}\wedge \Gamma^4J_{\hat S}
+\tfrac{\rm i}{2} J_Q\wedge \Gamma^4J_Q
+\tfrac{\rm i}{2} J_{\hat Q}\wedge \Gamma^4J_{\hat Q}\,,\label{L22}
\end{aligned}
\ee
where we used \eqref{FE} and the projection properties of the generators \eqref{PRO}. The explicit expression for the Neveu--Schwarz--Neveu--Schwarz 2-form $B_2  =  - {\rm i} E^1 \wedge \Gamma^{01234} E^2$ has been determined from the type IIB supergravity constraints corresponding to the particular background \eqref{flux5}.

\subsection{T-duality transformations}

In order to T-dualise along $x^m$ and $\theta^{\alpha}$, we carry out the procedure described in Section \ref{generic}. Upon introducing the auxiliary 1-form fields $A'^m$ and $A'^\alpha$ \eqref{A'} together with the dual variables $\tilde x^m$ and $\tilde\theta_\alpha$, the Lagrangian takes the form
\begin{subequations}
\begin{equation}\label{123}
\mathcal L\ =\ \mathcal L_1+\mathcal L_2+\mathcal L_3
\end{equation}
where
\begin{eqnarray}
\mathcal L_1\! &:=&\! \tfrac12{*A'_m}\wedge A'_n\eta^{mn}-\tfrac12A'^m\wedge A'^n\,{\mathcal M}_{mn}+A'^m\wedge \mathcal J_m~,\\
\mathcal L_2\! &:=&\! \tfrac{\rm i}{2}A'^{\alpha}\wedge A'^{\beta}N_{\alpha\beta}-\tfrac{\rm i}{2}{*A'^{\alpha}}\wedge A'^{\beta}(NL)_{\alpha\beta}+A'^{\alpha}\wedge \mathcal J_{\alpha}~,\label{L2}\\
\mathcal L_3\! &:=&\! \tfrac12{*J_{K_m}}\wedge J_{K_n}\eta_{mn}
+\tfrac12{*J_D^{(0)}}\wedge J_D^{(0)}
+\tfrac12{*J_{R_{\hat a}}^{(0)}}\wedge J_{R_{\hat a}}^{(0)}-\,\notag\\
&&\kern1cm
-\tfrac{\rm i}{2}J_S^{(1)}\wedge \Gamma^4J_S^{(1)}
-\tfrac{\rm i}{2}J_{\hat S}\wedge \Gamma^4J_{\hat S}
+\tfrac{\rm i}{2}J_{\hat Q}^{(0)}\wedge \Gamma^4J_{\hat Q}^{(0)}~.\label{L3}
\end{eqnarray}
with
\begin{eqnarray}
\mathcal J_m\! &:=&\!
-{\rm d}\tilde x^n\big[{\rm e}^B{  P}_m{\rm e}^{-B}\big]_{  P_n}
-{\rm d}\tilde\theta_{\alpha}\big[{\rm e}^B{  P_m}{\rm e}^{-B}\big]_{Q_{\alpha}}
+{\rm i}J_{ \hat Q}^{(0)}\Gamma_m\xi
+*J_{  K_m}~,\label{Jm}\\
\mathcal J_{\alpha}\! &:=&\!
-{\rm d}\tilde x^m\big[{\rm e}^BQ_{\alpha}{\rm e}^{-B}\big]_{  P_m}
+{\rm d}\tilde\theta_{\beta}\big[{\rm e}^BQ_{\alpha}{\rm e}^{-B}\big]_{Q_{\beta}}-\,\notag \\
&&\kern1cm-{  {\rm i}}{*J_D^{(0)}}(\Gamma^4\xi)_{\alpha}
-{  {\rm i}}*J_{  R_{\ha}}^{(0)}(\Gamma^{\ha}\xi)_{\alpha}
-{\rm i}(J_S^{(1)}\Gamma^4{\mathcal S})_{\alpha}\label{JNL}
\end{eqnarray}
and
\begin{eqnarray}
{\mathcal M}_{AB}\ :=\ {\rm i}\xi\Gamma_{AB}\Gamma_4\xi~,\kern3.5cm   \label{MAB}\\
 N_{\alpha\beta}\ :=\ \big(\Gamma^4(1+\mathcal S^2)\big)_{\alpha\beta}~,\quad (NL)_{\alpha \beta}\ :=\ {\rm i}(\Gamma^4\xi)_{\alpha}(\Gamma^4\xi)_{\beta}+{\rm i}(\Gamma^{\ha}\xi)_{\alpha}(\Gamma_{\ha}\xi)_{\beta}\label{N}~.
\end{eqnarray}
\end{subequations}
Next, after some algebra, the equations of motion for the auxiliary fields ${A'}^m$ and ${A'}^\alpha$ that follow  from the Lagrangian \eqref{123} are given by
\be\label{neweq}
\begin{aligned}
A'^m\ &=\ {*\mathcal J^n}[(1-\mathcal M^2)^{-1}]_n{}^m+\mathcal J^n[\mathcal M(1-\mathcal M^2)^{-1}]_n{}^m~,\\
A'^{\alpha}\ &=\ {\rm i}[(1-L^2)^{-1}N^{-1}\mathcal J]^{\alpha}-{\rm i}[L(1-L^2)^{-1}N^{-1}*\mathcal J]^{\alpha}~.
\end{aligned}
\ee
Upon plugging these back into the Lagrangian \eqref{123}, we obtain the Lagrangian  of the dualised model
\begin{subequations}
\begin{equation}\label{tL}
\tilde{\mathcal L}\ =\ \tilde{\mathcal L}_1+\tilde{\mathcal L}_2+\mathcal L_3
\end{equation}
with
\begin{eqnarray}
\tilde{\mathcal L}_1\! &:=&\! \tfrac12{*\mathcal J^m}\wedge \mathcal J^n[(\eta-\mathcal M^2)^{-1}]_{mn}+\tfrac12\mathcal J^m\wedge\mathcal J^n[\mathcal M(1-\mathcal M^2)^{-1}]_{mn}~,\label{L1}\\
\tilde{\mathcal L}_2\! &:=&\!\tfrac{\rm i}{2}
\mathcal J\wedge (N-NL^2)^{-1}\mathcal J
+\tfrac {\rm i}{2}{*\mathcal J}\wedge L(N-NL^2)^{-1}\mathcal J~\label{tL2}\,.
\end{eqnarray}
\end{subequations}
and $\mathcal L_3$ being the same as in \eqref{L3}, since it is a function of currents that are not involved in the dualisation process.

We now notice that $\tilde{\mathcal L}_1$ can be cast in the following simpler form
\begin{equation}\label{L11}
\tilde{\mathcal L}_1\ =\ \tfrac12(1-\tfrac 14 \mathcal M_{\hat a\hb}\mathcal M^{\ha\hb})^{-1}\big({*\mathcal J^m}\wedge \mathcal J^n\eta_{mn}+\mathcal J^m\wedge \mathcal J^n\mathcal M_{mn}\big)\,.
\end{equation}
To arrive at \eqref{L11} one should use  the projection properties of $\xi$ and the Fierz identity
 \be\label{Fierz1}
(\Gamma^m\xi)_\alpha(\xi\Gamma_m)_\beta\ =\ -\tfrac 1{4} (\mathbb P_- \Gamma_{\hat a\hat b}\Gamma_4)_{\alpha\beta}(\xi\Gamma^{\hat a\hat b}\Gamma_4\xi)~,
\ee
to show that 
\be\label{M2}
\mathcal M^2_{mn}\ =\ \mathcal M^2_{nm}\ =\ \mathcal M_m{}^l\mathcal M_{lm}\ =\ \tfrac 14 \eta_{mn} \mathcal M_{\hat a\hb}\mathcal M^{\ha\hb}~.
\ee

To cast \eqref{tL} in a  form similar to the original Lagrangian, we first notice that, using the invariant form \eqref{eq:psuIF}, we can identify the quantities appearing in \eqref{Jm}, \eqref{JNL} as follows
\be\label{tildeK0}
\begin{aligned}
&{\rm d}\tilde x^n\big[{\rm e}^BP_m{\rm e}^{-B}\big]_{P_n}
+{\rm d}\tilde\theta_{\alpha}\big[{\rm e}^BP_m{\rm e}^{-B}\big]_{Q_{\alpha}}\ =\\
&\kern 1cm=\
{ \mathrm{Str}}\big\{-\tfrac12{\rm d}\tilde x^n{\rm e}^BP_m{\rm e}^{-B}K_n
+\tfrac{\rm i}{2}{\rm d}\tilde\theta_{\alpha}{\rm e}^BP_m{\rm e}^{-B}(\Gamma^4S)^{\alpha}\big\}\\
&\kern1cm =\ \big[{\rm e}^{-B}({\rm d}\tilde x^nK_n- {\rm i}{\rm d}\tilde\theta\Gamma^4S){\rm e}^B\big]_{K_m}\ =\
\tilde J_{K_m}^{(0)}~,\\
& \ {\rm i}J_{\hat Q}^{(0)}\Gamma_m\xi\ =\ \tilde J^{(1)}_{P_m} ~, \\
& {{ {\rm d}\tilde x^m\big[{\rm e}^B Q_{\alpha}{\rm e}^{-B}\big]_{P_m}
-{\rm d}\tilde\theta_{\beta}\big[{\rm e}^B Q_{\alpha}{\rm e}^{-B}\big]_{Q_{\beta}}= }}\\
&\kern 1cm=\
{{{ \mathrm{Str}}\big\{-\tfrac12{\rm d}\tilde x^n{\rm e}^B Q_\alpha {\rm e}^{-B}K_n
-\tfrac{\rm i}{2}{\rm d}\tilde\theta_{\beta}{\rm e}^B Q_\alpha {\rm e}^{-B}(\Gamma^4 S)^{\beta}\big\} }} \\
&\kern1cm =  {{\ -{\rm i}\Gamma^4_{\alpha\beta}\big[{\rm e}^{-B}({\rm d}\tilde x^nK_n- {\rm i}{\rm d}\tilde\theta\Gamma^4 S){\rm e}^B\big]_{S_\beta}\ =\
 -{\rm i}(\Gamma^4\tilde J^{(0)}_S)_{\alpha} }}~.
\end{aligned}
\ee

As a consequence, the quantities \eqref{Jm}, \eqref{JNL} appearing in the dualised Lagrangian take the form
\begin{equation}\label{Jm1}
\begin{aligned}
\mathcal J_m\ &=\ \tilde J^{(1)}_{P_m}-\tilde J^{(0)}_{K_m}+*J_{K_m}~,\\
\mathcal J_{\alpha}\ &=\
{\rm i}(\Gamma^4\tilde J^{(0)}_S)_{\alpha}
-{\rm i}J_{S_\beta}^{(1)}\Gamma^4_{\beta\gamma}\mathcal S^\gamma{}_\alpha
-{\rm i}{*J_D^{(0)}}(\Gamma^4\xi)_{\alpha}
-{\rm i}{*J_{  R_{\ha}}^{(0)}}(\Gamma_{ \ha}\xi)_{\alpha}\,.
\end{aligned}
\end{equation}
The expressions \eqref{tildeK0} can be interpreted as Maurer--Cartan forms coming from a different (dual) choice of the coset representative, whose appropriate form turns out to be
\begin{equation}\label{tildeg}
\tilde J\ =\ \tilde g^{-1}{\rm d}\tilde g~,\quad
\tilde g\ :=\ {\rm e}^{\tilde x^nK_n-{\rm i}\tilde\theta\Gamma^4S}{\rm e}^B{\rm e}^{-(\xi Q+S_\alpha\mathcal S^\alpha{}_\beta \xi^\beta)(1-\frac 14\mathcal M_{\hat a\hat b}\mathcal M^{\hat a\hat b})^{-1}}~,
\end{equation}
where the factor $(1-\frac 14\mathcal M_{\hat a\hat b}\mathcal M^{\hat a\hat b})^{-1}$ appearing in the last exponent is the same as the scaling factor in \eqref{L11}.

Let us now show that upon a complicated change of variables, the dual coset element \eqref{tildeg} can be brought to a form similar to the original coset element \eqref{g22}, that is, with the last exponent to the right being ${\rm e}^{-\xi' Q}$ only. This is achieved by writing the following chain of equalities where we use the Baker--Campbell--Hausdorff formula and the $\mathfrak{psu}(2,2|4)$ superalgebra \eqref{eq:PSU22-12-Bos} and \eqref{eq:PSU22-12},
\bea\label{chain2}
\tilde g \! &:=&\! {\rm e}^{\tilde x^nK_n-{\rm i}\tilde\theta\Gamma^4S}{\rm e}^B{\rm e}^{-(\xi Q+S_\alpha\mathcal S^\alpha{}_\beta \xi^\beta)(1-\frac 14\mathcal M_{\hat a\hat b}\mathcal M^{\hat a\hat b})^{-1}}\nonumber\\
\!&=&\! {\rm e}^{\tilde x^nK_n-{\rm i}\tilde\theta\Gamma^4S}{\rm e}^B{\rm e}^{-\frac{\rm i}{2} D(\xi\Gamma^4\mathcal S\xi+\mathcal O(\xi^8))-\frac{\rm i}{2} R_{\hat a}\,\xi\Gamma^{\hat a}\mathcal S\xi}{\rm e}^{-S_\alpha\mathcal S^\alpha{}_\beta \xi^\beta+\mathcal O(\xi^7)}{\rm e}^{-\xi' Q}{\rm e}^{H}~,
\eea
where $\xi'=\xi +\mathcal O(\xi^5, \xi^7)$ and $H=-\frac{\rm i}{4} M_{mn}\xi\Gamma^{mn4}\mathcal S\xi-\frac {\rm i}{4}R_{\hat a\hat b}\xi\Gamma^{\hat a\hat b4}\mathcal S\xi$ takes values in the stability subalgebra $\mathfrak{so}(1,3)\oplus\mathfrak{so}(5)$. The last exponent in \eqref{chain2} can be removed by a corresponding local gauge transformation of the supercoset element, so we have
\bea\label{chain22}
\tilde g\! &\simeq &\!
{\rm e}^{\tilde x^nK_n-{\rm i}\tilde\theta\Gamma^4S}{\rm e}^B{\rm e}^{-\frac i2 D(\xi\Gamma^4\mathcal S\xi+\mathcal O(\xi^8))-\frac i2 R_{\hat a}(\xi\Gamma^{\hat a}\mathcal S\xi)}{\rm e}^{-S_\alpha\mathcal S^\alpha{}_\beta \xi^\beta+\mathcal O(\xi^7)}{\rm e}^{-\xi' Q}\nonumber\\
\!&=&\!{\rm e}^{\tilde x^nK_n-{\rm i}\tilde\theta\Gamma^4S}{\rm e}^{B'}{\rm e}^{-S_\alpha\mathcal S^\alpha{}_\beta \xi^\beta+\mathcal O(\xi^7)}{\rm e}^{-\xi' Q}\nonumber\\
\!&=&\!{\rm e}^{\tilde x^nK_n-{\rm i}\tilde\theta\Gamma^4S}\left({\rm e}^{B'}{\rm e}^{-S_\alpha\mathcal S^\alpha{}_\beta \xi^\beta+O(\xi^7)}{\rm e}^{-B'}\right){\rm e}^{B'}{\rm e}^{-\xi' Q}\nonumber \\
\!&=&\!{\rm e}^{\tilde x^nK_n-{\rm i}\tilde\theta\Gamma^4S}\,{\rm e}^{f^n(y,\hat\theta,\hat\xi,\xi) K_n -{\rm i}f_\alpha(y,\hat\theta,\hat\xi,\xi) (\Gamma^4S)^\alpha}{\rm e}^{B'}\,{\rm e}^{-\xi' Q}\nonumber\\
\!&=&\!{\rm e}^{{\tilde x}'^{n}K_n-{\rm i}{\tilde\theta}'\Gamma^4S}\,{\rm e}^{B'}\,{\rm e}^{-\xi' Q}\,,
\eea
where ${\rm e}^{B'}={\rm e}^{B}\,{\rm e}^{-\frac{\rm i}{2} D(\xi\Gamma^4\mathcal S\xi+\mathcal O(\xi^8))-\frac {\rm i}{2} R_{\hat a}(\xi\Gamma^{\hat a}\mathcal S\xi)}$, ${\tilde x}'^{n}=\tilde x^n+f^n(y,\hat\theta,\hat\xi,\xi)$,
${\tilde\theta}'_\alpha={\tilde\theta}_\alpha+f_\alpha(y,\hat\theta,\hat\xi,\xi)$ and $f^n$ and $f_\alpha$ are certain functions of the coordinates $y^{\hat a}$ of $S^5$, the radial direction $|y|$ of AdS$_5$ and the Gra{\ss}mann-odd coordinates $\hat\theta$, $\hat\xi$, and $\xi$.

The choices \eqref{tildeg}  and \eqref{chain22} of the dual coset element are associated with the $\mathbb{Z}_4$-automorphism
\be\label{auto}
\begin{gathered}
P_m\ \leftrightarrow\ K_m~,\quad
D\ \rightarrow\ -D~, \quad
{  R}_{\hat a}\ \rightarrow\ - {  R}_{\hat a}~,\\
S\ \rightarrow\ - {\rm i}Q~, \quad
\hat S\ \rightarrow\ -{\rm i}\hat Q~, \quad
Q\ \rightarrow\ - {\rm i}S~,  \quad
\hat Q \rightarrow\ - {\rm i}\hat S~
\end{gathered}
\ee
of the $\mathfrak{psu}(2,2|4)$ Lie superalgebra \eqref{eq:PSU22-12-Bos} and \eqref{eq:PSU22-12} together with field re-definitions that can be read off by comparing \eqref{g22} with \eqref{tildeg} and \eqref{chain22}. Instead of mapping the Lie algebra generators as above one can use the induced transformation on the currents
\be\label{Z4}
\begin{gathered}
J_{M_{mn}}\ \leftrightarrow\ J_{M_{mn}}~,\quad
J_{R_{\hat a\hat b}}\ \leftrightarrow\ J_{R_{\hat a\hat b}}~,\quad
J_{P_m}\ \leftrightarrow\ J_{K_m}~,\quad
J_D\ \leftrightarrow\ - J_D~,\quad
J_{R_{\hat a}}\ \leftrightarrow\ - J_{R_{\hat a}}~,\\
J_Q\ \rightarrow\ -{\rm i}J_S~,\quad
J_S\ \rightarrow\ -{\rm i}J_Q~,\quad
J_{\hat Q}\ \rightarrow\ -{\rm i}J_{\hat S}~,\quad
J_{\hat S}\ \rightarrow\ -{\rm i}J_{\hat Q}~.
\end{gathered}
\ee

{ {The choice \eqref{tildeg} of the dual element $\tilde g$,  implies that the $\xi$-independent components $\tilde J_{\hat Q}^{(0)},\tilde J_{\hat S}^{(0)},\tilde J_{D}^{(0)},\tilde J_{R_{\hat a}}^{(0)},\tilde J_{R_{\hat a\hat b}}^{(0)}$ and $\tilde J_{M_{mn}}^{(0)}$ of the currents are the same as the ones without tilde, whereas the full expressions for the dual currents $\tilde J_{P_{m}}$, $\tilde J_{K_m}$, $\tilde J_{\hat S}$, and $\tilde J_{\hat Q}$ are }}
\begin{subequations}\label{eq:tildeJ}
\begin{eqnarray}
\tilde J_{P_{m}}\! &=&\! \tilde J^{(1)}_{P_{m}}\big(1-\tfrac 14 \mathcal M_{\ha\hb}\mathcal M^{\ha\hb}\big)^{-\frac 12}~,\label{tildeJp}\\
\tilde J_{K_m}\! &=&\! (\tilde J_{K_m}^{(0)}-J_{K_l}\mathcal M_{l}{}^{m})\big(1-\tfrac 14 \mathcal M_{\ha\hb}\mathcal M^{\ha\hb}\big)^{-\frac 12}~,\label{tildeJk}\\
\tilde J_{\hat S}\! &=&\! \tilde J^{(0)}_{\hat S}+\tilde J_{K_m}(\Gamma_4\Gamma_{m}\xi)+\frac 12 J_{K_n}\mathcal M_{nm}(\Gamma^4\Gamma^{m}\xi)\big(1+\tfrac 1{16}\mathcal M_{\ha\hb}\mathcal M^{\ha\hb}\big)~,\label{tildeJs}\\
\tilde J_{\hat Q}\! &=&\! \tilde J_{\hat Q}^{(0)}+\tfrac 12 \tilde J^{(1)}_{P_{m}}(\Gamma^4\Gamma^n\xi)\mathcal M_{nm}\big(1+\tfrac 3{16}\mathcal M_{\ha\hb}\mathcal M^{\ha\hb}\big)~.\label{tildeJq}
\end{eqnarray}
Likewise, the dual currents $\tilde J_{R_{\underline a}}=(\tilde J_D,\tilde J_{R_{\hat a}})$, $\tilde J_Q$, and $\tilde J_S$ read
\begin{eqnarray}
\tilde J_{R_{\underline a}}\! &=&\! \tilde J^{(0)}_{R_{\underline a}}-{\rm i}(\tilde J^{(0)}_S\Gamma^4-J^{(1)}_S\Gamma^4\mathcal S)(N-NL^2)^{-1}\Gamma^{\underline a}\xi+\tfrac {\rm i}{2} J^{(0)}_{R_{\underline b}}\xi\Gamma^{\underline a}L(N-NL^2)^{-1}\Gamma_{\underline b}\xi\,-\notag\\
&& -\tfrac 12 (\tilde J^{(0)}_S+\mathcal SJ^{(1)}_S)\Gamma_{\underline b}\xi(\xi\Gamma^{\underline b}L\Gamma^4\Gamma^{\underline a}\xi)\label{DR}
+\tfrac 18 J^{(0)}_{R_{\underline c}}\xi\Gamma^{\underline b}L\Gamma^4\Gamma^{\underline a}\xi(\xi\Gamma_{\underline b}L\Gamma^4\Gamma_{\underline c}\xi)~,\\
\tilde J_{Q}\! &=&\!\big[N^2(1-L^2)\big]^{-\frac 12}(\tilde J^{(1)}_{Q}-\mathcal S \tilde J^{(0)}_{S})~, \quad \tilde J^{(1)}_{Q}\ =\ -J^{(1)}_{S}-J^{(0)}_{R_{\underline a}}(\Gamma^4\Gamma_{\underline a}\xi)~,\label{Jq}\\
\tilde J_{S}\! &=&\! \big[N^2(1-L^2)\big]^{-\frac 12}\big[\tilde J^{(0)}_{S}-\mathcal S \tilde J^{(1)}_{Q}+({ 2}\mathcal S-\Gamma^4NL){  J^{(1)}_S}\big]~.\label{Js}
\end{eqnarray}
\end{subequations}

Next, upon substituting \eqref{tildeJp} and \eqref{tildeJk} into \eqref{Jm1}, we get the following expression for $\mathcal J_m$
\be\label{Jm2}
\mathcal J^m\ =\ (\tilde J_{P_m}-\tilde J_{K_m})\big(1-\tfrac 14 \mathcal M_{\ha\hb}\mathcal M^{\ha\hb}\big)^{\frac 12}+{*J_{K_m}}-{  J}_{K_n}\mathcal M_n{}^m~.
\ee
Also, by using \eqref{eq:tildeJ}, we find the  relations
\begin{subequations}\label{eq:jsjp}
\be\label{jsjs}
-\tfrac {\rm i}{2} \tilde J_{\hat S}\Gamma^4\tilde J_{\hat S}+\tfrac {\rm i}{2} \tilde J^{(0)}_{\hat S}\Gamma^4\tilde J^{(0)}_{\hat S}\ =\
\tilde J_{K_m}^{(0)}J_{K_n}\eta_{mn}+\tfrac 12 \tilde J_{K_m}\tilde J_{K_n}\mathcal M_{mn}-\tfrac 12 J_{K_m}J_{K_n}\mathcal M_{mn}
\ee
and
\be\label{jpjp}
\tfrac 12 \tilde J_{P_{m}}\tilde J_{P_{n}} \mathcal M_{mn}\ =\ \tfrac {\rm i}{2} \tilde J_{\hat Q}\Gamma^4\tilde J_{\hat Q}-\tfrac {\rm i}{2} \tilde J^{(0)}_{\hat Q}\Gamma^4\tilde J^{(0)}_{\hat Q}~.
\ee
\end{subequations}
Furthermore, upon combing \eqref{Jm2} and \eqref{eq:jsjp} with the Lagrangians $\tilde{\mathcal L}_1$ and $\mathcal L_3$ defined in \eqref{L3} and \eqref{L11}, we find the following form for $\tilde{\mathcal L}_1+\mathcal L_3$
\be\label{L1+L3}
\begin{aligned}
\tilde {\mathcal L}_1+\mathcal L_3\ &=\ \tfrac12{*(\tilde J_{P_m}-\tilde J_{K_m})}\wedge (\tilde J_{P_n}-\tilde J_{K_n})\eta_{mn}-\tfrac {\rm i}{2} \tilde J_ {\hat S}\wedge \Gamma^4\tilde J_{\hat S}+\tfrac{\rm i}{2}\tilde J_{\hat Q}\wedge\Gamma^4\tilde J_{\hat Q}\,+\\
&\kern1cm+\tfrac12{*J^{(0)}_D}\wedge J^{(0)}_D
+\tfrac12{*J^{(0)}_{R_{\ha}}}\wedge J^{(0)}_{R_{\ha}}
-\tfrac{\rm i}{2} J^{(1)}_{S}\wedge \Gamma^4J^{(1)}_{\ S}\,-\\
&\kern1cm-\tilde J_{K_m}\wedge \tilde J_{P_n}\mathcal M_{mn}-\tilde J^{(1)}_{P_m}\wedge J_{K_n}\eta_{mn}~.
\end{aligned}
\ee
In addition, with help of \eqref{DR}, the Lagrangian $\tilde{\mathcal L}_2$ as given in \eqref{tL2}  becomes
\be\label{tL21}
\begin{aligned}
\tilde {\mathcal L}_2\ &=\ \tfrac 12 {*\tilde J_{R_{\underline a}}}\wedge \tilde J_{R_{\underline a}}-\tfrac 12 {*J^{(0)}_{R_{\underline a}}}\wedge J^{(0)}_{R_{\underline a}}\,-\\
&\kern1cm-\tfrac {\rm i}{2}\big(\tilde J^{(0)}_{S}+\mathcal S J^{(1)}_{S}\big)\wedge(N-NL^2)^{-1}\big(\tilde J^{(0)}_{S}+\mathcal S J^{(1)}_{S}\big)\\
&\kern1cm+{\rm i}J^{(0)}_{R_{\underline a}}\wedge\big(\tilde J^{(0)}_{S}+\mathcal S J^{(1)}_{S}\big)\Gamma^4L(N-NL^2)^{-1}\Gamma_{\underline a}\xi\,+\\
&\kern1cm+\tfrac{\rm i}{2} J^{(0)}_{R_{\underline a}}\wedge J^{(0)}_{R_{\underline b}}\xi\Gamma_{\underline a}(N-NL^2)^{-1}\Gamma_{\underline b}\xi~.
\end{aligned}
\ee
Finally, summing up \eqref{L1+L3} with \eqref{tL21} and using \eqref{Jq} and \eqref{Js}, we obtain
\begin{subequations}
\be\label{fL}
\tilde {\mathcal L}\ =\ \tilde{\mathcal L}_1+\tilde {\mathcal L}_2+\mathcal L_3=\mathcal L_{\tilde g}+\mathcal L'+\mathcal L^{\prime\prime}~,
\ee
where
\be\label{ltg}
\begin{aligned}
\mathcal L_{\tilde g}\ &:=\
\tfrac12{*(\tilde J_{P_m}-\tilde J_{K_m})}\wedge(\tilde J_{P_n}-\tilde J_{K_n})\eta_{mn}
+\tfrac12{*\tilde J_D}\wedge\tilde J_D
+\tfrac12{*\tilde J_{R_{\ha}}}\wedge \tilde J_{R_{\ha}}\\
&\kern1cm-\tfrac{\rm i}{2}\tilde J_S\wedge\Gamma^4\tilde J_S
-\tfrac{\rm i}{2} \tilde J_{\hat S}\wedge\Gamma^4\tilde J_{\hat S}
+\tfrac{\rm i}{2}\tilde J_Q\wedge\Gamma^4\tilde J_Q
+\tfrac{\rm i}{2}\tilde J_{\hat Q}\wedge\Gamma^4\tilde J_{\hat Q}
\end{aligned}
\ee
is constructed in terms of the $\frac{PSU(2,2|4)}{SO(1,4)\times SO(5)}$ currents built from the dual coset element \eqref{tildeg} while $\mathcal L'$ and $\mathcal L''$ are given by
\be\label{L'}
\begin{aligned}
\mathcal L'\ &:=\ {\rm i} {  J}^{(1)}_{S}\wedge(N-NL^2)^{-1}N^{-1}\Gamma^4\tilde J^{(1)}_Q-{\rm i} {  J}^{(1)}_{S}\wedge\Gamma^4\mathcal S(N-NL^2)^{-1}\mathcal S\tilde J^{(1)}_Q\,+\\
&\kern1cm+J_{K_m}\wedge \tilde J^{(1)}_{P_n}\eta_{mn}(1+\tfrac 14 \mathcal M_{\hat a\hat b}\mathcal M^{\hat a\hat b})
\end{aligned}
\ee
and
\be\label{L''}
\begin{aligned}
\mathcal L''\ &:=\ {  {\rm i}} \tilde J^{(0)}_{S}\wedge (N-NL^2)^{-1}(4N^{-1}\Gamma^4\mathcal S -\mathcal S-L) J^{(1)}_S\,-\\
&\kern1cm-{\rm i}J^{(0)}_{R_{\underline a}}\wedge\tilde J^{(0)}_{S}(N-NL^2)^{-1}N^{-1}\Gamma^4(2\mathcal S-\Gamma^4NL)\Gamma^4\Gamma_{\underline a}\xi\,-\\
&\kern1cm-\tilde J^{(0)}_{K_m}\wedge\tilde J^{(1)}_{P_n}{ \mathcal M}_{mn}\big(1+\tfrac 14\mathcal  M_{\hat a\hat b}\mathcal M^{\hat a\hat b}\big)~.
\end{aligned}
\ee
\end{subequations}
We note that \eqref{ltg} has exactly the same form as the initial Lagrangian \eqref{L22}. Therefore, the action of the superstring on AdS$_5\times S^5$ is self-dual provided that $\mathcal L'+\mathcal L''$ is a total derivative. One can indeed show that this is true by performing quite involved computations, using Fierz identities and the Maurer--Cartan equation projected on the  $SO(1,3)\times SO(5)$ generators
\be
\begin{gathered}
({\rm d}J-J\wedge J)|_{R_{\hat a\hat b}}\ =\ ({\rm d}J-J\wedge J)|_{M_{mn}}\ =\ 0~,\\
({\rm d}\tilde J-\tilde J\wedge \tilde J)|_{R_{\hat a\hat b}}\ =\ ({\rm d}\tilde J-\tilde J\wedge \tilde J)|_{M_{mn}}\ =\ 0~.
\end{gathered}
\ee
The results are
\begin{subequations}
\be\label{L'1}
\mathcal L'\ =\ -{\rm i}\,{\rm d}\Big[J^{(1)}_S\Gamma^4\xi\big(1+\tfrac 14 \mathcal M_{\hat a\hat b}\mathcal M^{\hat a\hat b}\big)\Big]
\ee
and
\be\label{L''2}
\mathcal L''\ =\ -\tfrac 12 {\rm d} \Big[\tilde J^{(1)}_{M_{mn}}\mathcal M_{mn}\big(1+\tfrac 14 \mathcal M_{\hat a\hat b}\mathcal M^{\hat a\hat b}\big)\Big]~,
\ee
\end{subequations}
where the matrices $\mathcal M_{mn}$ and $\mathcal M_{\hat a\hat b}$ have been defined in \eqref{MAB}.

Altogether, $\int \mathcal L_{g}=\int \mathcal L_{\tilde g}$ and we have thus proved the self-duality of the AdS$_5\times S^5$ superstring action under the worldsheet duality transformation on  $(x^m, \theta^\alpha)$ coordinates, without gauge fixing kappa symmetry. The worldsheet T-duality transforms the superstring sigma model action \eqref{L22} constructed with the use of the supercoset representative \eqref{g22} into the action constructed using the supercoset element \eqref{tildeg}.

{ {The duality makes use of the automorphism \eqref{auto} of the $\mathfrak{psu}(2,2|4)$ Lie superalgebra, which in particular exchanges supercharges $Q$ and $S$. When dualising
the gauge-fixed action, one is indeed forced to change the kappa symmetry gauge when mapping the original lagrangian to the dual one \cite{Berkovits:2008ic,Beisert:2008iq}. }}

\section{Self-duality of AdS$_3\times S^3\times T^4$ superstrings}\label{sec:ads3s3}

The AdS$_3\times S^3\times T^4$ solutions of type IIB supergravity preserve 16 supersymmetries which generate the superisometries forming the $PSU(1,1|2)\times PSU(1,1|2)$ supergroup (see e.g. \cite{Claus:1998mw}). The AdS$_3\times S^3$ curvatures are taken to be
\be\label{ads3}
R_{_{{\rm AdS}_3}}^{ab}\ =\ -e^a\wedge e^b~,\quad R_{_{S^3}}^{{\ha}{\hb}}\ =\ e^{\hat a}\wedge e^{\hb}~,
\ee
where  $e^a=e^{a}(x)$ for $a,b,\ldots=0,1,4$ and $e^{\hat a}=e^{\hat a}(y)$ for $\hat a,\hat b,\ldots=5,6,7$ are the vielbeins of (unit radius) AdS$_3$ and $S^3$, respectively.\footnote{This particular choice of the tangent space indices associated with AdS$_3\times S^3\times T^4$ is related to the way we obtain these solutions by truncation of the $\mathfrak{psu}(2,2|4)$ superalgebra to $\mathfrak{psu}(1,1|2) \oplus \mathfrak{psu}(1,1|2)$, as we will see shortly.} One such background is supported by constant Ramond--Ramond 5-form flux
\be\label{F5}
F_5\ =\ { \tfrac 13} (\varepsilon_{cba} e^{a}\wedge e^b\wedge e^c +\varepsilon_{\hat c\hat b\hat a} e^{\hat a}\wedge e^{\hat b}\wedge e^{\hat c})\wedge ({\rm d}\varphi^2\wedge{\rm d}\varphi^3+{\rm d}\varphi^8\wedge {\rm d}\varphi^9)~,
\ee
where ${\rm d}\varphi^{a'}$ $(a',b',\ldots =2,3,8,9)$ are the flat vielbeins along $T^4$. Note the change in the form of the $F_5$-flux \eqref{F5} as compared with its value \eqref{F55} in the AdS$_5\times S^5$ solution. This difference results in changing the geometry of $D=10$ space-time and breaking half of the 32 supersymmetries.

Since only half of the maximal supersymmetry is preserved, the fermionic modes of the string in these backgrounds split into 16 fermions $\vartheta$ which are associated with the preserved supersymmetries and 16 fermions $\upsilon$ associated with the broken ones. Explicitly, the splitting is realized by using the additional projectors $\tfrac 12 (1\pm\Gamma^{2389})$ as follows
\be\label{tv}
\vartheta^i\ =\ \tfrac 12 (1-\Gamma^{2389})\Theta^i~,\quad \upsilon^i\ =\ \tfrac 12 (1+\Gamma^{2389})\Theta^i~.
\ee
Like in the AdS$_5\times S^5$ case, the fermions $ \vartheta$ can be regarded as Gra{\ss}mann-odd directions of the supercoset space $\frac{PSU(1,1|2)\times PSU(1,1|2)}{SO(1,2)\times SU(2)}$ which contains AdS$_3\times S^3$ as  the bosonic subspace. The $T^4$ directions and the non-supercoset fermions $\upsilon$ extend this supercoset to a full solution to the 10-dimensional type IIB supergravity constraints.

For certain classical string solutions in AdS$_3\times S^3\times T^4$ one can use kappa symmetry to gauge fix to zero all the non-supercoset fermions $\upsilon$.
In this gauge, modulo the Virasoro constraints, the oscillations of the string along $T^4$ decouple from the $\frac{PSU(1,1|2)\times PSU(1,1|2)}{SO(1,2)\times SU(2)}$ modes (see e.g. \cite{Babichenko:2009dk} for details). Consequently,  the superstring action reduces to its supercoset part, which can be obtained as a truncation of the AdS$_5 \times S^5$ action, once we select the $PSU(1,1|2)\times PSU(1,1|2)$ subgroup of $PSU(2,2|4)$ and reduce to it.

To this end, we first identify the 10-dimensional indices  2, 3, 8, and 9 as associated with the $T^4$ directions $(\varphi^{a'})=(\varphi^I,\varphi^{I'})$ $(I,J,\ldots=2,3;\,I',J',\ldots=8,9)$. Then, we remove from the algebra all the bosonic generators with the indices $m=2,3$ and ${\hat a}=8,9$, and halve the number of fermionic generators by acting on the original 32 generators defined in \eqref{psu22-0} with the additional projector  introduced in \eqref{tv}
\bea\label{PRO1}
\mathcal Q^i\ =\ \tfrac{1}{2}\mathcal Q^i(1-\Gamma^{2389})~,\qquad i\ =\ 1,2~.
\eea
Consequently, the generators $(Q,\hat Q, S,\hat S)$ defined in \eqref{PRO} are subject to the same projection.
The algebra $\mathfrak{psu}(1,1|2) \oplus \mathfrak{psu}(1,1|2)$ is then given by \eqref{eq:PSU22-12-Bos}--\eqref{psu22-2} with $m,n,\ldots =0,1$, $\hat{a}, \hat{b},\ldots= 5,6,7$ and the projectors $\mathbb P_{\pm}$ replaced by $\frac 12\mathbb{ P}_{\pm}(1-\Gamma^{2389})$.

From a geometrical point of view, this truncation corresponds to obtaining the AdS$_3\times S^3\times T^4$ background from AdS$_5\times S^5$ by formally compactifying  two directions of the 4-dimensional Minkowski boundary of AdS$_5$ and two directions of $S^5$ onto $T^4\cong T^2\times T^2$, as well as deforming the value of the $F_5$ flux as in \eqref{F5}.

\paragraph{Self-duality of the supercoset model.}
Given that the worldsheet sigma model on $\frac{PSU(1,1|2)\times PSU(1,1|2)}{SO(1,2)\times SU(2)}$ can be described as a truncation of the AdS$_5 \times S^5$ supercoset one, we can use the results of the previous section to easily show that it is self-dual under T-dualisation of the bosonic coordinates along the 2-dimensional Minkowski boundary of AdS$_3$ plus four fermionic directions associated with a commuting subalgebra of the $PSU(1,1|2)\times PSU(1,1|2)$ isometries.

In fact, the $\frac{PSU(1,1|2)\times PSU(1,1|2)}{SO(1,2)\times SU(2)}$ supercoset sigma model Lagrangian has the same form as \eqref{L22} with the currents constructed with the coset element having a form similar to \eqref{g22}. Proceeding exactly as in Section 2, the T-dualised Lagrangian turns out to be equal (modulo a total derivative) to the AdS$_3\times S^3$ sigma model Lagrangian constructed in terms of the supercoset element
\be\label{tildeg11}
\tilde J\ =\ \tilde g^{-1}{\rm d}\tilde g~,\qquad
\tilde g\ :=\ {\rm e}^{\tilde x^nK_n-{\rm i}\tilde\theta\Gamma^4S}{\rm e}^B{\rm e}^{-(\xi Q+S_\alpha\mathcal S^\alpha{}_\beta \xi^\beta)}~,
\ee
where $\mathcal S^\alpha{}_\beta$ was defined in \eqref{MS}. { {Since in this case at most $\xi^4$ powers can appear, the factor $(1-\frac 14\mathcal M_{\hat a\hat b}\mathcal M^{\hat a\hat b})^{-1}$ which was present in the dual supercoset element of the AdS$_5 \times S^5$ case does not enter the expression for $\tilde g$.}}

As in the AdS$_5\times S^5$ case, the dual element \eqref{tildeg11} can be brought to a form similar to the initial coset element \eqref{g22}, {\it i.e.}~with the last factor to the right being simply ${\rm e}^{-\xi Q}$. Specifically, the following chain of equalities hold\footnote{Note that in contrast to \eqref{chain2} no stability group elements $e^H$ appear and hence no compensating gauge transformation is needed due to the symmetry properties of the gamma-matrices in this case. For the same reason, no terms of the form  ${\rm e}^{R_{\hat a}\,\xi\Gamma^{\hat a}\mathcal S\xi}$ appear.}
\begin{eqnarray}\label{chain}
\tilde g\! & :=&\! {\rm e}^{\tilde x^nK_n-{\rm i}\tilde\theta\Gamma^4S}{\rm e}^B{\rm e}^{-(\xi Q+S_\alpha\mathcal S^\alpha{}_\beta \xi^\beta)}\nonumber\\
\!&=&\! {\rm e}^{\tilde x^nK_n-{\rm i}\tilde\theta\Gamma^4S}{\rm e}^B{\rm e}^{-\frac{\rm i}{2} D\xi\Gamma^4\mathcal S\xi}\,{\rm e}^{-S_\alpha\mathcal S^\alpha{}_\beta \xi^\beta}\,{\rm e}^{-\xi Q}\nonumber\\
\! &=&\! {\rm e}^{\tilde x^nK_n-{\rm i}\tilde\theta\Gamma^4S}{\rm e}^{B'}\,{\rm e}^{-S_\alpha\mathcal S^\alpha{}_\beta \xi^\beta}\,{\rm e}^{-\xi Q}\nonumber\\
\!&=&\!{\rm e}^{\tilde x^nK_n-{\rm i}\tilde\theta\Gamma^4S}\left({\rm e}^{B'}\,{\rm e}^{-S_\alpha\mathcal S^\alpha{}_\beta \xi^\beta}{\rm e}^{-B'}\right){\rm e}^{B'}\,{\rm e}^{-\xi Q}
\nonumber\\
\!&=&\!{\rm e}^{\tilde x^nK_n-{\rm i}\tilde\theta\Gamma^4S}\,{\rm e}^{f^n(y,\hat\theta,\hat\xi,\xi) K_n -{\rm i}f_\alpha(y,\hat\theta,\hat\xi,\xi) (\Gamma^4S)^\alpha}{\rm e}^{B'}\,{\rm e}^{-\xi Q}\nonumber\\
\!&=&\!{\rm e}^{{\tilde x}'^{n}K_n-{\rm i}{\tilde\theta}'\Gamma^4S}\,{\rm e}^{B'}\,{\rm e}^{-\xi Q}\,,
\end{eqnarray}
where ${\rm e}^{B'}={\rm e}^{B}\,{\rm e}^{-\frac i2 D\xi\Gamma^4\mathcal S\xi}$, ${\tilde x}'^{n}=\tilde x^n+f^n(y,\hat\theta,\hat\xi,\xi)$,
${\tilde\theta}'_\alpha={\tilde\theta}_\alpha+f_\alpha(y,\hat\theta,\hat\xi,\xi)$ and $f^n$ and $f_\alpha$ are certain functions of the coordinates $y^{\hat a}$ of $S^3$, the radial direction $|y|$ of AdS$_3$ and the Gra{\ss}mann-odd coordinates $\hat\theta$, $\hat\xi$, and $\xi$.

We have thus shown that in the kappa symmetry gauge in which the sixteen non-supercoset fermions $\upsilon$ are set to zero the AdS$_3\times S^3\times T^4$ superstring action is self-dual under the combined fermionic and bosonic T-duality. However, this gauge is not always admissible, for instance in the case when the classical string moves entirely in AdS$_3\times S^3$ \cite{Rughoonauth:2012qd}, so it is important to understand how T-dualisation works in different gauges or without fixing kappa symmetry. To this end we should know the structure of the AdS$_3\times S^3\times T^4$ superstring Lagrangian in the presence of the fermionic modes $\upsilon$. To all orders in $\upsilon$ this dependence is rather complicated, so in what follows we will restrict only to the quadratic order in $\upsilon$. In particular, we will see that the presence of the fermions $\upsilon$ requires that the T-dualisation is also performed along two of the torus directions.

\paragraph{Non-supercoset fermions.}
To derive the explicit form of the Green--Schwarz superstring Lagrangian \eqref{GSaction} in the AdS$_3\times S^3\times T^4$ background, we need expressions for the supervielbeins $\mathcal E^A(X,\vartheta,\upsilon)$ and the Neveu--Schwarz--Neveu--Schwarz 2-form $B_2(X,\vartheta,\upsilon)$ as series in powers of $\upsilon^i$, which can be determined in the same way as the $\Theta$-expansion derived in \cite{Wulff:2013kga} (see for example \cite{Wulff:2015mwa}). To quadratic order in $\upsilon$ one finds
\be\label{hat E1}
\mathcal E^A\ =\ E^A(X,\vartheta)-{\rm i}E\Gamma^A\upsilon-\tfrac {\rm i}2 {\mathcal D}\upsilon \Gamma^A\upsilon~,
\ee
and\footnote{The general expression for $B^{\rm coset}_2$ is
$$
B^{\rm coset}_2\ =\ -\tfrac{\rm i}{4}E\mathcal K\sigma^3E
$$
where $\mathcal K$ is $8$ times the inverse of the matrix that appears in $\slashed F$ of \eqref{S} (dropping the projector). For type IIA $\sigma^3$ is replaced by $\Gamma_{11}$. See \cite{Wulff:2015mwa}.}
\be\label{hat B1}
 B_2\ =\ B^{\rm coset}_2(x,y,\vartheta)-{\rm i} E^A\wedge E\Gamma_A\sigma^3\upsilon-\tfrac 12 E\Gamma^A\upsilon\wedge E\Gamma_A\sigma^3\upsilon-\tfrac {\rm i}2 E^A\wedge\mathcal D\upsilon\Gamma_A\sigma^3\upsilon~,
\ee
where $E^{\alpha i}(x,y,\vartheta)=\frac 12(1-\Gamma^{2389})^{\alpha}{}_{\beta}E^{\beta i}(x,y,\vartheta)$, $E^{a}(x,y,\vartheta)$, and $E^{\hat a}(x,y,\vartheta)$ are the supervielbeins constructed in terms of the $\frac{PSU(1,1|2)\times PSU(1,1|2)}{SO(1,2)\times SU(2)}$ supercoset currents, as in \eqref{FE}, while $E^{a'}={\rm d}\varphi^{a'}$ is the flat vielbein along $T^4$. The Pauli matrix $\sigma^3_{ij}$ contracts the indices $i,j=1,2$ of the spinors. Moreover, the Neveu--Schwarz--Neveu--Schwarz gauge potential $B_2^{\rm coset}$ has again the form as in \eqref{L22}, and for the background under consideration the covariant derivative $\mathcal D$ is given by
\be\label{cD}
\begin{aligned}
{\mathcal D}\upsilon\ =\  \nabla\upsilon{ -}\tfrac {\rm i}{16\cdot 5!}E^A F_{B_1\cdots B_5}\Gamma^{B_1\cdots B_5}\Gamma_A\sigma^2\upsilon\
=\ \nabla\upsilon{ -}\tfrac {\rm i}4 E^A(1-\Gamma^{2389})\Gamma^{01234}\Gamma_A\sigma^2\upsilon ~,
\end{aligned}
\ee
where $\nabla:= {\rm d}-\tfrac 14 \Omega^{AB}\Gamma_{AB}$ and $\Omega^{AB}(x,y,\vartheta)$ is the spin connection on $\frac{PSU(1,1|2)\times PSU(1,1|2)}{SO(1,2)\times SU(2)}$ defined in terms of the currents { {as in \eqref{BE}}}.

Upon substituting the expressions \eqref{hat E1} and \eqref{hat B1} into the string action \eqref{GSaction} and using the projector properties \eqref{tv} of $\vartheta$ and $\upsilon$, we find
\be\label{SSL11}
\begin{aligned}
\mathcal L\ &=\ \mathcal L_{\rm coset}+\tfrac 12{*{\rm d}\varphi^{a'}}\wedge{\rm d}\varphi^{a'}-{\rm i}{*{\rm d}\varphi^{a'}}\wedge E\Gamma_{a'}\upsilon -{\rm i}{\rm d}\varphi^{a'}\wedge E\Gamma_{a'}\sigma^{3}\upsilon-\tfrac 12 {*E}\Gamma^{a'}\upsilon\wedge E\Gamma_{a'}\upsilon\,-\\
&\kern1cm-\tfrac 12 E\Gamma^{a'}\upsilon\wedge E\Gamma_{a'}\sigma^3\upsilon-\tfrac {\rm i}2 {*E^A}\wedge\mathcal D\upsilon\Gamma_A\upsilon -\tfrac {\rm i}2E^A\wedge\mathcal D\upsilon\Gamma_A\sigma^{3}\upsilon~.
\end{aligned}
\ee
Here, $\mathcal L_{\rm coset}$ has the same form as \eqref{GHaction1} (with $\gamma=\Gamma^4$) and is similar to \eqref{L22}.
In the gauge in which $\upsilon$ are non-zero, the above Lagrangian contains a lot of $\upsilon$-dependent terms which contribute to the T-dualisation along the supercoset directions $(x^m,\theta)$.

\paragraph{Self-duality up to second order in the non-supercoset fermions.}
Since we have not used kappa symmetry to get rid of non-supersymmetric fermions, we are still free to use it to simplify the dualisation procedure. We find convenient to impose the gauge $\xi=0$ on fermions in the coset representative \eqref{g22}. Then, in view of \eqref{eq:cosetreppsu224}, the current components $J_{K_m}$ and $J_S$ vanish.

We begin by focusing on the part of the Lagrangian which is linear in $\upsilon$. Expressing the 10-dimensional geometric objects in terms of the coset currents (as done in \eqref{FE}, \eqref{L22}), replacing $E^{a'} = {\rm d} \varphi^{a'}$ and defining $\upsilon^\pm=\frac {1}{\sqrt{2}}(\upsilon^1\pm {\rm i}\upsilon^2)$, the Lagrangian at the linear order in $\upsilon$ reads
\be\label{L1v}
\begin{aligned}
\mathcal L\ &=\ \mathcal L_{\rm coset}+\tfrac 12{*{\rm d}\varphi^{a'}}\wedge{\rm d}\varphi^{a'}+{\rm i}{*{\rm d}\varphi^{a'}}\wedge J_{\hat Q}\Gamma_{a'}\upsilon^+ +{\rm i}{\rm d}\varphi^{a'}\wedge J_{\hat Q}\Gamma_{a'}\upsilon^-\,-\\
&\kern1cm-{\rm i}{*{\rm d}\varphi^{a'}}\wedge (J_{\hat S}-J_Q)\Gamma_{a'}\upsilon^- -{\rm i}{\rm d}\varphi^{a'}\wedge (J_{\hat S}-J_Q)\Gamma_{a'}\upsilon^+\,.
\end{aligned}
\ee
If we apply the general dualisation procedure described in Section \ref{generic}, this action gets mapped into its dual counterpart constructed with the use of the dual coset representative \eqref{tildeg11} at $\xi=0$, provided that we use the $\mathbbm{Z}_4$-automorphism similar to \eqref{Z4}, supplement the duality along  $x^m$ and $\theta$ with an extra T-transformation along the $T^2$ isometries parametrized by the coordinates $\varphi^I$ $(I=2,3)$ and make the following field re-definitions (which are complex in the case of the fermions)\footnote{The overall sign in these re-definitions is only fixed once we consider terms of the form ${\rm d}\varphi\wedge\upsilon {\rm d}\upsilon$, which arise for example in the AdS$_2\times S^2\times T^6$ case considered in the next section.}
\be\label{rups}
\varphi^{I'}\ \rightarrow\ -\varphi^{I'}~,\quad
\mathbb P_+v^{\pm}\ \rightarrow\ {\rm i}\mathbb P_+ v^{\pm}~,\quad
 \mathbb P_- v^{\pm}\ \rightarrow\ {\rm i}\mathbb P_-v^{\mp}~,
\ee
where $\mathbb P_\pm$ are given in \eqref{eq:Projector}.

The technical reason requiring the T-dualisation of $\varphi^I$ is the following. The gamma matrices $\Gamma^2$ and $\Gamma^3$ appearing in terms like $J\Gamma^{I}\upsilon$ of \eqref{L1v} do not commute with the projectors $\mathbb P_\pm$ which define the supercoset currents in accordance with \eqref{PRO}, namely
\be\label{JP}
{ \mathbb P_+J_Q}\ =\ J_Q~,\quad { \mathbb P_-J_{\hat Q, \hat S}}\ =\ J_{\hat Q,\hat S}~,\quad \mathbb P_\pm \Gamma^I\ =\ \Gamma^I\mathbb P_\mp~.
 \ee
As a result, the corresponding terms do not map to themselves upon dualisation along the supercoset directions. This gets corrected by the further dualisation of the $T^2$ coordinates $\varphi^I$. Note that these $T^2$ directions are exactly the `descendants' of the Minkowski boundary of AdS$_5$ which we T-dualised in the AdS$_5 \times S^5$ case. On the other hand, if we were to choose the gauge in which all the non-supercoset fermions are put to zero ($\upsilon=0$) we would not directly see the need to T-dualise half of the torus directions, since in this gauge the $T^4$ sector almost decouples from the supercoset sigma model (the two sectors are only related via coupling to the intrinsic worldsheet metric, or, equivalently, via the Virasoro constraints \cite{Babichenko:2009dk}). The need to T-dualise half of the $T^4$ directions is also in agreement with the results of \cite{OColgain:2012ca} on the T-self-duality of the AdS$_3\times S^3 \times T^4$ backgrounds (see Section 7).

We then proceed by considering the part of the Lagrangian which is quadratic in $\upsilon$. From \eqref{SSL11}, again expressing the 10-dimensional geometric objects in terms of the supercoset currents, we have
\begin{subequations}
\begin{equation}\label{L222}
\begin{aligned}
\mathcal L^{(2)}\ &=\
\tfrac12{*(J_{\hat S}-J_Q)}\wedge W^{--}(J_{\hat S}-J_Q) +\tfrac12{*J_{\hat Q}}\wedge W^{++}J_{\hat Q} -{*(J_{\hat S}-J_Q)}\wedge W^{-+}J_{\hat Q}\,-\\
&\kern1cm
-\tfrac12(J_{\hat S}-J_Q)\wedge W^{+-}(J_{\hat S}-J_Q) +\tfrac12J_{\hat Q}\wedge W^{+-}J_{\hat Q}\, +\\
&\kern1cm +\tfrac12(J_{\hat S}-J_Q)\wedge (W^{++}-W^{--})J_{\hat Q}
-\tfrac{\rm i}{2}{*E^A}\wedge\nabla\upsilon^+\Gamma_A\upsilon^-\,-\\
&\kern1cm
-\tfrac{i}{2}{*E^A}\wedge\nabla\upsilon^-\Gamma_A\upsilon^+
-\tfrac{\rm i}{2}E^A\wedge \nabla\upsilon^+\Gamma_A\upsilon^+
-\tfrac{\rm i}{2}E^A\wedge\nabla\upsilon^-\Gamma_A\upsilon^-\,-\\
&\kern1cm
-\tfrac{1}{8}{*E^A}\wedge E^B\,(\upsilon^++\upsilon^-)\Gamma_B(1-\Gamma^{2389})\Gamma^{01234}\Gamma_A(\upsilon^+-\upsilon^-)\, +\\
&\kern1cm+\tfrac{1}{8}E^A\wedge E^B\,(\upsilon^++\upsilon^-)\Gamma_B(1-\Gamma^{2389})\Gamma^{01234}\Gamma_A(\upsilon^+-\upsilon^-)
\end{aligned}
\end{equation}
where
\begin{equation}
 W^{\pm\pm}\ :=\ (\Gamma^A\upsilon^\pm)\,(\upsilon^\pm\Gamma_A)~, \quad W^{\pm\mp}\ :=\ (\Gamma^A\upsilon^\pm)\,(\upsilon^\mp\Gamma_A)~.
\end{equation}
\end{subequations}

Upon a somewhat lengthy calculation, one can prove that this expression maps into itself when we dualise along $x$, $\theta$, $\varphi^2$, and $\varphi^3$, and use the mapping \eqref{rups}.

This proves the self-duality of the type IIB AdS$_3\times S^3 \times T^4$ superstring sigma model with $F_5$-flux under a T-duality transformation that involves two coordinates $x^m$ of the Minkowski boundary of AdS$_3$, four fermionic directions $\theta$ and two torus directions $\varphi^I$.

To conclude we have checked that the Green-Schwarz string action is T-self-dual also in the case in which the AdS$_3\times S^3 \times T^4$ background is supported not by the $F_5$-flux \eqref{F5} but by the Ramond--Ramond 3-form flux
\be\label{F3}
F_3\ =\ { \tfrac 13}(\varepsilon_{cba} e^{a}\wedge e^b\wedge e^c +\varepsilon_{\hat c\hat b\hat a} e^{\hat a}\wedge e^{\hat b}\wedge e^{\hat c})~.
\ee
In this case, the fermionic vielbeins of the supercoset split as follows
\be\label{FE3}
 E^1\ =\ \tfrac 1{\sqrt 2}(J_S+J_{\hat S}-J_{Q}-J_{\hat Q})~, \quad  E^2\ =\ \tfrac {\rm i}{\sqrt 2}\Gamma^{23}(J_S-J_{\hat S}+J_{Q}-J_{\hat Q})\,.
\ee
This is consistent with the fact that the two backgrounds are related by toroidal T-dualities. Moreover, by T-dualising an odd number of toroidal directions of the type IIB background, one gets superstring sigma models on type IIA AdS$_3\times S^3\times T^4$ backgrounds with an $F_4$-flux, which are also invariant under the combined fermionic and bosonic T-dualities.

\section{Self-duality of AdS$_2\times S^2\times T^6$ superstrings}\label{ADS2}

AdS$_2\times S^2\times T^6$ backgrounds are solutions of type IIA and type IIB supergravity related by T-duality \cite{Tseytlin:1996bh,Klebanov:1996mh,Gauntlett:1996pb,Duff:1998us,Boonstra:1998yu,Lee:1999yu}. They preserve only 1/4, that is, 8 of 32 supersymmetries in ten dimensions that generate the $PSU(1,1|2)$ isometries.
As above, we set the radii of AdS$_2$ and $S^2$ to one, so their curvatures are
\be
 R^{ab}_{{\rm AdS}_2}\ =\ - e^{  a}\wedge e^{  b}~,\quad
R^{\hat a\hat b}_{S^2}\ =\ e^{\hat a}\wedge e^{\hat b}~.
\ee
The complete Green--Schwarz AdS$_2\times S^2\times T^6$ superstring action is not a $\frac{PSU(1,1|2)}{SO(1,1)\times U(1)}$ supercoset sigma model, since, in addition to four bosonic and eight fermionic supercoset modes it also contains non-trivially coupled bosonic modes associated with the $T^6$ directions and 24 fermionic modes associated with broken supersymmetries \cite{Sorokin:2011rr}. Note that in this case 16 independent kappa symmetry transformations are not enough to put the 24 non-supercoset fermions to zero. The theory reduces to the $\frac{PSU(1,1|2)}{SO(1,1)\times U(1)}$ supercoset sigma model plus the decoupled $T^6$-sector if we put these string fermionic modes to zero by hand.

We will consider T-dualisation both in type IIA and type IIB backgrounds. While self-duality of type IIB solutions easily follow from self-duality of the master AdS$_5 \times S^5$ model as before, the type IIA case requires a separated treatment.

\subsection{Type IIB backgrounds}\label{IIB}

One of the examples of a type IIB superstring in AdS$_2\times S^2\times T^6$ can be obtained by a formal `compactification' to $T^2$ of two spatial coordinates of the AdS$_3\times S^3\times T^4$ background, the coordinate $x^1$ of the AdS$_3$ Minkowski boundary and the coordinate $y^7$ of $S^3$.
This background is supported by the following Ramond--Ramond 5--form flux
\be\label{F52}
F_5\ =\ \tfrac 12 (1+*)\varepsilon_{ab}e^a\wedge e^b\wedge {\rm Re}({\rm Vol}_3)~,
\ee
where $a,b=0,4$ and $\hat a,\hat b=5,6$ are now the AdS$_2 \times S^2$ indices and  ${\rm Vol}_3={\rm d}(\varphi^1+{\rm i}\varphi^7)\wedge {\rm d} (\varphi^2+{\rm i}\varphi^8)\wedge {\rm d}(\varphi^3-{\rm i}\varphi^9)$ is the holomorphic 3-form on $T^6$.

To split the 32 fermionic modes \eqref{tv} into 8 supercoset and 24 non-supercoset ones we use the additional projectors $\frac 12(1\pm \Gamma^{1278})$ and write
\be\label{tv2}
\vartheta^i \ =\ \tfrac 14 (1+ \Gamma^{1278})(1-\Gamma^{2389})\Theta^i\ =\ \mathcal{P}_4\Theta~,\quad \upsilon^i\ =\ (1-\mathcal P_4)\Theta^i~,
\ee
where $\mathcal P_4$ is a projector of rank 4 and $(1-\mathcal P_4)$ is its complementary of rank 12.

The $\mathfrak{psu}(1,1|2)$ superalgebra can be obtained by a truncation of the $\mathfrak{psu}(1,1|2)\oplus\mathfrak{psu}(1,1|2)$ algebra, exploiting a procedure similar to the one that gives $\mathfrak{psu}(1,1|2)\oplus\mathfrak{psu}(1,1|2)$ as a sub-superalgebra of $\mathfrak{psu}(2,2|4)$. We select the AdS$_3$ direction associated with the index $m=1$ and the $S^3$ direction associated with $\hat a=7$ as those to be compactified on an extra $T^2$. Then we remove from the $\mathfrak{psu}(1,1|2)\oplus\mathfrak{psu}(1,1|2)$ superalgebra all the bosonic generators carrying the indices $m=1$ and $\hat a = 7$ and halve the number of fermionic generators by projecting the fermionic $\mathfrak{psu}(1,1|2)\oplus\mathfrak{psu}(1,1|2)$ generators with the additional projector $\tfrac12 (1 + \Gamma^{1278})$. We are then left with eight supercharges $\mathcal Q=(Q,\hat Q, S, \hat S)$ associated with $\vartheta$, obeying the projection property
\be\label{QP8}
\mathcal Q\ =\ \mathcal Q\mathcal P_4~,
\ee
with ${\cal P}_4$ given in \eqref{tv2}. The algebra turns out to have the same form as \eqref{eq:PSU22-12-Bos}--\eqref{psu22-2} where we set $m,n,\ldots=0$ ($\eta_{00}=-1$), $\hat a , \hat b, \ldots= 5,6$ and project the supercharges with the ${\cal P}_4$ projector. For the reader's convenience, we give the explicit form of the algebra in Appendix \ref{ads2gamma0}.

\paragraph{Verification of self-duality and dual representative.}
The action for the type IIB superstring in AdS$_2\times S^2 \times T^6$ with Ramond--Ramond flux given in \eqref{F52} has exactly the same form as \eqref{SSL11} in which the fermionic fields are now assumed to be projected as in \eqref{tv2}, ${\cal L}^{\rm coset}$ is still given by \eqref{L22}, while the covariant derivative $\mathcal D$ takes the form
\be\label{cD2}
\mathcal D\upsilon\ :=\ {\rm d}\upsilon+\tfrac 12(J_P+J_K)\Gamma^0\Gamma^4\upsilon+J_{R_{56}}\Gamma^{56}\upsilon-\tfrac {\rm i}2 E^A\mathcal P_4\Gamma^{01234}\Gamma_A\sigma^2\upsilon~.
\ee
Because of the same structure, the proof of the self-duality of the $\frac{PSU(1,1|2)}{SO(1,1)\times U(1)}$ supercoset part of the AdS$_2\times S^2 \times T^6$ action is exactly the same as that of the AdS$_5\times S^5$ and AdS$_3\times S^3$ actions. Since now $\xi$ is a two-component Gra{\ss}mann-odd spinor, the supercoset element \eqref{tildeg} which is the building block of the dual AdS$_2\times S^2$ supercoset action simplifies to
\be\label{tildeg2}
\tilde g\ :=\ {\rm e}^{\tilde x^0 K-{\rm i}\tilde\theta \Gamma^4 S}{\rm e}^B{\rm e}^{-\xi Q}~.
\ee
Similarly, the analysis of the T-dualisation of the type IIB AdS$_2\times S^2 \times T^6$ action in the presence of the non-supercoset fermions proceeds in the same way as for the AdS$_3\times S^3 \times T^4$ action \eqref{SSL11} (again in the gauge $\xi=0$). At the first order in $\upsilon$, the AdS$_2\times S^2 \times T^6$ Lagrangian has the form \eqref{L1v}, in which $\varphi^{a'}$ ($a'=1,2,3,7,8,9$) parametrize $T^6$. Together with the supercoset sector, it is not hard to see that this part of the string action transforms into itself under combined T-duality of the time direction $x^0$, two fermionic directions $\theta$ and three torus directions $\varphi^I$ ($I=1,2,3$), upon the re-definition \eqref{rups}. Then, one can check that the self-duality also persists when the second order terms in $\upsilon$ are taken into account in the type IIB AdS$_2\times S^2 \times T^6$ superstring action.

\subsection{Type IIA backgrounds}

For completeness, we will now consider  examples of strings in  type IIA  AdS$_2\times S^2\times T^6$ backgrounds.

\paragraph{\mathversion{bold}Background with $F_2$- and $F_4$-flux.}
Let us start with the background which is supported by the following Ramond--Ramond 2-form and 4-form fluxes (see \cite{Sorokin:2011rr,Cagnazzo:2011at} for more details, modulo signs)
\be\label{RRF}
F_2\ =\ \tfrac{1}{2}\varepsilon_{ab}e^b\wedge e^a~,\quad
F_4\ =\ \tfrac{1}{2}\varepsilon_{\hat a\hat b}e^{\hat b}\wedge e^{\hat a}\wedge J_2~,
\ee
where as above $a,b=0,4$, $\hat a,\hat b=5,6$
and $a', b'=1,2,3,7,8,9$ are the AdS$_2$, $S^2$ and $T^6$ indices, respectively, and
$J_2=\frac{1}{2}{\rm d}\varphi^{b'}\wedge{\rm d}\varphi^{a'}J_{a'b'}$ is the K\"ahler form  on $T^6$. For our purposes it is convenient to choose a basis in $T^6$ in which \footnote{See Appendix \ref{A} for the realization of 10-dimensional ($32\times 32$)-matrices $\Gamma^A$ used in this section.}
\be\label{Jbasis}
J_{a'b'}\Gamma^{a'b'}\ =\ 2(\Gamma^{17}+\Gamma^{28}-\Gamma^{39})\,.
\ee
The type IIA AdS$_2\times S^2\times T^6$ superstring action is obtained by substituting into the generic form of the Green--Schwarz action \eqref{GSaction} the worldsheet pullbacks of the supervielbeins $\mathcal E^A(X,\Theta)$ and the Neveu--Schwarz--Neveu--Schwarz 2-form $B_2(X,\Theta)$ that describe the type IIA AdS$_2\times S^2\times T^6$ background. As in the type IIB cases, to find an explicit form of $\mathcal E^A(X,\Theta)$ and $B_2(X,\Theta)$ we split the 32-component Majorana spinor $\Theta^{\hat\alpha}$ into an 8-component spinor $\vartheta$ corresponding to eight supersymmetries preserved by the AdS$_2\times S^2\times T^6$ background and parametrizing the Gra{\ss}mann-odd directions of the coset $\frac{PSU(1,1|2)}{SO(1,1)\times U(1)}$, and a 24-component spinor $\upsilon$ corresponding to the broken supersymmetries as follows
\begin{subequations}
\begin{equation}\label{8+24}
\vartheta\ =\ {\mathcal P_8}\,\Theta~,\quad \upsilon\ =\ (1-\mathcal P_{8})\,\Theta\ =\ \mathcal P_{24}\Theta
\end{equation}
where
\be\label{P8}
\mathcal P_8\ :=\ \tfrac 18(2-{\rm i}J_{a'b'}\Gamma^{a'b'}\gamma^{(7)})\quad\mbox{with}\quad \gamma^{(7)}\ =\ {\rm i}\Gamma^{123789}
\ee
\end{subequations}
is the rank-8 projection matrix.

The fermionic 32-component supervielbeins $\mathcal E^{\hat \alpha}$ split accordingly as follows
\be\label{fsv}
\mathcal E_8(X,\vartheta,\upsilon)\ =\  \mathcal P_8\mathcal E ~, \quad \mathcal E_{24}(X,\vartheta,\upsilon)\ =\  \mathcal P_{24}\mathcal E~.
\ee
If we set $\upsilon=0$ in \eqref{fsv}, the supervielbeins $\mathcal E_{24}$ vanish, while the 8-component
$\mathcal E_8|_{\upsilon=0}= E(x,y,\vartheta)$ describe, together with the bosonic supervielbeins $\mathcal E^{a}|_{\upsilon=0}=E^{a}(x,y,\vartheta)$,
 the geometry of the supercoset $\frac{PSU(1,1|2)}{SO(1,1)\times U(1)}$. Finally, at $\upsilon=0$ the vielbeins $\mathcal E^{a'}$ along the $T^6$ directions reduce  to the differentials of the $T^6$-coordinates $\mathcal E^{a'}|_{\upsilon=0}=E^{a'}={\rm d}\varphi^{a'}$.
As a result, when the non-supersymmetric fermionic modes $\upsilon$ are set to zero the $T^6$ sector completely decouples and the Green--Schwarz action reduces to the $\frac{PSU(1,1|2)}{SO(1,1)\times U(1)}$ sigma model.

We begin by discussing the T-dualisation of the supercoset action ($\upsilon = 0$). As a consequence of the particular definition of the $PSU(1,1|2)$ generators given in Appendix \ref{ads2gamma},  the $PSU(1,1|2)$ currents are related to the supervielbeins $E^a$, $E^{\hat\alpha}$ and the components of the spin connection $\Omega^{ab}$ as
\begin{subequations}
\be\label{bosonic+fermionic}
\begin{gathered}
{\rm AdS}_2: \qquad J_P\ =\ \tfrac 12 (E^0+\Omega^{04})~, \quad J_K\ =\ -\tfrac 12 (E^0-\Omega^{04})~,\quad J_D\ =\ E^4~,\\
S^2:\qquad J_{R_{\hat a}}\ =\ E^{\hat a}~, \quad J_{R_{56}}\ =\ -{ \tfrac12}\Omega^{56}~,\quad (\hat a=5,6)~,\\
J_Q\ =\  - \tfrac {1}{\sqrt 2} \mathbb P_+ (E^1-{\rm i}\Gamma^{123} E^2)~, \quad J_{\hat Q}\ =\  -\tfrac {1}{\sqrt 2}  \mathbb P_- (E^1+{\rm i}\Gamma^{123}  E^2)~,\\
J_S\ =\   \tfrac {1}{\sqrt 2} \mathbb P_+ (E^1+{\rm i}\Gamma^{123} E^2)~, \quad J_{\hat S}\ =\  \tfrac {1}{\sqrt 2}  \mathbb P_- (E^1-{\rm i}\Gamma^{123}  E^2)~.
\end{gathered}
\ee
Here the fermionic vielbeins $E^{i}$ with $(i=1,2)$ are Majorana--Weyl spinors with opposite chiralities
\be\label{MW}
E^1\ :=\ \tfrac 12 (1+\Gamma^{11})E~, \quad E^2\ :=\  \tfrac 12 (1-\Gamma^{11})E
\ee
\end{subequations}
and $\mathbb P_\pm$ are the projectors given in \eqref{eq:Projector}.

Note that each of the fermionic currents has two independent spinorial components only, and upon inverting the fermionic part in  \eqref{bosonic+fermionic}, we get
\be\label{E}
E^1\ =\ \tfrac{1}{\sqrt{2}}(J_S+J_{\hat S}-J_Q-J_{\hat Q})~,\quad E^2\ =\ \tfrac{\rm i}{\sqrt{2}}\Gamma^{123}(J_Q-J_{\hat Q}+J_S - J_{\hat S})~.
\ee
Since the identification of the $PSU(1,1|2)$ currents as given in Appendix \ref{ads2gamma} corresponds to the form of the $\mathfrak{psu}(1,1|2)$ superalgebra in the type IIB case, eqs. \eqref{psu11-1} and \eqref{psu11-2}, the AdS$_2\times S^2$ supercoset Lagrangian $\mathcal L_{\rm coset}$ has the form $\eqref{GHaction1}$ with $\gamma=\Gamma^4$ (similar to that in the type IIB cases \eqref{L22}). Therefore, as in the other AdS$_d\times S^d$ cases, the proof of the T-self-duality of the $PSU(1,1|2)$ supercoset model is carried out using the generic procedure described in Section 2 with the dual coset element \eqref{tildeg1} being of the form \eqref{tildeg2}.

We now turn to the complete superstring action ($\upsilon \neq 0$). In order to find out how the $T^6$ sector couples to the supercoset sector via $\upsilon$, one should know the explicit dependence of $\mathcal E^A$, $\mathcal E^{\hat \alpha}$  and $B_2$ on $\upsilon$. For the case under consideration the expression of $\mathcal E^A$ and $\mathcal E^{\hat \alpha}$ up to the quadratic order in $\upsilon$ was derived in \cite{Cagnazzo:2011at} and the explicit form of $B_2$ can be easily computed following \cite{Wulff:2013kga}. We thus find that the expressions coincide with \eqref{hat E1}  and \eqref{hat B1} with the matrix $\sigma^3$ in \eqref{hat B1} substituted by $\Gamma^{11}$ and the covariant differential of $\upsilon$ given by
\be\label{cD1}
\begin{aligned}
\mathcal D\upsilon\ &=\  \nabla \upsilon +\tfrac 1{8} E^A(\tfrac 12 F_{BC}\Gamma^{BC}\Gamma^{11}+\tfrac 1{4!}F_{BCDE}\Gamma^{BCDE})\Gamma_A\upsilon\\
&=\ \nabla\upsilon-\tfrac 12 E^{a'}{\mathcal P_8}\Gamma^0\Gamma^4\Gamma^{11}\Gamma_{a'} \upsilon~.
\end{aligned}
\ee
Moreover, we have
\be
B^{\rm coset}_2(x,y,\vartheta)\ =\ \tfrac {\rm i}2 E\Gamma^0\Gamma^4\mathcal P_8 E~.
\ee
Upon substituting the expressions for the supervielbeins into the action \eqref{GSaction}, we get the following Lagrangian for the type IIA AdS$_2\times S^2 \times T^6$ superstring to the second order in the non-supercoset fermionic modes $\upsilon$
\be\label{SSL1}
\begin{aligned}
\mathcal L\ &=\ \mathcal L_{\rm coset}+\tfrac 12{*{\rm d}\varphi^{a'}}\wedge{\rm d}\varphi^{a'}-{\rm i}{*{\rm d}\varphi^{a'}}\wedge E\Gamma_{a'}\upsilon -{\rm i}{\rm d}\varphi^{a'}\wedge E\Gamma_{a'}\Gamma^{11}\upsilon -\tfrac 12 {*E}\wedge \Gamma^{a'}\upsilon\, E\Gamma_{a'}\upsilon \\
& \kern1cm -\tfrac 12 E\wedge \Gamma^{a'}\upsilon\,E\Gamma_{a'}\Gamma^{11}\upsilon\, -\tfrac {\rm i}2 {*E^A}\wedge \mathcal D\upsilon\Gamma_A\upsilon\, -\tfrac {\rm i}2E^A\wedge\mathcal D\upsilon\Gamma_A\Gamma^{11}\upsilon\,.
\end{aligned}
\ee
\if{}
&&\\
&=&L^{coset}+\frac 12\eta^{\mu\nu}\partial_\mu\varphi^{a'}\partial_\nu\varphi^{a'}-i\partial_\mu\varphi^{a'}\,E_\nu\Gamma_{a'}(\eta^{\mu\nu}-\varepsilon^{\mu\nu}\Gamma^{11})\upsilon
\nonumber\\
&& -\frac 12 E_{\mu}\Gamma^{a'}\upsilon\, E_{\nu}\Gamma_{a'}(\eta^{\mu\nu}-\varepsilon^{\mu\nu}\Gamma^{11})\upsilon-\frac i2 E_\mu^A\,\mathcal D_\nu\upsilon\Gamma_A(\eta^{\mu\nu}-\varepsilon^{\mu\nu}\Gamma^{11})\upsilon\,.\nonumber
\eea
\fi
Remarkably, this expression has the same form of the Lagrangian \eqref{SSL11} for the AdS$_3 \times S^3 \times T^4$ and type IIB AdS$_2 \times S^2 \times T^6$ cases, with $\sigma^3 \to \Gamma^{11}$. Therefore, the T-dualisation of the part of the action which includes the fermions $\upsilon$ is carried out in a way similar to those cases. In particular, the projectors \eqref{eq:Projector} commute in a different way with the two sets of the gamma matrices along $T^6$, namely
$\mathbb P_\pm\Gamma^{1,2,3}=\Gamma^{1,2,3}\mathbb P_\mp$ and $\mathbb P_\pm\Gamma^{7,8,9}=\Gamma^{7,8,9}\mathbb P_\pm$, and for the action to be mapped to itself, the combined bosonic and fermionic T-duality has to include the T-dualisation of three torus directions $\varphi^I$ ($I=1,2,3$).

In conclusion, the type IIA AdS$_2 \times S^2 \times T^6$ background supported by 2-form and 4-form flux \eqref{RRF} is self-dual under a suitable combination of bosonic and fermionic T-transformations. This is consistent with the fact that this background is related to the previous type IIB one by a bosonic T-duality transformation along torus directions.

We would like to point out that the choice of the relevant $\mathbbm{Z}_4$-automorphism of the $\mathfrak{psu}(1,1|2)$ superalgebra (see Appendix \ref{ads2gamma}) associated with the appropriate splitting \eqref{E} of the supercoset currents is crucial  for the proof of the self-duality of the string actions in the presence of the non-supercoset fermions $\upsilon$. A different (inappropriate) choice of $\mathbbm{Z}_4$-grading would make the proof of the self-duality of the complete superstring actions in AdS$_2\times S^2\times T^6$ much more complicated if at all possible. For instance, if the projectors $\mathbb P_\pm$ were to commute in the same way with all the $T^6$ gamma matrices one would encounter serious problems with the proper T-dualisation of the non-supercoset part of the string action.

\paragraph{\mathversion{bold}$F_4$-flux background.}
Finally,  let us mention the type IIA AdS$_2\times S^2\times T^6$ background which can be obtained by T-dualisation along $\varphi^1$ of the type IIB background with the $F_5$-flux \eqref{F52}. This type IIA background is supported by the $F_4$-flux
\be\label{F4}
F_4\ =\
\tfrac{1}{2}\varepsilon_{ab}e^be^a\wedge ({\rm d}\varphi^3\wedge {\rm d}\varphi^2-{\rm d}\varphi^8\wedge {\rm d}\varphi^9)
+\tfrac{1}{2}\varepsilon_{\hat a\hat b}e^{\hat b}e^{\hat a}\wedge ({\rm d}\varphi^9\wedge {\rm d}\varphi^2+{\rm d}\varphi^8\wedge {\rm d}\varphi^3)~.
\ee
In this background, the covariant differential \eqref{cD1} of the fermionic modes $\upsilon$ and the supercoset part $B^{\rm coset}_2$ of the Neveu--Schwarz--Neveu--Schwarz form take the following form
\be
\mathcal D\upsilon\ =\ \nabla\upsilon-\tfrac 12 E^{a'}{\mathcal P_8}\Gamma^{0234}\Gamma_{a'} \upsilon~,
 \quad  B^{\rm coset}_2(x,y,\vartheta)\ =\ -\tfrac {\rm i}2 E\Gamma^{0234}\Gamma^{11}\mathcal P_8 E
\ee
while the supercoset parts of the fermionic supervielbeins split into the currents $J_{Q,\hat Q, S,\hat S}$ as follows
\be\label{E5}
\begin{gathered}
E^1\ =\ \tfrac 12(1+\Gamma^{11})E\ =\ \tfrac{1}{\sqrt{2}}(J_S+J_{\hat S}-J_Q-J_{\hat Q})~,\\
E^2\ =\ \tfrac 12(1-\Gamma^{11})E\ =\  \tfrac{\rm i}{\sqrt{2}}\,\Gamma^{1}(J_Q-J_{\hat Q}+J_S - J_{\hat S}).
\end{gathered}
\ee
With these definitions, the proof of the invariance of the string action in this background under the combined bosonic and fermionic T-dualities proceeds exactly as in the previous cases.

\section{Self-duality of AdS$_d\times S^d\times S^d\times T^{10-3d}$ superstrings}\label{sec:d21}

In this section, we will extend the previous discussion to the cases of superstrings on AdS$_d \times S^d \times S^d\times T^{10-3d}$ $(d=2,3)$. As their AdS$_d \times S^d \times T^{10-2d}$ counterparts, these backgrounds preserve $1/4$ and $1/2$ of the 10-dimensional supersymmetry and can be supported by  either Neveu--Schwarz--Neveu--Schwarz or Ramond--Ramond fluxes \cite{Cowdall:1998bu,Boonstra:1998yu,Gauntlett:1998kc,deBoer:1999rh,Gukov:2004ym}. Here we will consider the latter ones. For instance, a type IIB   AdS$_3 \times S^3 \times S^3\times S^{1}$  background can be supported by the following $F_3$ flux
\be\label{ExF3}
F_3\ =\ { \tfrac 13}\Big(\varepsilon_{cba} e^{a}\wedge e^b\wedge e^c +\frac {R_{AdS}}{R_+}\varepsilon_{\hat c\hat b\hat a} e^{\hat a}\wedge e^{\hat b}\wedge e^{\hat c}+\frac {R_{AdS}}{R_-}\varepsilon_{c'b'a'} e^{a'}\wedge e^{b'}\wedge e^{c'}\Big),
\ee
where $\hat a$ and $a'$ are, respectively the tangent space indices of the two three-spheres and $R_{\pm}$ are their radii.

Upon the T-dualization of the above background along the $S^1$ one gets the type IIA AdS$_3 \times S^3 \times S^3\times S^{1}$  with the $F_4$-flux
\be\label{ExF4}
F_4\ =\ {\rm d}\varphi^9\wedge { \tfrac 13}\Big(\varepsilon_{cba} e^{a}\wedge e^b\wedge e^c +\frac {R_{AdS}}{R_+}\varepsilon_{\hat c\hat b\hat a} e^{\hat a}\wedge e^{\hat b}\wedge e^{\hat c}+\frac {R_{AdS}}{R_-}\varepsilon_{c'b'a'} e^{a'}\wedge e^{b'}\wedge e^{c'}\Big).
\ee

Because of the technical complexity, in these cases we will put the `non-supersymmetric' fermionic modes of the string to zero ($\upsilon=0$) by fixing a kappa symmetry gauge in the $d=3$ case and `by hand' in the $d=2$ case. Then the $T^{10-3d}$-sector decouples (modulo the Virasoro constraints) and we may concentrate on the AdS$_d \times S^d \times S^d$ sectors described by supercoset sigma models with the isometries governed by the exceptional Lie supergroups $D(2,1;\alpha)$ (for $d=2$) and $D(2,1;\alpha)\times D(2,1;\alpha)$ (for $d=3$). In particular, we shall show that they are also T-self-dual under combined bosonic and fermionic T-dualities, provided that T-dualisation involves one of the spheres $S^d$, the latter causing some additional technical difficulties.

\subsection{Self-duality for ${\rm AdS}_2\times S^2\times S^2$ }

\paragraph{Supercoset structure.}
The sigma model on ${\rm AdS}_2\times S^2\times S^2$ is based on the supercoset
\begin{equation}\label{eq: supercosetstructureforads2s2m2}
\frac{\text D(2,1;\alpha)}{SO(1,1)\times SO(2)\times SO(2)}~.
\end{equation}
To construct the corresponding action and analyse its T-duality properties, let us discuss the Lie superalgebra $\mathfrak{d}(2,1;\alpha)$ of $D(2,1;\alpha)$. For general properties of the exceptional Lie superalgebra $\mathfrak{d}(2,1;\alpha)$ see {\it e.g.}~\cite{Frappat:1996pb,Ivanov:2015iia}. For the 10-dimensional supergravity solutions under consideration, the values of the parameter $\alpha$ are restricted to the interval $[0,1]$.\footnote{For superbackgrounds whose isometries are governed by $\mathfrak{d}(2,1;\alpha)$ with other values of $\alpha$ see {\it e.g.}~\cite{Bandos:2002nn,Butter:2015tra}.} They determine the relation between the radii of ${\rm AdS}_2\times S_+^2\times S_-^2$,
\begin{equation}
\alpha\ =\ \frac{R^2_{\rm AdS}}{R^2_-} \quad\mbox{and}\quad  {1-\alpha}\ =\ \frac {R^2_{\rm AdS}}{R^2_+}~.
\end{equation}
In order to avoid confusion between the parameter $\alpha$  and the spinor index $\alpha$, in what follows we will set  $\alpha:=\cos^2(\tau) := c^2$ and $1-\alpha:=\sin^2(\tau):=s^2$, respectively.

\paragraph{\mathversion{bold}Lie superalgebra $\mathfrak{d}(2,1;c^2)$.}
The maximal Gra{\ss}mann-even subalgebra of the Lie superalgebra $\mathfrak{d}(2,1;c^2)$  is $\mathfrak{sl}(2,\mathbbm{R})\oplus\mathfrak{su}(2)\oplus\mathfrak{su}(2)$, and we set $\mathfrak{sl}(2,\mathbbm{R}):=\langle P,K,D\rangle$, $\mathfrak{su}(2):=\langle L_a\rangle$, and $\mathfrak{su}(2):=\langle {R^\alpha}_\beta\rangle$, respectively, for $a,b,\ldots=1,2,3$ and $\alpha,\beta,\ldots=1,2$. The corresponding commutation relations are\footnote{One could also start from the 10-dimensional form of the algebra analogous to \eqref{psu22-0} for the $\mathfrak{psu}(2,2|4)$ case. This form follows directly from the general construction of the symmetric space superisometry algebras of \cite{Wulff:2015mwa} upon inserting the form of the fluxes.}
\begin{subequations}\label{eq:d21alg}
\begin{equation}
\begin{gathered}
 [D,P]\ =\ P~,\quad [D,K]\ =\ -K~,\quad [P,K]\ =\ 2D~,\\
 [L_+,L_-]\ =\ -2{\rm i}L_3~,\quad [L_3,L_\pm]\ =\ \pm{\rm i}L_\pm~,\quad L_\pm\ :=\ {\rm i}L_1\pm L_2~,\\
 [{R^\alpha}_\beta,{R^\gamma}_\delta]\ =\ {\rm i}({\delta^\gamma}_\beta {R^\alpha}_\delta-{\delta^\alpha}_\delta {R^\gamma}_\beta)~.
\end{gathered}
\end{equation}
Furthermore,  $\mathfrak{d}(2,1;c^2)$ contains eight fermionic generators which we denote by $Q_\alpha$, $\hat Q_\alpha$, $S_\alpha$, and $\hat S_\alpha$, respectively.  Letting $\sigma^{1,2,3}_{\alpha\beta}$ be the Pauli matrices\footnote{We lower and raise Greek indices using $\epsilon_{\alpha\beta}={\rm i}\sigma^2_{\alpha\beta}$ and $\epsilon^{\alpha\beta}={\rm i}\sigma^{2\,\alpha\beta}$ with $\epsilon^{\alpha\gamma}\epsilon_{\gamma\beta}={\delta^\alpha}_\beta$, so that {\it e.g.}~$ {\sigma^{1\,\alpha}}_\beta\ :=\ {\rm i}\sigma^{2\,\alpha\gamma}\sigma^1_{\gamma\beta}\ =\ -{\rm i}\sigma^{1\,\alpha\gamma}\sigma^2_{\gamma\beta}$.}, the remaining non-vanishing (anti-)commutation relations of $\mathfrak{d}(2,1;c^2)$ are given by
\begin{equation}\label{eq:gb}
\begin{gathered}
 \{Q_\alpha,\hat Q_\beta\}\ =\ -\sigma^2_{\alpha\beta} P~,~~
  \{S_\alpha,\hat S_\beta\}\ =\ -\sigma^2_{\alpha\beta} K~,\\
  \{Q_\alpha,\hat S_\beta\}\ =\ -c^2\sigma^2_{\alpha\beta} L_+~,~~
  \{\hat Q_\alpha,S_\beta\}\ =\ c^2\sigma^2_{\alpha\beta} L_-~,\\
  \{Q_\alpha,S_\beta\}\ =\  -\sigma^2_{\alpha\beta}(D+{\rm i} c^2L_3)-{\rm i} s^2 \sigma^2_{\alpha\gamma}{R^\gamma}_\beta~,\\
  \{\hat Q_\alpha,\hat S_\beta\}\ =\  \sigma^2_{\alpha\beta}(D-{\rm i}c^2L_3)+{\rm i} s^2 \sigma^2_{\alpha\gamma}{R^\gamma}_\beta~,\\
  [P,S_\alpha]\ =\ -\hat Q_\alpha~,~~
  [P,\hat S_\alpha]\ =\ -Q_\alpha~,~~
  [K,Q_\alpha]\ =\ -\hat S_\alpha~,~~
  [K,\hat Q_\alpha]\ =\ -S_\alpha~,\\
  [D,Q_\alpha]\ =\ \tfrac12 Q_\alpha~,~~
  [D,\hat Q_\alpha]\ =\ \tfrac12 \hat Q_\alpha~,~~
  [D,S_\alpha]\ =\ -\tfrac12 S_\alpha~,~~
  [D,\hat S_\alpha]\ =\ -\tfrac12 \hat S_\alpha~,\\
  [L_3,Q_\alpha]\ =\ \tfrac{\rm i}{2} Q_\alpha~,~~
  [L_3,\hat Q_\alpha]\ =\ -\tfrac{\rm i}{2} \hat Q_\alpha~,~~
  [L_3,S_\alpha]\ =\ -\tfrac{\rm i}{2} S_\alpha~,~~
  [L_3,\hat S_\alpha]\ =\ \tfrac{\rm i}{2} \hat S_\alpha~,\\
  [L_+,S_\alpha]\ =\ \hat S_\alpha~,~~
    [L_-,\hat S_\alpha]\ =\ - S_\alpha~,~~
  [L_-,Q_\alpha]\ =\ \hat Q_\alpha~,~~
    [L_+,\hat Q_\alpha]\ =\ - Q_\alpha~,\\
  [{R^\alpha}_\beta,T_\gamma]\ =\ -{\rm i}({\delta^\alpha}_\gamma T_\beta-\tfrac12 {\delta^\alpha}_\beta T_\gamma)~,~~\mbox{for}~~ T_\alpha\ \in\ \{Q_\alpha,\hat Q_\alpha,S_\alpha,\hat S_\alpha\}~.
\end{gathered}
\end{equation}
\end{subequations}
The bosonic generators $P$, $K$, $D$, and $L_a$ are skew-Hermitian while $({R^1}_1)^\dagger={R^2}_2$ and $({R^1}_2)^\dagger=-{R^2}_1$. The fermionic generators enjoy the reality conditions $Q_1^\dagger = \hat Q_2$, $Q_2^\dagger=-\hat Q_1$ and $S_1^\dagger = -\hat S_2$, $S_2^\dagger=\hat S_1$. It is straightforward to check that the superalgebra \eqref{eq:d21alg} is invariant under these reality conditions. Notice also that in the limit  $c^2\to 0$, we recover the superalgebra $\mathfrak{psu}(1,1|2)$ discussed in Section \ref{IIB} (modulo some obvious re-definitions) since in that limit the generators $L_\pm$ and $L_3$ decouple.

Furthermore, the non-vanishing components of the invariant form of $\mathfrak{d}(2,1;c^2)$ that is compatible with the above choice of the basis are
\begin{equation}\label{eq:IFD21}
\begin{gathered}
 {\rm Str}(PK)\ =\ 2~,~~{\rm Str}(DD)\ =\ 1~,\\
 {\rm Str}(L_+L_-)\ =\ -\tfrac{2}{c^2}~,~~{\rm Str}(L_3L_3)\ =\ \tfrac{1}{c^2}~,\\
 {\rm Str}({R^\alpha}_\beta{R^\gamma}_\delta)\ =\ \tfrac{2}{s^2}\big({\delta^\alpha}_\delta {\delta^\gamma}_\beta-\tfrac12 {\delta^\alpha}_\beta {\delta^\gamma}_\delta\big)~,\\
 {\rm Str}(Q_\alpha S_\beta)\ =\ -2\sigma^2_{\alpha\beta}~,~~
  {\rm Str}(\hat Q_\alpha \hat S_\beta)\ =\ 2\sigma^2_{\alpha\beta}~.
 \end{gathered}
\end{equation}

\paragraph{\mathversion{bold}$\mathbbm{Z}_4$-grading and order-4 automorphism.}
In order to formulate the supercoset action based on \eqref{eq: supercosetstructureforads2s2m2}, we need to fix a $\mathbbm{Z}_4$-grading of the superalgebra $\mathfrak{d}(2,1;c^2)\otimes\mathbbm{C}\cong\bigoplus_{m=0}^3\mathfrak{g}_{(m)}$.
\begin{subequations}
In view of \eqref{eq:GenZ4} we choose the following decomposition
\begin{equation}\label{eq:D21Z4}
\begin{gathered}
 \mathfrak{g}_{(0)}\ :=\ \big\langle P+K,L_++L_-,\sigma^1_{\gamma[\alpha}{R^\gamma}_{\beta]}\big\rangle~,\\
 \mathfrak{g}_{(1)}\ :=\ \big\langle Q_\alpha  - {\sigma^{1\,\beta}}_\alpha S_\beta, \hat Q_\alpha-{\sigma^{1\,\beta}}_\alpha \hat S_\beta\big\rangle~,\\
  \mathfrak{g}_{(2)}\ :=\ \big\langle P-K,D,L_+-L_-,L_3,\sigma^1_{\gamma(\alpha}{R^\gamma}_{\beta)}\big\rangle~,\\
   \mathfrak{g}_{(3)}\ :=\ \big\langle Q_\alpha+{\sigma^{1\,\beta}}_\alpha S_\beta, \hat Q_\alpha+{\sigma^{1\,\beta}}_\alpha \hat S_\beta\big\rangle~,\end{gathered}
\end{equation}
\end{subequations}
where brackets (respectively, parentheses) indicate normalised anti-symmetrisation (respectively, symmetrisation) of the enclosed indices. Notice that we have indeed $\mathfrak{g}_{(0)}\cong \mathfrak{so}(1,1)\oplus \mathfrak{so}(2)\oplus \mathfrak{so}(2)$. The order-4 automorphism $\Omega:\mathfrak{d}(2,1;c^2)\to \mathfrak{d}(2,1;c^2)$ associated with this $\mathbbm{Z}_4$-grading is given explicitly by
\begin{equation}
\begin{gathered}
 \Omega(P)\ =\ K~,~~\Omega(K)\ =\ P~,~~\Omega(D)\ =\ -D~,\\
 \Omega(L_3)\ =\ -L_3~,~~\Omega(L_\pm)\ =\ L_\mp~,~~
 \Omega({R^\alpha}_\beta)\ =\ \sigma^{1\,\alpha\gamma}\sigma^1_{\beta\delta} {R^\delta}_\gamma~,\\
  \Omega(Q_\alpha)\ =\ -{\rm i}{\sigma^{1\,\beta}}_\alpha S_\beta~,~~
  \Omega(\hat Q_\alpha)\ =\ -{\rm i}{\sigma^{1\,\beta}}_\alpha \hat S_\beta~,\\
   \Omega(S_\alpha)\ =\ -{\rm i}{\sigma^{1\,\beta}}_\alpha Q_\beta~,~~
  \Omega(\hat S_\alpha)\ =\ -{\rm i}{\sigma^{1\,\beta}}_\alpha \hat Q_\beta~.
\end{gathered}
\end{equation}

Furthermore,
\begin{equation}\label{eq:IFD21Z4}
\begin{gathered}
 {\rm Str}\big[(P\pm K)(P\pm K)\big]\ =\ \pm 4~,~~{\rm Str}(DD)\ =\ 1~,\\
 {\rm Str}\big[(L_+\pm L_-)(L_+\pm L_-)\big]\ =\ \mp\tfrac{4}{c^2}~,~~{\rm Str}(L_3L_3)\ =\ \tfrac{1}{c^2}~,\\
 {\rm Str}\big[(\sigma^1_{\mu[\alpha}{R^\mu}_{\beta]})(\sigma^1_{\nu[\gamma}{R^\nu}_{\delta]})\big]\ =\ -\tfrac{1}{s^2}\sigma^2_{\alpha\beta}\sigma^2_{\gamma\delta}~,\\
 {\rm Str}\big[(\sigma^1_{\mu(\alpha}{R^\mu}_{\beta)})(\sigma^1_{\nu(\gamma}{R^\nu}_{\delta)})\big]\ =\ -\tfrac{1}{s^2}(\sigma^1_{\alpha\beta}\sigma^1_{\gamma\delta}-\sigma^1_{\alpha\gamma}\sigma^1_{\beta\delta}-\sigma^1_{\alpha\delta}\sigma^1_{\gamma\beta})~,\\
 {\rm Str}\big[(Q_\alpha\pm{\sigma^{1\,\gamma}}_\alpha S_\gamma)(Q_\beta\mp{\sigma^{1\,\delta}}_\beta S_\delta)\big]\ =\ \mp {\rm 4i}\sigma^1_{\alpha\beta}~,\\
 {\rm Str}\big[(\hat Q_\alpha\pm{\sigma^{1\,\gamma}}_\alpha \hat S_\gamma)(\hat Q_\beta\mp {\sigma^{1\,\delta}}_\beta \hat S_\delta)\big]\ =\ \pm{\rm 4i}\sigma^1_{\alpha\beta}
   \end{gathered}
\end{equation}
which follow from \eqref{eq:IFD21}.

\paragraph{Coset representative and associated current.}
Next, we need to choose a coset representative $g$ for the supercoset space \eqref{eq: supercosetstructureforads2s2m2}. In view of \eqref{eq:gb}, the generators $P$, $Q_\alpha$, and $L_+$ are in involution\footnote{Note that the maximal Abelian subalgebra of $\mathfrak{d}(2,1;c^2)$ has two bosonic and two fermionic generators.} and, consequently, are associated with the directions along which we will perform T-dualisation. Following our general discussion in Section \ref{generic}, an appropriate form of the coset representative is\footnote{\label{12} To obtain the coset representative for AdS$_2\times S^2\times T^2$ from the representative \eqref{eq:D21CosRep} in the limit $c\to 0$, one first needs to re-scale the coordinates $\lambda_+\to c\lambda_+$, $\lambda_3\to c\lambda_3$, and ${\rho^\alpha}_\beta\to s {\rho^\alpha}_\beta$ and then perform the limit. In this limit, the second sphere $S^2$, whose metric becomes flat, decouples from the AdS$_2\times S^2$ supercoset and re-compactifies into $T^2$ which is part of $T^6$ of the backgrounds discussed in Section 5.}
\begin{equation}\label{eq:D21CosRep}
\begin{gathered}
  g\ :=\ {\rm e}^{xP+\theta^\alpha Q_\alpha+\lambda_+L_+}\, {\rm e}^B\, {\rm e}^{\xi^\alpha S_\alpha}~,\\
  {\rm e}^B\ :=\ {\rm e}^{\hat \theta^\alpha\hat Q_\alpha+\hat\xi^\alpha\hat S_\alpha}\, { |y|^D}\,{\rm e}^{-\lambda_3 L_3}\,{\rm e}^{-{\rho^\beta}_{\alpha}{R^\alpha}_\beta}~.
  \end{gathered}
\end{equation}
Here, we assume that both $\lambda_+$ and $\lambda_3$ are complex. This is merely a technical assumption which will facilitate the T-duality transformations below. Hence, we are essentially dealing with the complexification $SL(2,\mathbbm{C})/\mathbbm{C}^*$ of the coset $SO(3)/SO(2)\cong SU(2)/U(1)\cong S^2$, and from the point of view of fermionic T-duality, such a complexification is rather natural (see \cite{Berkovits:2008ic} for a similar case in AdS$_5\times S^5$). Note that the resulting line element on $SL(2,\mathbbm{C})/\mathbbm{C}^*$ is
\begin{equation}\label{smetric-1}
 ({\rm d}s)^2\ =\ \tfrac{1}{4c^2} \big[({\rm d}\lambda_3)^2+ {\rm e}^{2{\rm i} \lambda_3}({\rm d}\lambda_+)^2\big]~.
\end{equation}
Upon performing the change of coordinates $(\lambda_+,\lambda_3)\mapsto (\varphi,\vartheta)$,
\begin{equation}
\begin{aligned}
  \lambda_+\ =\ \frac{2\tan(\frac{\vartheta}{2})\sin(\varphi)}{1+2{\rm i}\tan(\frac{\vartheta}{2})\cos(\varphi)-\tan^2(\frac{\vartheta}{2})}~,\\
  {\rm e}^{-{\rm i}\lambda_3}\ =\ \frac{1+\tan^2(\frac{\vartheta}{2})}{1+2{\rm i}\tan(\frac{\vartheta}{2})\cos(\varphi)-\tan^2(\frac{\vartheta}{2})}
  \end{aligned}
\end{equation}
for $\varphi,\vartheta\in\mathbbm{C}$, we find the line element
\begin{equation}\label{smetric-2}
 ({\rm d}s)^2\ =\ \tfrac{1}{4c^2} \big[({\rm d}\vartheta)^2+ \sin^2(\vartheta)\, ({\rm d}\varphi)^2\big]~,
\end{equation}
which, upon considering the real  slice $\varphi^*=\varphi$ and $\vartheta^*=\vartheta$, becomes the standard line element on the two-sphere $S^2$.

The Maurer--Cartan form $J=g^{-1}{\rm d}g$ corresponding to the coset representative \eqref{eq:D21CosRep} is of the form
\begin{equation}
\begin{aligned}
 J\ &=\ {\rm e}^{-\xi^\alpha S_\alpha} J^{(0)} {\rm e}^{\xi^\alpha S_\alpha}+{\rm d}\xi^\alpha S_\alpha\\
 \ &=\ J^{(0)}-\xi^\alpha \big[ S_\alpha,J^{(0)}\big]+\tfrac{\rm i}{4}\xi^2 \sigma^{2\,\alpha\beta} \big\{ S_\alpha,\big[ S_\beta,J^{(0)}\big]\big\}+{\rm d}\xi^\alpha S_\alpha~,
 \end{aligned}
\end{equation}
where, as before, $J^{(0)}$ does not depend on the fermionic coordinate $\xi^\alpha$, and we have set $\xi^2:={\rm i}\sigma^2_{\alpha\beta}\xi^\alpha\xi^\beta$. The explicit form of the components of the current $J$ is given in the Appendix \ref D.

Using the $\mathbbm{Z}_4$-grading \eqref{eq:D21Z4}, the coset current $J$ decomposes according to $J=J_{(0)}+J_{(1)}+J_{(2)}+J_{(3)}$ with
\begin{equation}\label{eq:D21JDec}
\begin{aligned}
  J_{(0)}\ &=\ \tfrac12 (J_P+J_K)(P+K)+\tfrac12 (J_{L_+}+J_{L_-})(L_++L_-)-J_{{R^\alpha}_\beta}\sigma^{1\,\alpha\gamma}\sigma^1_{\delta[\gamma}{R^\delta}_{\beta]}~,\\
  J_{(1)}\ &=\ \tfrac12 (J_{Q_\alpha}-{\sigma^{1\,\alpha}}_\beta J_{S_\beta})(Q_\alpha-{\sigma^{1\,\beta}}_\alpha S_\beta)\tfrac12 (J_{\hat Q_\alpha}-{\sigma^{1\,\alpha}}_\beta J_{\hat S_\beta})(\hat Q_\alpha-{\sigma^{1\,\beta}}_\alpha \hat S_\beta)~,\\
   J_{(2)}\ &=\ \tfrac12 (J_P-J_K)(P-K)+J_D D+\tfrac12 (J_{L_+}-J_{L_-})(L_+-L_-)+J_{L_3}L_3\,-\\
     &\kern1cm -J_{{R^\alpha}_\beta}\sigma^{1\,\alpha\gamma}\sigma^1_{\delta(\gamma}{R^\delta}_{\beta)}~,\\
     J_{(3)}\ &=\  \tfrac12 (J_{Q_\alpha}+{\sigma^{1\,\alpha}}_\beta J_{S_\beta})(Q_\alpha+{\sigma^{1\,\beta}}_\alpha S_\beta)+\tfrac12 (J_{\hat Q_\alpha}+{\sigma^{1\,\alpha}}_\beta J_{\hat S_\beta})(\hat Q_\alpha+{\sigma^{1\,\beta}}_\alpha \hat S_\beta)~.
  \end{aligned}
\end{equation}

\paragraph{Supercoset action.}
Upon using the $\mathbb{Z}_4$-grading \eqref{eq:D21Z4} together with the invariant form \eqref{eq:IFD21} and the currents \eqref{eq:D21JDec},  the sigma model action \eqref{GHaction1} becomes
\begin{equation}\label{eq:GSD21Action}
\begin{aligned}
 S\ &=\ -\tfrac T2\int_\Sigma \Big\{-{*(J_P-J_K)}\wedge (J_P-J_K)+{*J_D}\wedge J_D \,+\\[-4pt]
    &\kern1.5cm +\tfrac{1}{c^2}*(J_{L_+}-J_{L_-})\wedge (J_{L_+}-J_{L_-}) +\tfrac{1}{c^2}{*J_{L_3}}\wedge  J_{L_3}\,+\\[2pt]
    &\kern2cm +\tfrac{1}{s^2}( {*J_{{R^\alpha}_\beta}}\wedge  J_{{R^\beta}_\alpha}-\sigma^{1\,\alpha\gamma}\sigma^1_{\beta\delta}
    {*J_{{R^\alpha}_\beta}}\wedge J_{{R^\gamma}_\delta})\,-\\[3pt]
    &\kern2.5cm -{\rm i}\sigma^1_{\alpha\beta}\big(J_{Q_\alpha}\wedge J_{Q_\beta}+J_{S_\alpha}\wedge J_{S_\beta} -J_{\hat Q_\alpha}\wedge J_{\hat Q_\beta}-J_{\hat S_\alpha}\wedge J_{\hat S_\beta}\big)\Big\}~.
 \end{aligned}
\end{equation}
Note that as in the `non-exceptional' cases (see \eqref{L22}), the matrix $(-\sigma^1)$ can be identified with the matrix $\Gamma^4\mathbb P$ along the AdS$_2$ radial direction with $\mathbb P$ being the projector which singles out eight unbroken supersymmetries of the background under consideration.

\paragraph{T-dualisation.}
Now, performing the T-dualisation of the action \eqref{eq:GSD21Action} following the general procedure described in Section \ref{generic}, upon some technically involved algebra, a field re-definition and using the Maurer--Cartan equations one can check that the resulting dual action has the same form as the initial one but with the currents  (see Appendix \ref D) constructed with the different coset element
\begin{equation}\label{eq:D21DualCosRep}
\begin{gathered}
  \tilde g\ :=\ {\rm e}^{\tilde x K-{\rm i}\sigma^{2\,\alpha\beta}\tilde\theta_{\alpha} S_\beta+\tilde \lambda_+L_-}\, {\rm e}^{B}\, {\rm e}^{{\sigma^{1\beta}}_\alpha  \xi^\alpha Q_\beta}~,
    \end{gathered}
\end{equation}
where, ${\rm e}^B$ is the same as in \eqref{eq:D21CosRep}. Therefore, the supercoset sigma model on AdS$_2\times S^2\times S^2$ is self-dual under the combined T-dualities along $x$, $\theta^\alpha$, and $\lambda_+$.

In the limit $c^2\to 0$, upon an appropriate re-scaling of the $J_L$-currents, the action reduces to the $\frac {PSU(1,1|2)}{SO(1,1)\times U(1)}$ supercoset sigma model considered in Section \ref{IIB}. In this limit, the dualised sphere $S^2$ gets `decompactified' into a $T^2$ torus which completely decouples from the AdS$_2\times S^2$ and fermionic sector.

\subsection{Self-duality for ${\rm AdS}_3\times S^3\times S^3$ }\label{sec:d21d21}

Considering the subsector of the ${\rm AdS}_3\times S^3\times S^3 \times S^1$ theory in which the string moves only in AdS$_3\times S^3\times S^3$ while its non-supersymmetric fermionic modes are gauge fixed to zero and the $S^1$-fluctuations decouple from the rest (modulo the Virasoro constraints), the T-dualisation process is almost identical to the just-presented discussion in the AdS$_2\times S^2\times S^2$ case, though the explicit calculations are technically more involved. Therefore, we refrain from giving any details here and instead, we just outline the basic steps and refer to Section \ref{sugra} for a supergravity treatment of this case.

The supercoset sigma model on ${\rm AdS}_3\times S^3\times S^3$ is based on the supercoset
\begin{equation}\label{eq: supercosetstructureforads3s3m3}
\frac{\text D(2,1;c^2)\times \text D(2,1;c^2)}{SO(1,2)\times SO(3)\times SO(3)}~.
\end{equation}
The Lie superalgebra  $\mathfrak{d}(2,1;c^2)\oplus \mathfrak{d}(2,1;c^2)$ (whose 10-dimensional form analogous to \eqref{psu22-0} can be found in \cite{Sundin:2012gc}) has  $\{P_m,D,K_m,L_a^\pm,{R^{\pm\,i}}_j\}$ for $m=0,1$, $a=1,2,3$, and $i,j=1,2$ as its bosonic generators and $\{Q_{i\alpha},S_{i\alpha},\hat Q_{i\alpha},\hat S_{i\alpha}\}$ for $\alpha=1,2$ as its fermionic generators, respectively. Here, the $L_a^\pm$ and ${R^{\pm\,i}}_j$ are the generators of $\mathfrak{so}(3)\oplus\mathfrak{so}(3)\oplus\mathfrak{so}(3)\oplus\mathfrak{so}(3)$. Furthermore, the generators $\{P_m, Q_{i\alpha},L^\pm:={\rm i}L^\pm_1+L^\pm_2\}$ are in involution\footnote{Note that the maximal Abelian subalgebra of $\mathfrak{d}(2,1;c^2)\oplus\mathfrak{d}(2,1;c^2)$ has four bosonic and four fermionic generators.} so that the coset representative  \eqref{g} will have the left factor of the form ${\rm e}^{x^m P_m+\theta^{i\alpha}Q_{i\alpha}+\lambda_+ L^++\lambda_- L^-}$.\footnote{See also our general discussion given in Section \ref{generic}.} The coordinates $x^m$ parametrize the 2-dimensional Minkowski boundary of AdS$_3$. Furthermore, as in the AdS$_2\times S^2\times S^2$ case, we shall work with the complexification $SO(4,\mathbbm{C})/SO(3,\mathbbm{C})$ of $SO(4)/SO(3)\cong [SU(2)\times SU(2)]/SU(2)\cong SU(2)\cong S^3$ and consequently, the coordinates $\lambda_\pm$ are assumed to be complex. The resulting line element on $SO(4,\mathbbm{C})/SO(3,\mathbbm{C})$ will be of the form
\begin{equation}\label{smetric-3}
 ({\rm d}s)^2\ =\ \tfrac{1}{4c^2} \big[({\rm d}\lambda_3)^2+ {\rm e}^{2{\rm i} \lambda_3}({\rm d}\lambda_+)^2+{\rm e}^{2{\rm i} \lambda_3}({\rm d}\lambda_-)^2\big]~.
\end{equation}
Upon choosing an appropriate $\mathbbm{Z}_4$-grading for \eqref{eq: supercosetstructureforads3s3m3}, T-duality is then performed along the bosonic directions $x^m$ and $\lambda_\pm$ and the fermionic directions $\theta^{i\alpha}$ following the same steps as in the previous subsection. The T-self-duality of the supercoset sigma model on ${\rm AdS}_3\times S^3\times S^3$ then follows. We have explicitly checked this up to the second order in the four-component fermions $\xi^{i\alpha}$, like in the AdS$_2\times S^2 \times S^2$ case. We believe that the invariance holds to the highest (4th-order) in $\xi^{i\alpha}$. This is supported by the fact that at $\alpha=0$, the model reduces to the AdS$_3 \times S^3$ supercoset sigma model times the torus sector, which have proved to be duality invariant. In the next section, we will also give additional evidence for the T-self-duality of the complete  AdS$_3\times S^3\times S^3\times S^1$ theory by proving the invariance under the combined T-duality of its supergravity background.

\section{Combined bosonic-fermionic T-duality of the Ramond--Ramond AdS$_d\times S^d \times M^{10-2d}$ backgrounds}\label{sugra}

In this final section, we shall prove (without alluding to the superstring sigma models) the invariance under the combined bosonic and fermionic T-duality of the AdS$_d\times S^d \times M^{10-2d}$ superbackgrounds with Ramond--Ramond  fluxes by applying the T-duality rules directly to the corresponding supergravity component fields. We will thus extend earlier results of \cite{Berkovits:2008ic,Bakhmatov:2009be,Godazgar:2010ph,Bakhmatov:2010fp,OColgain:2012ca} to the whole class of the Ramond--Ramond AdS$_d\times S^d \times M^{10-2d}$ superbackgrounds.

\subsection{Rules for fermionic T-duality}

\paragraph{Killing spinors.}
The T-dualisation of the component supergravity fields along the bosonic directions is carried out following the conventional rules \cite{Buscher:1987sk,Buscher:1987qj,Simon:1998az}.\footnote{For the generalization of the T-dualisation rules to the whole superspace supergravity see \cite{Bandos:2003bz}.} The generalization of these rules to fermionic T-duality was given in \cite{Berkovits:2008ic}. Specifically, the fermionic T-duality acts on the dilaton $\Phi(X)$ and the Ramond--Ramond  $p$-forms but leaves the metric and the Neveu--Schwarz--Neveu--Schwarz 2-form invariant. The directions along which we dualise are specified by the (Gra{\ss}mann-even) Killing spinors, denoted by $\Xi_\mu(X)$ in the following (with $\mu$ labelling their number), that generate the Abelian superisometries. This implies that the Killing spinors ought to satisfy the additional condition
\be\label{EGE}
\Xi_\mu \Gamma_A\Xi_\nu\ =\ 0 \quad \mbox{for all}\quad A,\mu,\nu \quad \mbox{with}\quad A=0,1,\ldots,9~.
\ee
This condition has non-trivial solutions if the Killing spinors are complex, thus manifesting the fact that they are associated with complex Gra{\ss}mann-odd directions in superspace.

The Killing spinor conditions themselves have the following form
\be\label{KS}
\begin{gathered}
\partial_M \Xi-\tfrac 14 \Omega_M^{AB}(X)\Gamma_{AB} \Xi\ =\ -\tfrac 18 \slashed F {\mathcal E}_M^A(X)\Gamma_A\Xi~,\\
\tfrac 1{16}\Gamma_A \,\slashed F\Gamma^A\Xi\ =\ 0\,
\end{gathered}
\ee
where $\Omega_M^{AB}(X)$ and ${\mathcal E}_M^A(X)$ are the spin connection and the bosonic vielbeins of the 10-dimensional background, and $\slashed F$ denotes the contribution of the Ramond--Ramond fluxes
\begin{equation}\label{S}
\slashed F\ =\ \begin{cases}
{\rm e}^\Phi\big(\frac12F_{AB}^{(2)}\Gamma^{AB}\Gamma_{11}+\frac{1}{4!}F_{ABC D}^{(4)}\Gamma^{ABCD}\big) & \mbox{type IIA}\\
-\frac{{\rm e}^\Phi}{2}(1+\Gamma^{11})\big({\rm i}F_A^{(1)}\Gamma^A\sigma^2+\frac{1}{3!}F_{ABC}^{(3)}\Gamma^{ABC}\sigma^1+\frac{\rm i}{2\cdot5!}F_{A\cdots E}^{(5)}\Gamma^{A\cdots E}\sigma^2\big) & \mbox{type IIB}
\end{cases}
\end{equation}
Note that the equations in \eqref{KS} are obtained by requiring that the supersymmetry transformations of the gravitino and the dilatino vanish and are determined by the geometry of the background and the values of the Ramond--Ramond fluxes.  The requirement of the integrability of the first equation determines a projector ${\mathcal P}_{8(d-1)}$ singling out the $8(d-1)$ fermionic isometries of the backgrounds of interest, as we discussed from the superalgebra perspective in the previous sections. The second equation in \eqref{KS} is then identically satisfied. Incidentally, in the backgrounds having non-zero $F_5$-flux only, the second equation in \eqref{KS} is actually identically zero because of gamma matrix identities.

\paragraph{Fermionic T-duality rules.}
Upon solving for the Killing spinor equations \eqref{KS}, one can derive from \footnote{Equations \eqref{C} determine the components $H_{M\mu\nu}$ of the field strength $H_3={\rm d}B_2$ of the Neveu--Schwarz--Neveu--Schwarz 2-form for the superbackgrounds under consideration.}
\begin{equation}\label{C}
\partial_M{\mathcal C}_{\mu\nu}\ =\
\begin{cases}
E_M^{A}\bar \Xi_\mu\Gamma_A\Gamma^{11}\Xi_\nu & \mbox{type IIA}\\
E_M^{A}\bar\Xi_\mu\Gamma_A\sigma^3\Xi_\nu& \mbox{type IIB}
\end{cases}
\end{equation}
the matrix ${\mathcal C}=({\mathcal C}_{\mu\nu}(X))$ which is formed by the components of the Neveu--Schwarz--Neveu--Schwarz 2-form $B_2$ along the Abelian fermionic isometries, that is,
\be\label{CB}
{\rm d}\theta^\mu\wedge {\rm d}\theta^{\nu} B_{\mu\nu}(X,\Theta)|_{\Theta=0}\ :=\  {\rm d}\theta^\mu\wedge {\rm d}\theta^{\nu} {\mathcal C}_{\mu\nu}(X)~.
\ee
Knowing the matrix ${\mathcal C}_{\mu\nu}$, one obtains the shift of the dilaton under the fermionic T-duality
\be\label{dPhi}
\Delta\Phi\ =\ \Phi'-\Phi\ =\ \tfrac 12 \log (\det{\mathcal C})\,
\ee
and of the Ramond--Ramond fields, which in our conventions is
\be\label{dF1}
\Delta F\ =\ \slashed F'-\slashed F\ =\ 8\Xi_\mu({\mathcal C}^{-1})^{\mu\nu}\Xi_\nu\Gamma~,
\ee
where $\Gamma$ is a certain product of gamma-matrices which has been used to split the fermionic $E^{(1,2)}$ currents into four pieces corresponding to the splitting of the superalgebra generators $\mathcal Q$  into $Q$, $\hat Q$, $S$, and $\hat S$, respectively. In particular, for the backgrounds with only $F_5$-flux, we have $\Gamma=1$. For the AdS$_d\times S^d\times M^{10-2d}$ (with $d=2,3$) backgrounds with $F_3$-flux (see \eqref{F3} and \eqref{ExF3}), we have $\Gamma=-\Gamma^{23}$. This can be read off \eqref{FE3}. For backgrounds with both $F_2$- and $F_4$-fluxes, as in \eqref{RRF} and \eqref{E}, we have $\Gamma=\Gamma^{11}\Gamma^{123}$, while for backgrounds with $F_4$-flux only, as in \eqref{F4} and \eqref{E5}, we have $\Gamma=\Gamma^1$.

\paragraph{Explicit form of the Killing spinors.}
As explained in \cite{Berkovits:2008ic}, a direct way to get a form of the Killing spinors, associated with the anti-commuting fermionic isometries along which one performs the fermionic T-duality of the supergravity backgrounds, is to read them off from the corresponding components of the fermionic currents $J_Q$ \eqref{eq:FormalCurrents} associated with the generators $Q$ of the superisometry algebra \eqref{conf}--\eqref{qhats}. Concretely, the Killing spinors, which by construction satisfy the defining relations \eqref{EGE} and \eqref{KS}, are the components of the matrix $J_\alpha{}^{\beta}(|y|,y^{\hat a},\lambda_3)$ in
\be\label{KillingJ}
J_{Q_\alpha}|_{\Theta=0}\ =\ {\rm d}\theta^{\mu}J_{\mu}{}^{\alpha}(|y|,y,\lambda_3)\ =\ {\rm d}\theta^{\mu}{\rm e}^{-B}Q_\mu {\rm e}^{B}|_{Q_\alpha,\hat\theta=\hat\xi=0}\ \overset{!}{=}\ {\rm d}\theta^\mu\Xi_\mu{}^\alpha~,
\ee
where ${\rm e}^B$ was defined in \eqref{g}, \eqref{g22}, and \eqref{eq:D21CosRep}.

The Killing spinor condition, which is a particular form of \eqref{KS}, is obtained by simple differentiation of \eqref{KillingJ}
\be\label{KillingEq}
\begin{gathered}
{\rm d}\Xi_\mu+\big[{\rm e}^{-B}{\rm d}{\rm e}^B,\Xi_\mu\big]|_{\Theta=0}\ =\ 0~,\\
 {\rm e}^{-B}{\rm d}{\rm e}^B|_{\Theta=0}\ =\ \Omega^{\hat a\hat b} (y/|y|) R_{\hat a\hat b}+J_D(|y|)D+J_{L_3}(\lambda_3)L_3~.
\end{gathered}
\ee
Note that the index $\mu$ should be regarded as an external one, labelling the number of the Killing spinors.

In view of the structure of the coset element ${\rm e}^{B(|y|,y,\lambda_3)}$ and the commutation relations $[D,Q]=\frac 12 Q$, $[R_{\hat a},Q]=-\frac {s^2}{2} Q\Gamma_{\hat a}\Gamma^4\mathbb P$, and $[L_3,Q]=\frac {\rm i}{2} Q$, we have the following generic form of the Killing spinors in question\footnote{To have a smooth limit from AdS$_d \times S^d \times S^d\times T^{10-3d}$ to AdS$_d \times S^d \times T^{10-2d}$ at $c\to 0$, we have rescaled the coordinates $\lambda_{\pm}$ and $\lambda_3$ of the second sphere as explained in footnote \ref{12}.}
\be\label{KillingJ1}
\Xi_\mu{}^\alpha\ =\ J_\mu{}^{\alpha}(|y|,y,\lambda_3)\ =\ |y|^{-\frac 12}{\rm e}^{\frac {\rm i}{2} c\lambda_3}\mathcal O_\mu{}^\alpha(y^{\hat a}/|y|)~,
\ee
where $\mathcal O_\mu{}^{\alpha}(y^{\hat a}/|y|):=({\rm e}^{s\mathbb P\Gamma_{\hat a}\Gamma_4 \,y^{\hat a}/(2|y|)})_\mu{}^{\alpha}$ is a $Spin(d+1)$-matrix associated with the coset $S^d\cong SO(d+1)/SO(d)$ and $\mathbb P:=\mathbb P_+\mathcal P_{8(d-1)}$ is the projector matrix which singles out the $2(d-1)$ anti-commuting isometries $Q=Q\mathbb P$ for each case of AdS$_d\times S^d \times M^{10-2d}$, as was described in the previous sections (see \eqref{eq:Projector} for the form of $\mathbb P_+$). By definition, we have
\be\label{OOT}
{\mathcal O}^T\Gamma_4{\mathcal O}\ =\ \Gamma_4 \mathbb P~,
\ee
The structure of the matrix ${\mathcal C}_{\mu\nu}$, see \eqref{CB}, is immediately read from the form of the WZ-term of the Green--Schwarz superstring action \eqref{GHaction1}, which in our conventions has the generic form
\be\label{B2C}
 B_{\mu\nu}|_{\Theta=0}\ =\ {\rm i} J_\mu{}^{\gamma}\Gamma^4_{\gamma\delta}J_\nu {}^{\delta}(|y|,y,\lambda_3)\ \overset{!}{=}\  {\mathcal C}_{\mu\nu}~.
\ee
Using \eqref{KillingJ1} and \eqref{OOT}, we find that
\begin{subequations}
\be\label{Cmn}
{\mathcal C}_{\mu\nu}\ =\ {\rm i}|y|^{-1}{\rm e}^{{\rm i} c\lambda_3}(\Gamma^4 \mathbb P)_{\mu\nu}
\ee
and its inverse is
\be\label{C-1}
 {\mathcal C}^{-1\mu\nu}\ =\ -{\rm i}|y|{\rm e}^{-{\rm i} c\lambda_3}(\mathbb P\Gamma^4)^{\mu\nu} ~.
\ee
\end{subequations}
From \eqref{Cmn} we can read off the shift \eqref{dPhi} of the dilaton
\be\label{dPhi1}
\Delta \Phi\ =\ \tfrac 12 \log(\det {\mathcal C})\ =\ - (d-1)\log |y| + {\rm i}(d-1)c\lambda_3~
\ee
and from \eqref{C-1} we read off the change \eqref{dF1} in the Ramond--Ramond fluxes upon the fermionic T-duality for all the considered cases
\be\label{dF}
\Delta F\ =\ 8J_\mu {\mathcal C}^{-1\mu\nu}J_\nu\Gamma \ =\ -8{\rm i} \mathcal {\mathbb P}\Gamma^4\Gamma\ =\ -(1+{\rm i}\Gamma^{0123})\,\slashed F\,.
\ee

\paragraph{Explicit form of the Ramond--Ramond fluxes.}
For completeness, let us give more details on the form of the Ramond--Ramond fluxes characterized by \eqref{dF} in some of the exceptional AdS$_d\times S^d \times S^d\times T^{10-3d}$ cases:

\begin{itemize}
\item[(i)] For AdS$_{3}\times S^{3}\times S^{3}\times S^{1}$, we can consider
the type IIB theory with an $F_3$-flux \eqref{ExF3} as {\it e.g.}~in \cite{Forini:2012bb},
\begin{equation}
\slashed F_{3}\ =\ 2\left(\Gamma^{014}+\sqrt{\alpha}\Gamma^{823}+\sqrt{1-\alpha}\Gamma^{567}\right)
 \ =\ 4\mathcal{P}_{16}\Gamma^{014}~.
\end{equation}
In this case, $\Gamma=-\Gamma^{23}$ as in the corresponding non-exceptional $\alpha=0$ case. Alternatively, we can T-dualise this background along the $S^1$-coordinate $\varphi^{9}$ to get the IIA background with only $F_{4}$-flux \eqref{ExF4} as written in \cite{Wulff:2014kja}, and use the same $\mathcal{P}_{16}$ with $\Gamma=\Gamma^{239}$.
\item[(ii)] For AdS$_{2}\times S^{2}\times S^{2}\times T^{4}$ with $F_4$-flux, we can write the
corresponding projector of rank 8 as a product of two rank-16 projectors, $\mathcal{P}_8=P_{1}P_{2}$, as in \cite{Wulff:2014kja}. Re-numbering the $F_4$-components of \cite{Wulff:2014kja} such that $0,\ldots,3$ are the directions along which we dualise (with $2,3,8,9$ the $T^{4}$ directions, one sphere being parametrised by $x^{7}=\lambda_{3}$ and $x^{1}=\lambda_{+}$, and the other sphere directions labeled by $5,6$) this reads
\begin{equation}
\begin{gathered}
\slashed F_{4}\ =\ 4P_1 P_2\Gamma^{04\,92}~,\\
P_{1}\ :=\ \tfrac{1}{2}(1+\Gamma^{9238})~,\qquad
P_{2}\ :=\ \tfrac{1}{2}(1+\sqrt{\alpha}\Gamma^{04\,71\,23}+\sqrt{1-\alpha}\Gamma^{04\,56\,98})
\end{gathered}
\end{equation}
and $\Gamma=\Gamma^{239}$.
\end{itemize}

In all the cases under consideration, the shifts \eqref{dPhi1} and \eqref{dF} are undone by the corresponding bosonic T-dualities, as we shall show next for the considered examples of the AdS$_d\times S^d\times M^{10-2d}$ backgrounds.

\subsection{Compensating bosonic T--duality}

\paragraph{General case.}
The complete Buscher rules for bosonic T-duality are part of the $O(D,D)$ symmetry of generalised geometry \cite{Alvarez:1994dn}. However, for the cases of interest here the antisymmetric Neveu--Schwarz--Neveu--Schwarz $B$-field vanishes and the metric is diagonal, which simplifies the rules greatly. Letting $\cal I$ be the set of directions along which we dualise, the new metric has the components
\begin{equation}
G'_{tt}\ =\ \frac{1}{G_{tt}}~,\qquad t\ \in\ \cal I
\end{equation}
and remains unchanged in all other directions. The shift in the dilaton is given by (minus half the log of) the determinant of this block, that is,
\be\label{BTD}
\Delta\Phi\ =\ -\tfrac{1}{2}\log\det G_{MN}\ =\ -\tfrac{1}{2}\sum_{t\in\cal I}\log G_{tt}~.
\ee
Allowing the slight abuse of notation that $t$  refers to flat directions here, we can write the change in the Ramond--Ramond forms simply as
\be\label{BTF}
\slashed F''\ =\ \left(\prod_{t\in \cal I}c_{t}\;\Gamma^{t}\right)\slashed F'~,\qquad c_{t}\ :=\ \begin{cases}
-{\rm i} & \text{for } t=0\\
1 & \text{else},
\end{cases}
\ee
where $\slashed F'$ is the result of the fermionic T-duality (see \eqref{dF1}).

This change was written in terms of the potentials in \cite{Fukuma:1999jt} (where the need to include factors ${\rm e}^{\Phi}$ if the dilaton is non-trivial was also noted) and in \cite{Hassan:1999bv} in terms of the
field strengths (whose formul{\ae} simplify here because we still assume the $B$-field vanishes and the metric is diagonal). Time-like T-duality always leads to imaginary forms \cite{Hull:1998vg}. The overall sign
is not physical, and clearly depends on the order in which we perform the dualities.

Now let us apply this to the backgrounds of our interest. The T-dualisation was performed along the directions labeled by $t\in\{0,\ldots,3\}$. So the T-dualised Ramond--Ramond fluxes \eqref{BTF} take the form
\be
\slashed F''\ =\ -{\rm i}\Gamma^{0123}\slashed F'~.
\ee
Substituting into the above equation $\slashed F'=\slashed F+\Delta F$ with the shift $\Delta F$ produced by the fermionic T-duality as in \eqref{dF}, we see that the combined bosonic-fermionc T-duality leaves the Ramond--Ramond fluxes intact, {\it i.e.}~$\slashed F''=\slashed F$.

\paragraph{\mathversion{bold}AdS$_d$ metric.}
Let us now consider the T-dualisation of the background metric and the dilaton. With our choice of the coset element and corresponding AdS$_d \times S^d$ metric, as in \eqref{AdsxS}, the effect of the dualising
along all $d-1$ boundary directions of AdS$_{d}$ on the line element on AdS$_d$ is\newcommand{\mody}{\lvert y \rvert}
\begin{subequations}
\bea
({\rm d}s)^{2}\!&=&\!\frac{-({\rm d}x^0)^{2}+\sum_{i=1}^{d-2}{\rm d}x^{i}{\rm d}x^{i}+\sum_{r=1}^{d+1} {\rm d}y^{r}{\rm d}y^{r}}{\mody^{2}}\nonumber\\
\!&\to &\!
\mody^{2}\Big[-({\rm d}x^0)^{2}+{\textstyle \sum_{i=1}^{d-2}}{\rm d}x^{i}{\rm d}x^{i}\Big]+\frac{\sum_{r=1}^{d+1} {\rm d}y^{r}{\rm d}y^{r}}{\mody^{2}}
\eea
and the dilaton shift is
\begin{equation}
\Delta_{\rm AdS}\Phi\ =\ (d-1)\log\mody~.
\end{equation}
\end{subequations}
We can return the metric to its original form by defining $y'^{r}=y^{r}/\mody$ which sends\linebreak
$\mody=\sqrt{\sum_{r=1}^{d+1} y^ry^r}\rightarrow \frac {1}{\mody}$.
Dualising along some torus directions has no effect on the metric
or the dilaton.

\paragraph{\mathversion{bold}$S^d$ metric.}
In the exceptional cases AdS$_{d}\times S^{d}\times S^{d}\times T^{10-3d}$ of Section \ref{sec:d21}, we also dualise along some directions of one of the spheres: $\lambda_{+}$ for $d=2$, and $\lambda_{\pm}$ for $d=3$. The effect on the line element on $S^d$ (see \eqref{smetric-1}, \eqref{smetric-2}, and \eqref{smetric-3}), rescaled as in footnote \ref{12},  is
\begin{subequations}
\bea
({\rm d}s)^{2}\!& =&\! \tfrac{1}{4}\Big[({\rm d}\lambda_{3})^{2}+{\rm e}^{2{\rm i}c\lambda_{3}}({\rm d}\lambda_{+})^{2}+\underbrace{{\rm e}^{2{\rm i}c\lambda_{3}}({\rm d}\lambda_{-})^{2}}_{\text{only for }d\, =\, 3}\Big]\nonumber\\
\! &\to&\! \tfrac{1}{4}\Big[({\rm d}\lambda_{3})^{2}+{\rm e}^{-2{\rm i}c\lambda_{3}}({\rm d}\lambda_{+})^{2}+\underbrace{{\rm e}^{-2{\rm i}c\lambda_{3}}({\rm d}\lambda_{-})^{2}}_{\text{only for }d\, =\, 3}\Big]
\eea
and we recover the original metric by defining $\lambda_{3}'=-\lambda_{3}$.
The effect on the dilaton is
\begin{equation}
\Delta_{S}\Phi\ =\ -{\rm i}(d-1)c\lambda_{3}+(d-1)\log2~.
\end{equation}
\end{subequations}

\if{}
Turning to the Ramond--Ramond forms, using $\mathbb{P}=\mathcal{P}\mathbb{P}_{+}$
with (3.5),%
\footnote{Recall that $\mathbb{P}_{+}$ always projects onto $\theta$ (rather
than $\hat{\theta}$), and for $d<5$ the additional factors $\mathcal{P}$
project out the non-coset directions $\xi,\hat{\xi}$. We dualise
only along $\theta$ directions. %
} the fermionic change (7.14) reads
\[
\Delta F=\not F-\not F'=8i\Gamma^{4}\mathbb{P}\Gamma=4i\Gamma^{4}\mathcal{P}\Gamma-4\mathcal{P}\Gamma^{01234}\Gamma\,.
\]
where $\Gamma$ varies for the different cases. Assume that second
term to cancels the original flux (i.e. $\not F=4\mathcal{P}\Gamma^{01234}\Gamma$)
leaving only the first term. The effect of bosonic T-duality along
directions $0,1,2,3$ on this term is then to produce
\[
\not F''=-i\Gamma^{0123}\left(4i\,\Gamma^{4}\mathcal{P}\Gamma\right)
\]
which clearly (by assumption) returns us to the original flux.

It remains to check that $\not F=4\mathcal{P}\Gamma^{01234}\Gamma$
in all cases:
\begin{itemize}
\item For the type IIB $AdS_{d}\times S^{d}\times T^{10-2d}$ cases with
only $F_{5}$ flux, $\Gamma=1$ and projector is
\[
\mathcal{P}=\begin{cases}
1, & d=5\\
\frac{1}{2}(1-\Gamma^{2389}), & d=3\text{, using (4.3)}\\
\mathcal{P}_{4}=\frac{1}{4}(1+\Gamma^{1278})(1-\Gamma^{2389}) & d=2\text{, using (5.3)}.
\end{cases}
\]

\item For IIA $AdS_{2}\times S^{2}\times T^{6}$ with $F_{2}$ and $F_{4}$,
the relevant projector is $\mathcal{P}_{8}$ of (5.9), and $\Gamma=\Gamma^{123}$.
\fi

\vspace{10pt}

Together $\Delta_{\rm AdS}\Phi+\Delta_{S}\Phi$ cancels the fermionic
duality's shift in the dilaton, (\ref{dPhi1}), modulo the constant
term which can be ignored, since its only contribution is as an overall factor multiplying the action in the path integral and hence will not affect the classical supergravity argument. In summary, we have thus shown that the AdS$_d\times S^d \times M^{10-2d}$ backgrounds with $(d=2,3,5)$ are invariant under the combined fermionic-bosonic T-duality.

\section{Conclusions and outlook}

In this paper, we have proved the self-duality of the supercoset sigma models describing strings in AdS$_d \times S^d$ backgrounds ($d=2,3,5$) and in an AdS$_d\times S^d\times S^d$ $(d=2,3)$ under a combined T-duality along Abelian bosonic and fermionic isometries of these backgrounds without gauge-fixing kappa symmetry of the sigma model actions.

When $d=2$ and $d=3$, the corresponding sigma models describe only subsectors of the complete superstring theories in AdS$_d\times S^d \times M^{10-2d}$ backgrounds which also include $8(5-d)$ non-supercoset fermionic modes associated with the 10-dimensional supersymmetries that are broken in these backgrounds. In the $d=3$ case, the 16 non-supercoset fermions can be put to zero by gauge fixing kappa symmetry, though this gauge is not admissible for all classical string configurations. For instance, this gauge is not admissible if the string moves entirely in AdS$_3 \times S^3$. In the $d=2$ case, there are not enough kappa symmetries to remove all the 24 non-supercoset fermions. So one should prove the  invariance of the  AdS$_d\times S^d \times M^{10-2d}$ superstring actions under the bosonic and fermionic T-dualities in the presence of the non-supercoset fermions $\upsilon$. We have shown that for the type IIB AdS$_d\times S^d \times T^{10-2d}$ backgrounds supported by $F_5$- or $F_3$-fluxes the actions remain invariant (to the second order in $\upsilon$) in the gauge $\xi=0$ for coset fermions but without imposing any kappa symmetry gauge on $\upsilon$. The presence of the $\upsilon$-fermions in the string actions requires to perform the T-dualisation not only along (anti-)commuting bosonic and fermionic isometries of AdS$_d\times S^d$ but also along half of the torus directions.  This provides evidence for the invariance of the complete type IIB superstring actions on AdS$_d\times S^d \times T^{10-2d}$ under the combined bosonic and fermionic T-duality. We have also demonstrated that the T-duality invariance naturally persists also in the type IIA AdS$_d\times S^d \times T^{10-2d}$ backgrounds, which are T-dual to their type IIB counterparts.  The proof of the self-duality requires the appropriate choice of the $\mathbbm{Z}_4$-grading of the generators of the superisometry algebra (descending from the structure of 10-dimensional supergravity constraints, see {\it e.g.}~\cite{Sorokin:2011rr,Wulff:2014kja}) which `feels' the presence of the non-supercoset sector, {\it i.e.}~the torus directions and the fermionic directions along which the supersymmetry is broken.

In this respect, let us note that there is a hybrid formulation of the AdS$_d\times S^d \times M^{10-2d}$ superstrings for $d=2,3$ \cite{Berkovits:1999zq} in which the AdS$_d\times S^d$ supercoset sector is completely decoupled from the $M^{10-2d}$ sector and the non-supercoset fermions get replaced with the Ramond--Neveu--Schwarz spinning string variables. The supercoset sectors of the hybrid sigma models differ from the kappa symmetric Green--Schwarz supercoset models considered here. The former are similar to the structure of the pure spinor action for the AdS$_5\times S^5$ superstring which has proved to be invariant under the T-duality \cite{Berkovits:2008ic}. We thus expect that also the hybrid models of the AdS$_d\times S^d \times M^{10-2d}$ superstrings will be self-dual, but their complete equivalence to the Green--Schwarz superstrings is yet to be proved (see {\it e.g.}~\cite{Cagnazzo:2012uq} and references therein).

Having acquired an experience of working with contributions of the non-coset fermionic modes of the string, it would be of interest to see whether and how the presence of these modes determine the combined bosonic-fermionic T-dualisation of the AdS$_4\times\mathbbm{C}P^3$ background and corresponding string sigma model. In particular, whether the T-dualisation of complexified $\mathbbm{C}P^3$ directions is required.

It would also be of interest to address an important problem of the combined fermionic and bosonic T-duality of AdS$_d\times S^d\times M^{10-2d}$ backgrounds for $d=2,3$ in the presence of Neveu--Schwarz--Neveu--Schwarz flux, as well as to find a manifestation of this T-duality on the CFT sides of the correspondences for the Ramond--Ramond (and Neveu--Schwarz--Neveu--Schwarz) $d=2,3$ backgrounds.

We hope to address these issues in future work.

\acknowledgements

We would like to thank Nathan Berkovits, Amit Dekel, Jan Gutowski, Emanuel Malek, Bogdan Stefanski, Mario Tonin, Alessandro Torrielli and Arkady Tseytlin for fruitful discussions and comments. Furthermore, we would like to thank Lorenzo Gerotto and Alessandro Torrielli for collaboration on early stages of this project.

M.C.A.~was supported by an NRF (South Africa) Innovation Fellowship. J.M.~is supported through the NRF CPRR program under GUN 87667. The work of S.P.~was partially supported by MIUR and INFN Special Initiative SSG. A.P.~was supported in part by the EPSRC under the grant EP/K503186/1. D.S.~was partially supported by the INFN Special Initiative ST \& FI and by the Russian Science Foundation grant 14--42--00047 in association with the Lebedev Physical Institute.  P.S.~was supported by a MIUR grant and the INFN, Special Initiative SSG. M.W.~was supported in part by the STFC Consolidated Grant ST/L000490/1 {\it Fundamental Implications of Fields, Strings, and Gravity}. L.W.~was supported by the ERC Advanced grant No.~290456 {\it Gauge Theory --- String Theory Duality: Maximally Symmetric Case and Beyond}.  S.P., P.S., and D.S.~would like to thank the Quantum Gravity and Strings Group at the University of Cape Town for their kind hospitality during the final stage of this work. S.P. thanks Galileo Galilei Institute for Theoretical Physics for their hospitality and the INFN for partial support during work in progress.  This paper has been partially supported by COST Actions MP1210 {\it The String Theory Universe} and MP1405 {\it Quantum Structure of Space-Time}.

\datamanagement

No additional research data beyond the data presented and cited in this work are needed to validate the research findings in this work.

\appendices

\subsection{Clifford algebra in ten dimensions}\label{A}

In this section we briefly list our conventions on gamma matrices, both for the type IIA and type IIB superspace.

\paragraph{Type IIA superspace.}
Let $\Theta$ be a 32-component Majorana spinor in 10-dimensional type IIA superstring theory. The Clifford algebra is defined as
\be\label{CA}
\{\Gamma^{A},\Gamma^B\}\ =\ 2\eta^{AB}~,
\ee
where $(\eta_{AB})={\rm diag}(-1,1,\ldots,1)$.

We use the realization of the 10-dimensional ($32\times 32$) gamma matrices $\Gamma^{A}$ for $A,B,\ldots =0,\ldots, 9$ and $\Gamma^{11}=-\Gamma^{0}\cdots\Gamma^9$ in which  $\Gamma^A$ are real and $C\Gamma^A$ are symmetric (with $C$ being a charge conjugation matrix used to lower (or raise) the spinor indices):
\be\label{Gamma}
C\Gamma^AC^{-1}\ =\ -(\Gamma^A)^T~.
\end{equation}
For our purposes it is convenient to choose a realization  in which the AdS$_2\times S^2\times T^6$ structure becomes manifest. The re-distribution of the 10-dimensional indices $A=(\underline a,a')$ among the AdS$_2\times S^2$ directions $(\underline a=0,4,5,6)$  and  $T^6$ directions $(a'=1,2,3,7,8,9)$ has been chosen in accordance with the type IIB notation of Section \ref{IIB}.

\paragraph{Type IIB superspace.}
In this case the 10-dimensional fermionic coordinates $\Theta^{\alpha i}$, $i=1,2$, are  Majorana--Weyl spinors whose indices $\alpha,\beta,\ldots$ take 16 values. The $(16\times 16)$-matrices $\Gamma^A_{\alpha\beta}$ and $\Gamma^{A\,\alpha\beta}$ obey the Clifford algebra \eqref{CA} and share the following symmetry properties
\bea\label{gammaCli}
\begin{gathered}
\Gamma_{\alpha\gamma}^A\Gamma^{B\,\gamma\beta}+\Gamma_{\alpha\gamma}^B\Gamma^{A\,\gamma\beta}\ =\ 2\delta_\alpha^\beta\, \eta^{AB}~,\\
\Gamma^A_{\alpha\beta}\ =\ \Gamma^{  A}_{\beta\alpha}~,\quad \Gamma^{A\alpha\beta}\ =\ \Gamma^{{  A}\beta\alpha}~, \quad \Gamma^{ABCDE}_{\alpha\beta}\ =\ \Gamma^{ABCDE}_{\beta\alpha},\quad \Gamma_{ABCDE}^{\alpha\beta}\ =\ \Gamma_{ABCDE}^{\beta\alpha}~,\\
\Gamma^{ABC}_{\alpha\beta}\ =\ -\Gamma^{ABC}_{\beta\alpha}~,\quad \Gamma^{ABC\,\alpha\beta}\ =\ -\Gamma^{ABC\,\beta\alpha}~,\\
(\Gamma^{ABC})^{\alpha\beta}\ =\ { \tfrac 1{7!}}\varepsilon^{ABCD_1\cdots D_7}\Gamma_{D_1\cdots D_7}^{\alpha\beta}~,\quad (\Gamma^{ABC})_{\alpha\beta}\ =\ { -}\tfrac 1{7!}\varepsilon^{ABCD_1\cdots D_7}(\Gamma_{D_1\cdots D_7})_{\alpha\beta}~.
\end{gathered}
\eea

\subsection{The $\mathfrak{psu}(1,1|2)$ algebra for type IIB AdS$_2\times S_2 \times T^6$ background}\label{ads2gamma0}

As explained in the text, in this case the $\mathfrak{psu}(1,1|2)$ algebra can be obtained as a truncation of the  $\mathfrak{psu}(1,1|2)\oplus\mathfrak{psu}(1,1|2)$ one, or directly as a truncation of the  $\mathfrak{psu}(2,2|4)$ algebra. Its explicit form can then be read from eqs. \eqref{eq:PSU22-12-Bos}--\eqref{psu22-2} where we set $m,n,\ldots=0$ ($\eta_{00}=-1$), $\hat a , \hat b, \ldots= 5,6$ and project the supercharges with the ${\cal P}_4$ projector (see \eqref{QP8}, \eqref{tv2}). For the reader's convenience we list here the resulting non-vanishing commutators ($P \equiv P_0$ and $K \equiv K_0$)
\begin{subequations}
\be\label{psu11-1}
\begin{gathered}
[D,P]\ =\ P~,\quad [D,K]\ =\ -K~, \quad [P,K]\ =\ 2D~,\\
[R_{\hat a},R_{\hat b}]\ =\ -\varepsilon_{\hat a\hat b}R_{56}~,\quad [R_{56},R_{\hat a}]\ =\ -\varepsilon_{\hat a\hat b}R_{\hat b}~,\\
[D,Q_{\alpha }]\ =\ \tfrac12Q_{\alpha }~,\quad[D,S_{\alpha }]\ =\ -\tfrac12S_{\alpha }\,,\quad[K,Q_{\alpha }]\ =\ (\hat S\Gamma_{04})_{\alpha }~,\quad[P,S_{\alpha }]\ =\ (\hat Q\Gamma_{04})_{\alpha }~,\\
[R_{\ha},S_{\alpha }]\ =\ \tfrac12(S\Gamma_{\ha 4})_{\alpha }~,\quad [R_{\ha},Q_{\alpha }]\ =\ -\tfrac12(Q\Gamma_{\ha 4})_{\alpha }~,\\
[R_{56},S_{\alpha }]\ =\ \tfrac12(S\Gamma_{56})_{\alpha }~,\quad[R_{56},Q_{\alpha }]\ =\ \tfrac12(Q\Gamma_{56})_{\alpha }~,
\end{gathered}
\ee
and the same for $(Q,S)\leftrightarrow(\hat Q,\hat S)$. Furthermore,
\be\label{psu11-2}
\begin{gathered}
\{\hat Q_{\alpha },Q_{\beta }\}\ =\ {\rm i}(\Gamma^0\mathbb P_+ \mathcal P_4)_{\alpha \beta }\,P~,\quad
\{\hat S_{\alpha },S_{\beta }\}\ =\ -{\rm i}(\Gamma^0\mathbb P_+ \mathcal P_4)_{\alpha \beta }\,K~,\\
\{S_{\alpha },Q_{\beta }\}\ =\
-{\rm i}(\Gamma^4\mathbb P_+ \mathcal P_4)_{\alpha \beta }\,D
-{\rm i}(\Gamma^{\ha}\mathbb P_+ \mathcal P_4)_{\alpha \beta }\,R_{\ha}
-{\rm i}(\Gamma^{56}\Gamma^4\mathbb P_+ \mathcal P_4)_{\alpha \beta }\,R_{56}~,\\
\{\hat S_{\alpha },\hat Q_{\beta }\}\ =\
-{\rm i}(\Gamma^4\mathbb P_- \mathcal P_4)_{\alpha \beta }\,D
-{\rm i}(\Gamma^{\ha}\mathbb P_- \mathcal P_4)_{\alpha \beta }\,R_{\ha}
-{\rm i}(\Gamma^{56}\Gamma^4\mathbb P_- \mathcal P_4)_{\alpha \beta }\,R_{56}~,
\end{gathered}
\ee
\end{subequations}
where $\mathbb P_\pm$ have been defined in \eqref{eq:Projector}.

\subsection{The $\mathfrak{psu}(1,1|2)$ algebra for type IIA AdS$_2\times S_2 \times T^6$ background}\label{ads2gamma}

To perform the worldsheet T-duality of the type IIA AdS$_2\times S^2$ supercoset sigma model it is convenient to choose the $\mathfrak{psu}(1,1|2)$ Lie superalgebra in the form similar to that of the type IIB case \eqref{psu11-1}--\eqref{psu11-2}, where the indices $0,4$ label the AdS$_2$ tangent space directions and $\hat a,\hat b,\ldots=5,6$ label those of $S^2$.

We consider the derivation of the form \eqref{psu11-1}--\eqref{psu11-2} of the $\mathfrak{psu}(1,1|2)$ Lie superalgebra from its 10-dimensional type IIA counterpart used in \cite{Sorokin:2011rr} to construct the superstring action in type IIA AdS$_2\times S^2 \times T^6$ background with $F_2$- and $F_4$-flux.

The bosonic
$\mathfrak{so}(2,1)\oplus\mathfrak{so}(3)$ subalgebra is generated by translations $(P_{\underline a})=(P_a, P_{\hat a})$  and $\mathfrak{so}(1,1)\oplus\mathfrak{so}(2)$ rotations $(M_{\underline{ab}})=(M_{ab},M_{\hat a\hat b})$ in AdS$_2\times S^2$
  \be\label{so}
[P_{\underline a}, P_{\underline b}]\ =\ -\tfrac{1}{2}R_{{\underline a}{\underline
b}}{}^{{\underline c}{\underline d}}\,M_{{\underline c}{\underline
d}}~,\quad[M_{{\underline a}{\underline b}},P_{\underline
c}]\ =\ \eta_{{\underline a}{\underline c}}P_{\underline b}-\eta_{{\underline
b}{\underline c}}P_{\underline a}~,\quad
{}[M_{{\underline a}{\underline b}}, M_{{\underline c}{\underline
d}}]\ =\ 0~,
\ee
where
\begin{equation}\label{Rua}
(R_{\underline{ab}}{}^{\underline{cd}})\ =\ (R_{ab}{}^{cd},R_{\hat a\hat b}{}^{\hat
c\hat d})\ =\ \big({2}\delta_{[a}^c\delta_{b]}^d,-2\delta_{[\hat
a}^{\hat c}\delta_{\hat b]}^{\hat d}\big)
\end{equation}
is the AdS$_2\times S^2$ curvature of unit radius. The fermionic part of $\mathfrak{psu}(1,1|2)$ is generated by eight Gra{\ss}mann-odd operators $\mathcal Q=\mathcal P_8\mathcal Q$ associated with eight fermionic degrees of freedom $\vartheta=\mathcal P_8 \Theta$ of the string (see \eqref{8+24}). They satisfy the following (anti-)commutation relations
\be\label{QQ}
\begin{gathered}
[P_{\underline a}, \mathcal Q]\ =\ \tfrac 12 \mathcal Q \Gamma^{04}\Gamma_{\underline a}\Gamma^{11}\,,
\qquad [M_{{\underline a}{\underline b}},\mathcal Q]\ =\ -\tfrac{1}{2}\mathcal Q\Gamma_{{\underline
a}{\underline b}}~,\\
\{\mathcal Q,\mathcal Q\}\ =\
2{\rm i}\,C\Gamma^{\underline a}\mathcal P_8\,P_{\underline a}
{ -}\tfrac{\rm i}{2}\,C\Gamma^{{\underline a}{\underline
b}}\,\Gamma^{04}\Gamma^{11}\mathcal P_8\,R_{{\underline a}{\underline b}}{}^{{\underline c}{\underline
d}}M_{{\underline c}{\underline d}}~.
\end{gathered}
\ee
Now let us make the form of the algebra \eqref{QQ} closer to that of the AdS$_5\times S^5$ case \eqref{psu22-0}. To this end let us split $\mathcal Q$ onto the chiral and anti-chiral parts
\be\label{ch}
\mathcal Q^1\ =\ \tfrac 12 \mathcal Q(1+\Gamma^{11})~,\quad \mathcal Q^2\ =\ \tfrac 12\mathcal Q(1-\Gamma^{11})~.
\ee
For these generators the (anti-)commutation relations are
\be\label{PSUcQ}
\begin{gathered}
[P_{\underline a}, \mathcal Q^i]\ =\ \tfrac 12 \varepsilon^{ij}\mathcal Q^{j} \Gamma^{04}\Gamma_{\underline a}~,
\quad [M_{{\underline a}{\underline b}},\mathcal Q^i]\ =\ -\tfrac{1}{2}\mathcal Q^i\Gamma_{{\underline
a}{\underline b}}~,\\
\{\mathcal Q^1,\mathcal Q^1\}\ = \
{\rm i}C\Gamma^{\underline a}(1+\Gamma^{11})\mathcal P_8\,P_{\underline a}~,\qquad \{\mathcal Q^2,\mathcal Q^2\}\ =\
{\rm i}C\Gamma^{\underline a}(1-\Gamma^{11})\mathcal P_8\,P_{\underline a}~,\\
\{\mathcal Q^1,\mathcal Q^2\}\ =\ {\rm i}C(1-\Gamma^{11})\mathcal P_8 \,M_{04}{ -}{\rm i}C\Gamma^{0456}(1-\Gamma^{11})\mathcal P_8\,M_{56}~.
\end{gathered}
\ee
We now convert the anti-chiral spinor $\mathcal Q^2$ into chiral one by multiplying it with $\Gamma^{123}$ (note that $\Gamma^{123}$ commutes with the projector \eqref{P8} with $J_{a'b'}$ defined as in \eqref{Jbasis})
\be\label{Qconvert}
\tilde {\mathcal Q}^2\ \equiv \ \tfrac 1{\sqrt 2}{\mathcal Q}^2\Gamma^{123}\ =\ { \tfrac 12} \tilde{\mathcal Q}^2(1+\Gamma^{11})~.
\ee
For these generators (with $\tilde{\mathcal Q}^1=\frac 1{\sqrt 2}\mathcal Q^1$) the algebra takes the form similar to \eqref{psu22-0}
\be\label{PSUcQt}
\begin{gathered}
[P_{\underline a}, \tilde{\mathcal Q}^i]\ =\ \tfrac 12 \varepsilon^{ij}\tilde{\mathcal Q}^{j} \Gamma^{01234}\Gamma_{\underline a}~,
\quad [M_{{\underline a}{\underline b}},\tilde{\mathcal Q}^i]\ =\ -\tfrac{1}{2}\tilde{\mathcal Q}^i\Gamma_{{\underline
a}{\underline b}}~,\\
\{\tilde{\mathcal Q}^i,\tilde{\mathcal Q}^j\}\ =\
{\rm i}\delta^{ij}\,C\Gamma^{\underline a} \tilde{\mathcal P}_4\,P_{\underline a}
-\tfrac {\rm i}2\varepsilon^{ij} \,C\Gamma^{a}\Gamma^{01234}\Gamma^b \tilde{\mathcal P}_4 \,M_{ab}
{ -}\tfrac {\rm i}2\varepsilon^{ij}\,C\Gamma^{\hat a}\Gamma^{01234}\Gamma^{\hat b}\mathcal P_4\,M_{\hat a\hat b}~,
\end{gathered}
\ee
where we have defined the projector
\be\label{P4}
\tilde{\mathcal P}_4\ :=\ \tfrac 12 (1+\Gamma^{11}) \mathcal P_8
\ee
which reduces the number of the components of the initial 32-component supercharge down to four. It is similar to the type IIB projector \eqref{tv2}.

Now we can relate the generators of \eqref{so}, \eqref{QQ} to those in \eqref{psu11-1}, \eqref{psu11-2}.
We start with the $AdS_2$ generators $P_a$ and $M_{ab}$ ($a=0,4$). From \eqref{so}, \eqref{Rua}, we deduce  their commutation relations
\be\label{AdS2}
[P_0,P_4]\ =\ -M_{04}~, \quad [P_0,M_{04}]\ =\ P_4~,\quad [P_4,M_{04}]\ =\ P_0~.
\ee
Therefore, we can identify the above generators with those in \eqref{psu11-1}--\eqref{psu11-2} as follows
\be\label{AdS2oz}
P_0\ =\ \tfrac 12(P-K)~,\quad P_4\ =\ D~, \quad M_{04}\ =\ \tfrac 12(P+K)~.
\ee
In the $SO(3)$ sector we have
\be\label{S2}
[P_5,P_6]\ =\ M_{56}~,\quad [P_{5},M_{56}]\ =\ -P_6~,\quad [P_6,M_{56}]\ =\ P_5~.
\ee
Thus,
\be\label{S2oz}
P_5\ =\ R_5~,\quad P_6\ =\ R_6~,\quad M_{56}\ =\ -R_{56}~.
\ee
Finally, in the fermionic sector we have
\be\label{Qr}
\begin{gathered}
Q\ =\ -\tfrac{1}{\sqrt2}(\tilde{\mathcal Q}^1-{\rm i}\tilde{\mathcal Q}^2)\mathbb P_+~,\quad
\hat Q\ =\ -\tfrac{1}{\sqrt2}(\tilde{\mathcal Q}^1+{\rm i}\tilde{\mathcal Q}^2)\mathbb P_-~,\\
S\ =\ \tfrac{1}{\sqrt2}(\tilde{\mathcal Q}^1+{\rm i}\tilde{\mathcal Q}^2)\mathbb P_+~,\quad
\hat S\ =\ \tfrac{1}{\sqrt2}(\tilde{\mathcal Q}^1-{\rm i}\tilde{\mathcal Q}^2)\mathbb P_-~,
\end{gathered}
\ee
where we have defined
\begin{equation}
\mathbb P_\pm\ :=\ \tfrac12(1\pm {\rm i}\Gamma^{0123})~.
\end{equation}
With this definition the corresponding currents have the form \eqref{bosonic+fermionic}. Accordingly, the resulting supercoset part of the type IIA AdS$_2\times S^2 \times T^6$ superstring action with the $F_2$- and $F_4$-flux has exactly the same form as its type IIB counterpart.

 \subsection{$D(2,1;\alpha)$ supercoset currents}\label{D}
Due to the structure \eqref{eq:d21alg} of the superalgebra $\mathfrak{d}(2,1;c^2)$ and the chosen coset element \eqref{eq:D21CosRep}, the coset currents $J^{(0)}$ and $J$ have the following components (see \eqref{eq:FormalCurrents}):
\begin{subequations}\label{eq:D21OrCur}
\begin{equation}
 \begin{gathered}
   J_P^{(0)}\ =\ \big[{\rm e}^{-B} \big({\rm d}xP +{\rm d}\theta^\alpha Q_\alpha+{\rm d}\lambda_+L_+\big) {\rm e}^{B}\big]_P~,~~
    J_K^{(0)}\ =\ 0~,~~
    J_D^{(0)}\ =\ \big[{\rm e}^{-B}{\rm d} {\rm e}^{B}\big]_{D}~,\\
    J_{L_+}^{(0)}\ =\ \big[{\rm e}^{-B} \big({\rm d}xP +{\rm d}\theta^\alpha Q_\alpha+{\rm d}\lambda_+L_+\big) {\rm e}^{B}\big]_{L_+}~,~~J_{L_-}^{(0)}\ =\ 0~,~~
    J_{L_3}^{(0)}\ =\ \big[{\rm e}^{-B}{\rm d} {\rm e}^{B}\big]_{L_3}~,\\
    J_{{R^\alpha}_\beta}^{(0)}\ =\ \big[{\rm e}^{-B}{\rm d} {\rm e}^{B}\big]_{{R^\alpha}_\beta}~,\\
    J_{Q_\alpha}^{(0)}\ =\ \big[{\rm e}^{-B} \big({\rm d}xP +{\rm d}\theta^\alpha Q_\alpha+{\rm d}\lambda_+L_+\big) {\rm e}^{B}\big]_{Q_\alpha}~,~~
    J_{\hat Q_\alpha}^{(0)}\ =\ \big[{\rm e}^{-B}{\rm d} {\rm e}^{B}\big]_{\hat Q_\alpha}~,\\
    J^{(0)}_{S_\alpha}\ =\ 0~,~~J_{\hat S_\alpha}^{(0)}\ =\ \big[{\rm e}^{-B}{\rm d} {\rm e}^{B}\big]_{\hat S_\alpha}
 \end{gathered}
\end{equation}
and
\begin{equation}
\begin{gathered}
 J_P\ =\ J_P^{(0)}~,~~
 J_{Q_\alpha}\ =\ J_{Q_\alpha}^{(0)}~,~~
  J_{L_+}\ =\  J_{L_+}^{(0)}~,\\
  J_{D}\ =\  J_{D}^{(0)}+\sigma^2_{\alpha\beta}J_{Q_{\alpha}}^{(0)}\xi^\beta ~,~~
  J_{L_3}\ =\  J_{L_3}^{(0)}+{\rm i}c^2\sigma^2_{\alpha\beta}J_{Q_{\alpha}}^{(0)}\xi^\beta ~,\\
  J_{{R^\alpha}_\beta}\ =\ J_{{R^\alpha}_\beta}^{(0)}-{\rm i}s^2\sigma^2_{\alpha\gamma}J_{Q_{\gamma}}^{(0)}\xi^\beta -\tfrac{\rm i}{2}s^2\sigma^2_{\gamma\delta}{\delta^\alpha}_\beta J^{(0)}_{Q_\gamma}\xi^\delta  ~,\\
  J_{\hat Q_\alpha}\ =\ J_{\hat Q_\alpha}^{(0)}-J_P^{(0)}\xi^\alpha~,~~
   J_{\hat S_\alpha}\ =\ J_{\hat S_\alpha}^{(0)} +J_{L_+}^{(0)}\xi^\alpha~,\\
   J_K\ =\ -\sigma^2_{\alpha\beta} J^{(0)}_{\hat S_\alpha}\xi^\beta +\tfrac{\rm i}{2} J^{(0)}_{L_+}\xi^2~,~~
    J_{L_-}\ =\ -c^2\sigma^2_{\alpha\beta}J^{(0)}_{\hat Q_\alpha}\xi^\beta-\tfrac{\rm i}{2}c^2J^{(0)}_P\xi^2~,\\
    J_{S_\alpha}\ =\ -\tfrac{1}{2}J^{(0)}_{D} \xi^\alpha -\tfrac{\rm i}{2}J^{(0)}_{L_3} \xi^\alpha-{\rm i}J^{(0)}_{{R^\beta}_\alpha} \xi^\beta+{\rm d}\xi^\alpha -\tfrac{\rm i}{2} s^2  J^{(0)}_{Q_\alpha}\xi^2~.
 \end{gathered}
\end{equation}
\end{subequations}
In these expressions, we have made all the $\xi$-dependence explicit.

\paragraph{Dual currents.}
The Maurer--Cartan form $\tilde J=\tilde g^{-1}{\rm d}\tilde g$ constructed from the dual coset representative \eqref{eq:D21DualCosRep} is of the form
\begin{equation}
\begin{aligned}
 \tilde J\ &=\ {\rm e}^{-{\sigma^{1\beta}}_\alpha  \xi^\alpha Q_\beta} \tilde J^{(0)} {\rm e}^{{\sigma^{1\beta}}_\alpha  \xi^\alpha Q_\beta}+{\sigma^{1\,\beta}}_\alpha  {\rm d}\xi^\alpha Q_\beta\\
 \ &=\ \tilde J^{(0)}-{\sigma^{1\,\beta}}_\alpha \xi^\alpha \big[ S_\beta,\tilde J^{(0)}\big]-\tfrac{\rm i}{4}\xi^2 \sigma^{2\,\alpha\beta} \big\{ S_\alpha,\big[ S_\beta,\tilde J^{(0)}\big]\big\}+{\sigma^{1\,\beta}}_\alpha  {\rm d}\xi^\alpha Q_\beta~,
 \end{aligned}
\end{equation}
where, as before, $\tilde J^{(0)}$ does not depend on the fermonic coordinate $\xi^\alpha$, and we have set $\xi^2:={\rm i}\sigma^2_{\alpha\beta}\xi^\alpha\xi^\beta$. A calculation similar to the one that led to \eqref{eq:D21OrCur} yields
\begin{subequations}\label{eq:D21DualCur}
\begin{equation}
 \begin{gathered}
   \tilde J_P^{(0)}\ =\ 0~,~~
   \tilde J_K^{(0)}\ =\ \big[ {\rm e}^{-B} \big({\rm d}\tilde x K  -{\rm i}\sigma^{2\,\alpha\beta}{\rm d}\tilde\theta_{\alpha} S_\beta+{\rm d}\tilde\lambda_+L_-\big) {\rm e}^{B}\big]_K~,~~
   \tilde J_D^{(0)}\ =\ \big[ {\rm e}^{-B} {\rm d} {\rm e}^{B}\big]_D~,\\
    \tilde J_{L_+}^{(0)}\ =\ 0~,~~
    \tilde J_{L_-}^{(0)}\ =\ \big[ {\rm e}^{-B} \big({\rm d}\tilde x K  -{\rm i}\sigma^{2\,\alpha\beta}{\rm d}\tilde\theta_{\alpha} S_\beta+{\rm d}\tilde\lambda_+L_-\big) {\rm e}^{B}\big]_{L_-}~,~~
    \tilde J_{L_3}^{(0)}\ =\ \big[ {\rm e}^{-B} {\rm d} {\rm e}^{B}\big]_{L_3}~,\\
    \tilde J_{{R^\alpha}_\beta}^{(0)}\ =\ \big[ {\rm e}^{-B} {\rm d} {\rm e}^{B}\big]_{{R^\alpha}_\beta}~,\\
    \tilde J_{Q_\alpha}^{(0)}\ =\ 0~,~~
    \tilde J_{\hat Q_\alpha}^{(0)}\ =\ \big[ {\rm e}^{-B} {\rm d} {\rm e}^{B}\big]_{\hat Q_\alpha}~,\\
    \tilde J^{(0)}_{S_\alpha}\ =\  \big[{\rm e}^{-B} \big({\rm d}\tilde x K  -{\rm i}\sigma^{2\,\alpha\beta}{\rm d}\tilde\theta_{\alpha} S_\beta+{\rm d}\tilde\lambda_+L_-\big) {\rm e}^{B}\big]_{S_\alpha}~,~~
    \tilde J_{\hat S_\alpha}^{(0)}\ =\ \big[ {\rm e}^{-B} {\rm d} {\rm e}^{B}\big]_{\hat S_\alpha}
 \end{gathered}
\end{equation}
and
\begin{equation}
\begin{gathered}
 \tilde J_K\ =\ \tilde J^{(0)}_K~,~~
  \tilde J_{S_\alpha}\ =\ \tilde J^{(0)}_{S_\alpha}~,~~
    \tilde J_{L_-}\ =\ \tilde J^{(0)}_{L_-}~,\\
  \tilde J_{D}\ =\  \tilde J^{(0)}_D-{\rm i}\sigma^1_{\alpha\beta}\tilde J^{(0)}_{S_\alpha}\xi^\beta~,~~
  \tilde J_{L_3}\ =\  \tilde J^{(0)}_{L_3}+c^2\sigma^1_{\alpha\beta}\tilde J^{(0)}_{S_\alpha}\xi^\beta~,\\
  \tilde J_{{R^\alpha}_\beta}\ =\   \tilde J^{(0)}_{{R^\alpha}_\beta}+s^2\big(\sigma^1_{\alpha\gamma}\tilde J^{(0)}_{S_\beta}-\tfrac12{\delta^\beta}_\alpha\sigma^1_{\gamma\delta}\tilde J^{(0)}_{S_\delta}\big)\xi^\gamma~,\\
  \tilde J_{\hat Q_\alpha}\ =\  \tilde J^{(0)}_{\hat Q_\alpha}+{\sigma^{1\,\alpha}}_\beta\tilde J^{(0)}_{L_-}\xi^\beta~,~~
   \tilde J_{\hat S_\alpha}\ =\ \tilde J^{(0)}_{\hat S_\alpha}-{\sigma^{1\,\alpha}}_\beta\tilde J^{(0)}_K\xi^\beta~,\\
    \tilde J_P\ =\  -{\rm i}\sigma^1_{\alpha\beta}\tilde J^{(0)}_{\hat Q_\alpha}\xi^\beta-\tfrac{\rm i}{2}\tilde J^{(0)}_{L_-}\xi^2~,~~
  \tilde J_{L_+}\ =\ -{\rm i}c^2\sigma^1_{\alpha\beta}\tilde J^{(0)}_{\hat S_\alpha}\xi^\beta+\tfrac{\rm i}{2}c^2\tilde J^{(0)}_K\xi^2~,\\
   \tilde J_{Q_\alpha}\ =\ {\sigma^{1\,\alpha}}_\beta\big(\tfrac12\tilde J^{(0)}_D\xi^\beta+\tfrac{\rm i}{2}\tilde J^{(0)}_{\tilde L_3}\xi^\beta+{\rm d}\xi^\beta\big)-{\rm i}{\sigma^{1\,\beta}}_\gamma \tilde J^{(0)}_{{R^\beta}_\alpha}\xi^\gamma+\tfrac{\rm i}{2}s^2\tilde J^{(0)}_{S_\alpha}\xi^2~.
 \end{gathered}
\end{equation}
\end{subequations}
As before, in these expressions, we have made all the $\xi$-dependence explicit. Note that $\tilde J_{\hat Q_\alpha}^{(0)}=J_{\hat Q_\alpha}^{(0)}$,  $\tilde J_{\hat S_\alpha}^{(0)}=J_{\hat S_\alpha}^{(0)}$,  $\tilde J_{D}^{(0)}=J_{D}^{(0)}$, $\tilde J_{L_3}^{(0)}=J_{L_3}^{(0)}$, and $\tilde J^{(0)}_{{R^\beta}_\alpha}=J^{(0)}_{{R^\beta}_\alpha}$, respectively.


\begin{thebibliography}{10}\parskip-1mm
{\small
\bibitem{Aharony:1999ti}
O.~Aharony, S.~S.~Gubser, J.~M.~Maldacena, H.~Ooguri, and Y.~Oz,
{\em {Large-$N$ field theories, string theory and gravity},}
\href{http://dx.doi.org/10.1016/S0370-1573(99)00083-6}{Phys. Rept. {\bf 323}
  (2000) 183} [{\tt
  \href{http://www.arxiv.org/abs/hep-th/9905111}{hep-th/9905111}}].

\bibitem{Maldacena:2003nj}
J.~M.~Maldacena,
{\em {TASI lectures on AdS/CFT},}
{\tt \href{http://www.arxiv.org/abs/hep-th/0309246}{hep-th/0309246}}.

\bibitem{Nastase:2007kj}
H.~Nastase,
{\em {Introduction to AdS/CFT},}
{\tt \href{http://www.arxiv.org/abs/0712.0689}{0712.0689 [hep-th]}}.

\bibitem{Beisert:2010jr}
N.~Beisert and ~others,
{\em {Review of AdS/CFT integrability: an overview},}
\href{http://dx.doi.org/10.1007/s11005-011-0529-2}{Lett. Math. Phys. {\bf 99}
  (2012)~3} [{\tt \href{http://www.arxiv.org/abs/1012.3982}{1012.3982
  [hep-th]}}].

\bibitem{'tHooft:1973jz}
G.~'t~Hooft,
{\em {A planar diagram theory for strong interactions},}
\href{http://dx.doi.org/10.1016/0550-3213(74)90154-0}{Nucl. Phys. B {\bf 72}
  (1974) 461}.

\bibitem{Maldacena:1997re}
J.~M.~Maldacena,
{\em {The large-$N$ limit of superconformal field theories and supergravity},}
\href{http://dx.doi.org/10.1023/A:1026654312961}{Int. J. Theor. Phys. {\bf 38}
  (1999) 1113} [{\tt
  \href{http://www.arxiv.org/abs/hep-th/9711200}{hep-th/9711200}}].

\bibitem{Minahan:2002ve}
J.~Minahan and K.~Zarembo,
{\em {The Bethe ansatz for $\mathcal{N}=4$ super Yang--Mills},}
\href{http://dx.doi.org/10.1088/1126-6708/2003/03/013}{JHEP {\bf 0303} (2003)
  013} [{\tt \href{http://www.arxiv.org/abs/hep-th/0212208}{hep-th/0212208}}].

\bibitem{Bena:2003wd}
I.~Bena, J.~Polchinski, and R.~Roiban,
{\em {Hidden symmetries of the AdS$_5\times S^5$ superstring},}
\href{http://dx.doi.org/10.1103/PhysRevD.69.046002}{Phys. Rev. D {\bf 69}
  (2004) 046002} [{\tt
  \href{http://www.arxiv.org/abs/hep-th/0305116}{hep-th/0305116}}].

\bibitem{Berkovits:2008ic}
N.~Berkovits and J.~Maldacena,
{\em {Fermionic T-duality, dual superconformal symmetry, and the
  amplitude/wilson loop connection},}
\href{http://dx.doi.org/10.1088/1126-6708/2008/09/062}{JHEP {\bf 09} (2008)
  062} [{\tt \href{http://www.arxiv.org/abs/0807.3196}{0807.3196 [hep-th]}}].

\bibitem{Beisert:2008iq}
N.~Beisert, R.~Ricci, A.~A.~Tseytlin, and M.~Wolf,
{\em {Dual superconformal symmetry from AdS$_5\times S^5$ superstring
  integrability},}
\href{http://dx.doi.org/10.1103/PhysRevD.78.126004}{Phys. Rev. D {\bf 78}
  (2008) 126004} [{\tt \href{http://www.arxiv.org/abs/0807.3228}{0807.3228
  [hep-th]}}].

\bibitem{Ricci:2007eq}
R.~Ricci, A.~A.~Tseytlin, and M.~Wolf,
{\em {On T-duality and integrability for strings on AdS backgrounds},}
\href{http://dx.doi.org/10.1088/1126-6708/2007/12/082}{JHEP {\bf 0712} (2007)
  082} [{\tt \href{http://www.arxiv.org/abs/0711.0707}{0711.0707 [hep-th]}}].

\bibitem{OColgain:2012si}
E.~O~Colgain,
{\em {Fermionic T-duality: a snapshot review},}
\href{http://dx.doi.org/10.1142/S0217751X12300323}{Int. J. Mod. Phys. A {\bf
  27} (2012) 1230032} [{\tt \href{http://www.arxiv.org/abs/1210.5588}{1210.5588
  [hep-th]}}].

\bibitem{Babichenko:2009dk}
A.~Babichenko, B.~Stefanski, Jr., and K.~Zarembo,
{\em {Integrability and the AdS$_3$/CFT$_2$ correspondence},}
\href{http://dx.doi.org/10.1007/JHEP03(2010)058}{JHEP {\bf 03} (2010) 058}
  [{\tt \href{http://www.arxiv.org/abs/0912.1723}{0912.1723 [hep-th]}}].

\bibitem{Sorokin:2010wn}
D.~Sorokin and L.~Wulff,
{\em {Evidence for the classical integrability of the complete
  AdS$_4\times\mathbbm{C}P^3$ superstring},}
\href{http://dx.doi.org/10.1007/JHEP11(2010)143}{JHEP {\bf 1011} (2010) 143}
  [{\tt \href{http://www.arxiv.org/abs/1009.3498}{1009.3498 [hep-th]}}].

\bibitem{Sorokin:2011rr}
D.~Sorokin, A.~Tseytlin, L.~Wulff, and K.~Zarembo,
{\em {Superstrings in AdS$_2\times S^2 \times T^6$},}
\href{http://dx.doi.org/10.1088/1751-8113/44/27/275401}{J. Phys. A A {\bf 44}
  (2011) 275401} [{\tt \href{http://www.arxiv.org/abs/1104.1793}{1104.1793
  [hep-th]}}].

\bibitem{Cagnazzo:2011at}
A.~Cagnazzo, D.~Sorokin, and L.~Wulff,
{\em {More on integrable structures of superstrings in
  AdS$_4\times\mathbbm{C}P^3$ and AdS$_2\times S^2\times T^6$
  superbackgrounds},}
\href{http://dx.doi.org/10.1007/JHEP01(2012)004}{JHEP {\bf 1201} (2012) 004}
  [{\tt \href{http://www.arxiv.org/abs/1111.4197}{1111.4197 [hep-th]}}].

\bibitem{Sundin:2012gc}
P.~Sundin and L.~Wulff,
{\em {Classical integrability and quantum aspects of the AdS$_3\times S^3\times
  S^3\times S^1$ superstring},}
\href{http://dx.doi.org/10.1007/JHEP10(2012)109}{JHEP {\bf 1210} (2012) 109}
  [{\tt \href{http://www.arxiv.org/abs/1207.5531}{1207.5531 [hep-th]}}].

\bibitem{Wulff:2014kja}
L.~Wulff,
{\em {Superisometries and integrability of superstrings},}
\href{http://dx.doi.org/10.1007/JHEP05(2014)115}{JHEP {\bf 1405} (2014) 115}
  [{\tt \href{http://www.arxiv.org/abs/1402.3122}{1402.3122 [hep-th]}}].

\bibitem{Wulff:2015mwa}
L.~Wulff,
{\em {On integrability of strings on symmetric spaces},}
\href{http://dx.doi.org/10.1007/JHEP09(2015)115}{JHEP {\bf 09} (2015) 115}
  [{\tt \href{http://www.arxiv.org/abs/1505.03525}{1505.03525}}].

\bibitem{Adam:2009kt}
I.~Adam, A.~Dekel, and Y.~Oz,
{\em {On integrable backgrounds self-dual under fermionic T-duality},}
\href{http://dx.doi.org/10.1088/1126-6708/2009/04/120}{JHEP {\bf 04} (2009)
  120} [{\tt \href{http://www.arxiv.org/abs/0902.3805}{0902.3805 [hep-th]}}].

\bibitem{Grassi:2009yj}
P.~A.~Grassi, D.~Sorokin, and L.~Wulff,
{\em {Simplifying superstring and D-brane actions in
  AdS$_4\times\mathbbm{C}P^3$ superbackground},}
\href{http://dx.doi.org/10.1088/1126-6708/2009/08/060}{JHEP {\bf 08} (2009)
  060} [{\tt \href{http://www.arxiv.org/abs/0903.5407}{0903.5407 [hep-th]}}].

\bibitem{Bakhmatov:2009be}
I.~Bakhmatov and D.~S.~Berman,
{\em {Exploring fermionic T-duality},}
\href{http://dx.doi.org/10.1016/j.nuclphysb.2010.01.026}{Nucl. Phys. B {\bf
  832} (2010)~89} [{\tt \href{http://www.arxiv.org/abs/0912.3657}{0912.3657
  [hep-th]}}].

\bibitem{Godazgar:2010ph}
H.~Godazgar and M.~J.~Perry,
{\em {Real fermionic symmetry in type II supergravity},}
\href{http://dx.doi.org/10.1007/JHEP01(2011)032}{JHEP {\bf 01} (2011) 032}
  [{\tt \href{http://www.arxiv.org/abs/1008.3128}{1008.3128 [hep-th]}}].

\bibitem{Bakhmatov:2010fp}
I.~Bakhmatov,
{\em {On AdS$_4\times\mathbbm{C}P^3$ T-duality},}
\href{http://dx.doi.org/10.1016/j.nuclphysb.2011.01.020}{Nucl. Phys. B {\bf
  847} (2011)~38} [{\tt \href{http://www.arxiv.org/abs/1011.0985}{1011.0985
  [hep-th]}}].

\bibitem{Huang:2010qy}
Y.~t.~Huang and A.~Lipstein,
{\em {Dual superconformal symmetry of $\mathcal{N}=6$ Chern--Simons theory },}
\href{http://dx.doi.org/10.1007/JHEP11(2010)076}{JHEP {\bf 1011} (2010) 076}
  [{\tt \href{http://www.arxiv.org/abs/1008.0041}{1008.0041 [hep-th]}}].

\bibitem{Gang:2010gy}
D.~Gang, Y.~t.~Huang, E.~Koh, S.~Lee, and A.~Lipstein,
{\em {Tree-level recursion relation and dual superconformal symmetry of the
  ABJM theory},}
\href{http://dx.doi.org/10.1007/JHEP03(2011)116}{JHEP {\bf 1103} (2011) 116}
  [{\tt \href{http://www.arxiv.org/abs/1012.5032}{1012.5032 [hep-th]}}].

\bibitem{Chen:2011vv}
W.~Chen and Y.~t.~Huang,
{\em {Dualities for loop amplitudes of $\mathcal{N}=6$ Chern--Simons matter
  theory},}
\href{http://dx.doi.org/10.1007/JHEP11(2011)057}{JHEP {\bf 1111} (2011) 057}
  [{\tt \href{http://www.arxiv.org/abs/1107.2710}{1107.2710 [hep-th]}}].

\bibitem{Bianchi:2011fc}
M.~Bianchi, M.~Leoni, A.~Mauri, S.~Penati, and A.~Santambrogio,
{\em {Scattering in ABJ theories},}
\href{http://dx.doi.org/10.1007/JHEP12(2011)073}{JHEP {\bf 1112} (2011) 073}
  [{\tt \href{http://www.arxiv.org/abs/1110.0738}{1110.0738 [hep-th]}}].

\bibitem{Bargheer:2010hn}
T.~Bargheer, F.~Loebbert, and C.~Meneghelli,
{\em {Symmetries of tree-level scattering amplitudes in $\mathcal{N}=6$
  superconformal Chern--Simons theory},}
\href{http://dx.doi.org/10.1103/PhysRevD.82.045016}{Phys. Rev. D {\bf 82}
  (2010) 045016} [{\tt \href{http://www.arxiv.org/abs/1003.6120}{1003.6120
  [hep-th]}}].

\bibitem{Bianchi:2011rn}
M.~Bianchi, M.~Leoni, A.~Mauri, S.~Penati, C.~Ratti, and A.~Santambrogio,
{\em {From correlators to Wilson loops in Chern--Simons matter theories},}
\href{http://dx.doi.org/10.1007/JHEP06(2011)118}{JHEP {\bf 1106} (2011) 118}
  [{\tt \href{http://www.arxiv.org/abs/1103.3675}{1103.3675 [hep-th]}}].

\bibitem{Bianchi:2011dg}
M.~Bianchi, M.~Leoni, A.~Mauri, S.~Penati, and A.~Santambrogio,
{\em {Scattering amplitudes/Wilson loop duality in ABJM theory},}
\href{http://dx.doi.org/10.1007/JHEP01(2012)056}{JHEP {\bf 1201} (2012) 056}
  [{\tt \href{http://www.arxiv.org/abs/1107.3139}{1107.3139 [hep-th]}}].

\bibitem{Gomis:2008jt}
J.~Gomis, D.~Sorokin, and L.~Wulff,
{\em {The complete AdS$_4\times\mathbbm{C}P^3$ superspace for the type IIA
  superstring and D-branes},}
\href{http://dx.doi.org/10.1088/1126-6708/2009/03/015}{JHEP {\bf 03} (2009)
  015} [{\tt \href{http://www.arxiv.org/abs/0811.1566}{0811.1566 [hep-th]}}].

\bibitem{Adam:2010hh}
I.~Adam, A.~Dekel, and Y.~Oz,
{\em {On the fermionic T-duality of the AdS$_4\times\mathbbm{C}P^3$
  sigma-model},}
\href{http://dx.doi.org/10.1007/JHEP10(2010)110}{JHEP {\bf 1010} (2010) 110}
  [{\tt \href{http://www.arxiv.org/abs/1008.0649}{1008.0649 [hep-th]}}].

\bibitem{Oda:2001zm}
I.~Oda and M.~Tonin,
{\em {On the Berkovits covariant quantization of GS superstring},}
\href{http://dx.doi.org/10.1016/S0370-2693(01)01131-5}{Phys. Lett. B {\bf 520}
  (2001) 398} [{\tt
  \href{http://www.arxiv.org/abs/hep-th/0109051}{hep-th/0109051}}].

\bibitem{Rughoonauth:2012qd}
N.~Rughoonauth, P.~Sundin, and L.~Wulff,
{\em {Near BMN dynamics of the AdS$_3\times S^3\times S^3\times S^1$
  superstring},}
\href{http://dx.doi.org/10.1007/JHEP07(2012)159}{JHEP {\bf 1207} (2012) 159}
  [{\tt \href{http://www.arxiv.org/abs/1204.4742}{1204.4742 [hep-th]}}].

\bibitem{OColgain:2012ca}
E.~O~Colgain,
{\em {Self-duality of the D1-D5 near-horizon},}
\href{http://dx.doi.org/10.1007/JHEP04(2012)047}{JHEP {\bf 1204} (2012) 047}
  [{\tt \href{http://www.arxiv.org/abs/1202.3416}{1202.3416 [hep-th]}}].

\bibitem{Buscher:1987sk}
T.~Buscher,
{\em {A symmetry of the string background field equations},}
\href{http://dx.doi.org/10.1016/0370-2693(87)90769-6}{Phys. Lett. B {\bf 194}
  (1987)~59}.

\bibitem{Buscher:1987qj}
T.~H.~Buscher,
{\em {Path integral derivation of quantum duality in non-linear sigma models},}
\href{http://dx.doi.org/10.1016/0370-2693(88)90602-8}{Phys. Lett. B {\bf 201}
  (1988) 466}.

\bibitem{Simon:1998az}
J.~Simon,
{\em {T-duality and effective D-brane actions},}
\href{http://dx.doi.org/10.1103/PhysRevD.61.047702}{Phys. Rev. D {\bf 61}
  (2000) 047702} [{\tt
  \href{http://www.arxiv.org/abs/hep-th/9812095}{hep-th/9812095}}].

\bibitem{Grisaru:1985fv}
M.~T.~Grisaru, P.~S.~Howe, L.~Mezincescu, B.~Nilsson, and P.~Townsend,
{\em {$\mathcal{N}=2$ superstrings in a supergravity background},}
\href{http://dx.doi.org/10.1016/0370-2693(85)91071-8}{Phys. Lett. B {\bf 162}
  (1985) 116}.

\bibitem{Wulff:2013kga}
L.~Wulff,
{\em {The type II superstring to order $\theta^4$},}
\href{http://dx.doi.org/10.1007/JHEP07(2013)123}{JHEP {\bf 1307} (2013) 123}
  [{\tt \href{http://www.arxiv.org/abs/1304.6422}{1304.6422 [hep-th]}}].

\bibitem{Serganova:1983vp}
V.~Serganova,
{\em {Classification of real simple Lie superalgebras and symmetric
  superspaces},}
\href{http://dx.doi.org/10.1007/BF01078102}{Funct. Anal. Appl. {\bf 17} (1983)
  200}.

\bibitem{Frappat:1996pb}
L.~Frappat, P.~Sorba, and A.~Sciarrino,
{\em {Dictionary on Lie superalgebras},}
{\tt \href{http://www.arxiv.org/abs/hep-th/9607161}{hep-th/9607161}}.

\bibitem{Metsaev:1998it}
R.~R.~Metsaev and A.~A.~Tseytlin,
{\em {Type IIB superstring action in AdS$_5\times S^5$ background},}
\href{http://dx.doi.org/10.1016/S0550-3213(98)00570-7}{Nucl. Phys. B {\bf 533}
  (1998) 109} [{\tt
  \href{http://www.arxiv.org/abs/hep-th/9805028}{hep-th/9805028}}].

\bibitem{Rahmfeld:1998zn}
J.~Rahmfeld and A.~Rajaraman,
{\em {The GS string action on AdS$_3 \times S^3$ with Ramond--Ramond charge},}
\href{http://dx.doi.org/10.1103/PhysRevD.60.064014}{Phys. Rev. D {\bf 60}
  (1999) 064014} [{\tt
  \href{http://www.arxiv.org/abs/hep-th/9809164}{hep-th/9809164}}].

\bibitem{Zhou:1999sm}
J.-G.~Zhou,
{\em {Super 0-brane and GS superstring actions on AdS$_2\times S^2$},}
\href{http://dx.doi.org/10.1016/S0550-3213(99)00462-9}{Nucl. Phys. B {\bf 559}
  (1999)~92} [{\tt
  \href{http://www.arxiv.org/abs/hep-th/9906013}{hep-th/9906013}}].

\bibitem{Berkovits:1999zq}
N.~Berkovits, M.~Bershadsky, T.~Hauer, S.~Zhukov, and B.~Zwiebach,
{\em {Superstring theory on AdS$_2\times S^2$ as a coset supermanifold},}
\href{http://dx.doi.org/10.1016/S0550-3213(99)00683-5}{Nucl. Phys. B {\bf 567}
  (2000)~61} [{\tt
  \href{http://www.arxiv.org/abs/hep-th/9907200}{hep-th/9907200}}].

\bibitem{Arutyunov:2008if}
G.~Arutyunov and S.~Frolov,
{\em {Superstrings on $AdS_4 \times \mathbbm{C}P^3$ as a coset sigma model},}
\href{http://dx.doi.org/10.1088/1126-6708/2008/09/129}{JHEP {\bf 09} (2008)
  129} [{\tt \href{http://www.arxiv.org/abs/0806.4940}{0806.4940 [hep-th]}}].

\bibitem{Stefanski:2008ik}
B.~Stefanski~Jr.,
{\em {Green--Schwarz action for type IIA strings on
  AdS$_4\times\mathbbm{C}P^3$},}
\href{http://dx.doi.org/10.1016/j.nuclphysb.2008.09.015}{Nucl. Phys. B {\bf
  808} (2009)~80} [{\tt \href{http://www.arxiv.org/abs/0806.4948}{0806.4948
  [hep-th]}}].

\bibitem{Zarembo:2010sg}
K.~Zarembo,
{\em {Strings on semi-symmetric superspaces},}
\href{http://dx.doi.org/10.1007/JHEP05(2010)002}{JHEP {\bf 1005} (2010) 002}
  [{\tt \href{http://www.arxiv.org/abs/1003.0465}{1003.0465 [hep-th]}}].

\bibitem{Dekel:2011qw}
A.~Dekel and Y.~Oz,
{\em {Self-duality of Green--Schwarz sigma models},}
\href{http://dx.doi.org/10.1007/JHEP03(2011)117}{JHEP {\bf 03} (2011) 117}
  [{\tt \href{http://www.arxiv.org/abs/1101.0400}{1101.0400 [hep-th]}}].

\bibitem{Schwarz:1992te}
A.~S.~Schwarz and A.~A.~Tseytlin,
{\em {Dilaton shift under duality and torsion of elliptic complex},}
\href{http://dx.doi.org/10.1016/0550-3213(93)90514-P}{Nucl. Phys. B {\bf 399}
  (1993) 691} [{\tt
  \href{http://www.arxiv.org/abs/hep-th/9210015}{hep-th/9210015}}].

\bibitem{Claus:1998mw}
P.~Claus, R.~Kallosh, J.~Kumar, P.~K.~Townsend, and A.~Van~Proeyen,
{\em {Conformal theory of M2, D3, M5 and D1-branes + D5-branes},}
\href{http://dx.doi.org/10.1088/1126-6708/1998/06/004}{JHEP {\bf 9806} (1998)
  004} [{\tt \href{http://www.arxiv.org/abs/hep-th/9801206}{hep-th/9801206}}].

\bibitem{Tseytlin:1996bh}
A.~A.~Tseytlin,
{\em {Harmonic superpositions of M-branes},}
\href{http://dx.doi.org/10.1016/0550-3213(96)00328-8}{Nucl. Phys. B {\bf 475}
  (1996) 149} [{\tt
  \href{http://www.arxiv.org/abs/hep-th/9604035}{hep-th/9604035}}].

\bibitem{Klebanov:1996mh}
I.~R.~Klebanov and A.~A.~Tseytlin,
{\em {Intersecting M-branes as four-dimensional black holes},}
\href{http://dx.doi.org/10.1016/0550-3213(96)00338-0}{Nucl. Phys. B {\bf 475}
  (1996) 179} [{\tt
  \href{http://www.arxiv.org/abs/hep-th/9604166}{hep-th/9604166}}].

\bibitem{Gauntlett:1996pb}
J.~P.~Gauntlett, D.~A.~Kastor, and J.~H.~Traschen,
{\em {Overlapping branes in M theory},}
\href{http://dx.doi.org/10.1016/0550-3213(96)00423-3}{Nucl. Phys. B {\bf 478}
  (1996) 544} [{\tt
  \href{http://www.arxiv.org/abs/hep-th/9604179}{hep-th/9604179}}].

\bibitem{Duff:1998us}
M.~J.~Duff, H.~Lu, and C.~N.~Pope,
{\em {AdS$_5\times S^5$ untwisted},}
\href{http://dx.doi.org/10.1016/S0550-3213(98)00464-7}{Nucl. Phys. B {\bf 532}
  (1998) 181} [{\tt
  \href{http://www.arxiv.org/abs/hep-th/9803061}{hep-th/9803061}}].

\bibitem{Boonstra:1998yu}
H.~J.~Boonstra, B.~Peeters, and K.~Skenderis,
{\em {Brane intersections, anti-de Sitter space-times and dual superconformal
  theories},}
\href{http://dx.doi.org/10.1016/S0550-3213(98)00512-4}{Nucl. Phys. B {\bf 533}
  (1998) 127} [{\tt
  \href{http://www.arxiv.org/abs/hep-th/9803231}{hep-th/9803231}}].

\bibitem{Lee:1999yu}
J.~Lee and S.~Lee,
{\em {Mass spectrum of $D=11$ supergravity on AdS$_2\times S^2\times T^7$ },}
\href{http://dx.doi.org/10.1016/S0550-3213(99)00598-2}{Nucl. Phys. B {\bf 563}
  (1999) 125} [{\tt
  \href{http://www.arxiv.org/abs/hep-th/9906105}{hep-th/9906105}}].

\bibitem{Cowdall:1998bu}
P.~Cowdall and P.~Townsend,
{\em {Gauged supergravity vacua from intersecting branes},}
\href{http://dx.doi.org/10.1016/S0370-2693(98)00445-6}{Phys. Lett. B {\bf 429}
  (1998) 281} [{\tt
  \href{http://www.arxiv.org/abs/hep-th/9801165}{hep-th/9801165}}].

\bibitem{Gauntlett:1998kc}
J.~P.~Gauntlett, R.~C.~Myers, and P.~Townsend,
{\em {Supersymmetry of rotating branes},}
\href{http://dx.doi.org/10.1103/PhysRevD.59.025001}{Phys. Rev. D {\bf 59}
  (1998) 025001} [{\tt
  \href{http://www.arxiv.org/abs/hep-th/9809065}{hep-th/9809065}}].

\bibitem{deBoer:1999rh}
J.~de~Boer, A.~Pasquinucci, and K.~Skenderis,
{\em {AdS / CFT dualities involving large $2-D$ $\mathcal{N}=4$ superconformal
  symmetry},}
Adv. Theor. Math. Phys. {\bf 3} (1999) 577 [{\tt
  \href{http://www.arxiv.org/abs/hep-th/9904073}{hep-th/9904073}}].

\bibitem{Gukov:2004ym}
S.~Gukov, E.~Martinec, G.~W.~Moore, and A.~Strominger,
{\em {The search for a holographic dual to AdS$_3\times S^3\times S^3\times
  S^1$},}
\href{http://dx.doi.org/10.4310/ATMP.2005.v9.n3.a3}{Adv. Theor. Math. Phys.
  {\bf 9} (2005) 435} [{\tt
  \href{http://www.arxiv.org/abs/hep-th/0403090}{hep-th/0403090}}].

\bibitem{Ivanov:2015iia}
E.~Ivanov, S.~Sidorov, and F.~Toppan,
{\em {Superconformal mechanics in $SU(2|1)$ superspace},}
\href{http://dx.doi.org/10.1103/PhysRevD.91.085032}{Phys. Rev. D {\bf 91}
  (2015) 085032} [{\tt
  \href{http://www.arxiv.org/abs/1501.05622}{1501.05622}}].

\bibitem{Bandos:2002nn}
I.~A.~Bandos, E.~Ivanov, J.~Lukierski, and D.~Sorokin,
{\em {On the superconformal flatness of AdS superspaces},}
\href{http://dx.doi.org/10.1088/1126-6708/2002/06/040}{JHEP {\bf 06} (2002)
  040} [{\tt \href{http://www.arxiv.org/abs/hep-th/0205104}{hep-th/0205104}}].

\bibitem{Butter:2015tra}
D.~Butter, G.~Inverso, and I.~Lodato,
{\em {Rigid 4D $\mathcal{N}=2$ supersymmetric backgrounds and actions},}
{\tt \href{http://www.arxiv.org/abs/1505.03500}{1505.03500}}.

\bibitem{Bandos:2003bz}
I.~A.~Bandos and B.~Julia,
{\em {Superfield T-duality rules},}
\href{http://dx.doi.org/10.1088/1126-6708/2003/08/032}{JHEP {\bf 0308} (2003)
  032} [{\tt \href{http://www.arxiv.org/abs/hep-th/0303075}{hep-th/0303075}}].

\bibitem{Forini:2012bb}
V.~Forini, V.~G.~M.~Puletti, and O.~Ohlsson~Sax,
{\em {The generalized cusp in AdS$_4\times\mathbbm{C}P^3$ and more one-loop
  results from semiclassical strings},}
\href{http://dx.doi.org/10.1088/1751-8113/46/11/115402}{J. Phys. A {\bf 46}
  (2013) 115402} [{\tt \href{http://www.arxiv.org/abs/1204.3302}{1204.3302
  [hep-th]}}].

\bibitem{Alvarez:1994dn}
E.~Alvarez, L.~Alvarez-Gaume, and Y.~Lozano,
{\em {An Introduction to T-duality in string theory},}
\href{http://dx.doi.org/10.1016/0920-5632(95)00429-D}{Nucl. Phys. Proc. Suppl.
  {\bf 41} (1995)~1} [{\tt
  \href{http://www.arxiv.org/abs/hep-th/9410237}{hep-th/9410237}}].

\bibitem{Fukuma:1999jt}
M.~Fukuma, T.~Oota, and H.~Tanaka,
{\em {Comments on T-dualities of Ramond--Ramond potentials on tori},}
\href{http://dx.doi.org/10.1143/PTP.103.425}{Prog. Theor. Phys. {\bf 103}
  (2000) 425} [{\tt
  \href{http://www.arxiv.org/abs/hep-th/9907132}{hep-th/9907132}}].

\bibitem{Hassan:1999bv}
S.~F.~Hassan,
{\em {T-duality, space-time spinors and RR fields in curved backgrounds},}
\href{http://dx.doi.org/10.1016/S0550-3213(99)00684-7}{Nucl. Phys. B {\bf 568}
  (2000) 145} [{\tt
  \href{http://www.arxiv.org/abs/hep-th/9907152}{hep-th/9907152}}].

\bibitem{Hull:1998vg}
C.~M.~Hull,
{\em {Time-like T-duality, de Sitter space, large-N gauge theories and
  topological field theory},}
\href{http://dx.doi.org/10.1088/1126-6708/1998/07/021}{JHEP {\bf 07} (1998)
  021} [{\tt \href{http://www.arxiv.org/abs/hep-th/9806146}{hep-th/9806146}}].

\bibitem{Cagnazzo:2012uq}
A.~Cagnazzo, D.~Sorokin, A.~A.~Tseytlin, and L.~Wulff,
{\em {Semi-classical equivalence of Green--Schwarz and pure-spinor/hybrid
  formulations of superstrings in AdS$_5\times S^5$ and AdS$_2\times S^2\times
  T^6$},}
\href{http://dx.doi.org/10.1088/1751-8113/46/6/065401}{J. Phys. A {\bf 46}
  (2013) 065401} [{\tt \href{http://www.arxiv.org/abs/1211.1554}{1211.1554
  [hep-th]}}].
}
\end{thebibliography}
\end{document}